%% file: BPFuzz.tex
\documentclass[conference]{IEEEtran}
\usepackage[T1]{fontenc}
\usepackage{fancyhdr}

\usepackage{cite}

\newcommand{\tofix}[1]{\textcolor{red}{[TO FIX: #1]}}

\newcommand{\alfred}[1]{\textcolor{red}{[Alfred: #1]}}

\newcommand{\ziwenres}[1]{\textcolor{blue}{[Ziwen's response: #1]}}

\newcommand{\original}[1]{}

\newcommand{\junjie}[1]{\textcolor{red}{[Junjie: #1]}}
\newcommand{\add}[1]{\textcolor{blue}{[#1]}}

\usepackage[utf8]{inputenc}
\usepackage{makecell}
\usepackage[english]{babel}
\usepackage[frozencache=true,cachedir=.]{minted}
\usepackage{graphicx}
\usepackage{booktabs}
\usepackage{verbatim}
\usepackage{listings}
\usepackage{multicol}
\usepackage{multirow}
\usepackage{cancel}
\usepackage{ulem}
\usepackage{url}
\usepackage{xcolor}
\usepackage{threeparttable}
\usepackage{longtable}

\usepackage[]{hyperref}
\hypersetup{  
  colorlinks,
  linkcolor={green!80!black},
  citecolor={red!70!black},
  urlcolor={blue!70!black}
}
\usepackage{amsmath}
\usepackage{breakurl}

\usepackage[font=footnotesize]{caption}

\definecolor{shockingpink}{rgb}{0.99, 0.06, 0.75}
\newcommand{\josh}[1]{{\color{shockingpink}(Josh: #1)}}
\definecolor{vividviolet}{rgb}{0.62, 0.0, 1.0}
\newcommand{\jhl}[1]{{\color{vividviolet}#1}}

\newcommand{\cut}[1]{}
\newcommand{\moved}[1]{}

\newcommand{\red}[1]{\textcolor{red}{#1}}

\iffalse
\newcommand{\newparts}[1]{{\color{blue}#1}}
\else
\newcommand{\newparts}[1]{{#1}}
\fi

\usepackage[english]{babel}

\usepackage{mdframed}
\mdfdefinestyle{rebuttalstyle}{innerleftmargin=3pt, innerrightmargin=3pt, innertopmargin=3pt, innerbottommargin=3pt}

\newcounter{response}[section]

\newcounter{revision}[section]

\newcounter{link}[section]

\newcounter{comments}[section]

\ifCLASSINFOpdf
\else
\fi
\iftrue

\newcommand{\nsection}[1]{\section{#1}}
\newcommand{\nsubsection}[1]{\subsection{#1}}
\newcommand{\nsubsubsection}[1]{\subsubsection{#1}}

\else

\newcommand{\nsection}[1]{\vspace{-0.1cm}\section{#1}\vspace{-0.1cm}}
\newcommand{\nsubsection}[1]{\vspace{-0.19cm}\subsection{#1}\vspace{-0.1cm}}
\newcommand{\nsubsubsection}[1]{\vspace{-0.2cm}\subsubsection{#1}\vspace{-0.1cm}}

\fi

\begin{document}
%
\title{Too Afraid to Drive: Systematic Discovery of Semantic DoS Vulnerability in Autonomous Driving Planning under Physical-World Attacks} 



%

\author{\IEEEauthorblockN{{\rm Ziwen Wan}$^{}$\quad {\rm Junjie Shen}$^{}$ \quad {\rm Jalen Chuang}$^{}$ \quad {\rm Xin Xia}$^{\dagger}$ \quad {\rm Joshua Garcia}$^{}$ \quad {\rm Jiaqi Ma}$^{\dagger}$ \quad {\rm Qi Alfred Chen}$^{}$}
\IEEEauthorblockA{$^{}$University of California, Irvine\quad $^{\dagger}$University of California, Los Angeles\\
$^{}$\{ziwenw8, junjies1, jzchuang, joshua.garcia, alfchen\}@uci.edu}\quad 
$^{\dagger}$\{x35xia, jiaqima\}@ucla.edu
}



\maketitle
\thispagestyle{fancy}
\fancyhead[C]{A version of this paper is in the Proceedings of \textit{the Network and Distributed System Security Symposium (NDSS) 2022}}

\input{abstract}
\input{introduction}

\input{background}
\input{motivation}

\input{design}
\input{evaluation}

\input{case_study}
\input{discussion}
\input{related_work}

\input{conclusion}
\input{acknowledge}


\bibliographystyle{ieeetr}{
\bibliography{bibs/AV_security.bib,bibs/planning_survey.bib,./bibs/general.bib, ./bibs/AV_testing.bib, ./bibs/vuln_discovery.bib, ./bibs/RV.bib, ./bibs/vuln_code.bib}
}

\input{appendix}

\end{document}

%% file: abstract.tex
\begin{abstract}
In high-level Autonomous Driving (AD) systems, behavioral planning is in charge of making high-level driving decisions such as cruising and stopping, and thus highly \newparts{security-critical}. In this work, we perform the first systematic study of semantic security vulnerabilities specific to \newparts{overly-conservative} AD behavioral planning behaviors, i.e., those that can cause failed or significantly-degraded mission performance, which can be critical for AD services such as robo-taxi/delivery. We call them semantic Denial-of-Service (DoS) vulnerabilities, which we envision to be most generally exposed in practical AD systems due to the tendency for conservativeness to avoid safety incidents. To achieve high practicality and realism, we assume that the attacker can only introduce seemingly-benign external physical objects to the driving environment, e.g., off-road dumped cardboard boxes.


To systematically discover such vulnerabilities, we design PlanFuzz, a novel dynamic testing approach that addresses various problem-specific design challenges. Specifically, we propose and identify planning invariants as novel testing oracles, and design new input generation to systematically enforce problem-specific constraints for attacker-introduced physical objects. We also design a novel behavioral planning vulnerability distance metric to effectively guide the discovery. We evaluate PlanFuzz on 3 planning implementations from practical open-source AD systems, and find that it can effectively discover 9 previously-unknown semantic DoS vulnerabilities without false positives. We find all our new designs necessary, \newparts{as without each design, statistically significant performance drops are generally observed.} We further perform exploitation case studies using simulation and real-vehicle traces. We discuss root causes and potential fixes.

\end{abstract}


%% file: introduction.tex
\nsection{Introduction}\label{sec:intro}

Today, various companies are developing high-level (e.g., Level-4~\cite{sae2018}) Autonomous Driving (AD) vehicles. Some of them, e.g., Google Waymo~\cite{waymo_service}, TuSimple~\cite{tusimple-truck}, and Pony.ai~\cite{ponyai_service}, are already providing services on public roads. To enable such highly-automated driving, after the environmental sensing steps such as perception and localization, the AD systems need to use the sensed information to make high-level driving decisions such as cruising, stopping, lane changing, etc., that not only are safe and efficient, but also conform to driving norms such as traffic rules. Such a decision-making process is commonly referred to as \textit{behavioral planning}, which is highly security critical as any mistakes made in it can directly lead to undesired driving behaviors such as driving too aggressively to cause collisions, or too conservatively to cause unnecessary emergency stops and road blocking. Various prior works studied security vulnerabilities in AD systems with such semantic consequences~\cite{junjie:usenix:2020, ningfei-msf-adv,cao2019adversarial, sun2020towards, zhao2018seeing}, but \newparts{they mostly focus on} environmental sensing errors (e.g., camera/LiDAR object detection~\cite{zhao2018seeing, ningfei-msf-adv,cao2019adversarial, sun2020towards}) instead of planning. \newparts{There are recent works able to discover semantic planning errors~\cite{fremont2020formal2}, but they (1) are designed for whole-system testing instead of being planning-specific; (2) only consider overly-aggressive behaviors, leaving the overly-conservative side unexplored; and (3) focus on safety instead of security (i.e., lack explicit threat model considerations).} 



To fill this research gap, in this paper we perform the first AD planning-specific semantic vulnerability discovery. We specifically choose to focus on the more under-explored \textit{overly-conservative} behavioral planning behaviors, especially those that can cause the victim AD vehicle to have a failed or significantly-degraded mission performance (e.g., permanent stop and never reach the destination). We refer to these as \textit{semantic Denial-of-Service (DoS) vulnerabilities} for behavioral planning. We envision that such vulnerabilities can be most generally exposed in practice, based on the hypothesis that behavioral planning in real-world AD systems, especially in production settings, will generally try to be as conservative as possible to avoid any possible safety incidents, which can cause great business reputation and financial damages as in recent fatal accidents for Uber and Tesla~\cite{uber_crash, tesla_crash_truck,tesla_crash_texas, tesla_mich}. In fact, such overly-conservative behaviors have already been observed for many production AD vehicles (e.g., Waymo, Uber, Volvo~\cite{youtube_stuck_1, youtube_stuck_2, youtube_stuck_3, mit_tech_review, uber_bully, volvo_bully}), causing troubles to AD service and traffic flow (\S\ref{subsec:attack_goal}).




Considering the realism and generality of such problems in practice, we set our goal to develop an automated system to systematically discover such vulnerabilities to most generally help address these problems at the AD system development stage. To achieve high practicality and realism, we assume that the attacker can only introduce seemingly-benign external physical objects to the driving environment, e.g., dumped cardboard boxes or parked bikes on \newparts{the road side}, attacker-driven vehicles, etc. \original{To achieve automated discovery, we adopt a dynamic testing approach, which shows high generality, effectiveness, and efficiency for semantic vulnerability discovery recently~\cite{petsios2017nezha, brubaker2014using, pei2017deepxplore,tian2018deeptest, kim2019rvfuzzer, choi2020cyber,kim2020control,kimpgfuzz}.} \newparts{Dynamic testing is a promising approach to achieve domain-specific vulnerability discovery in general~\cite{petsios2017nezha, brubaker2014using, pei2017deepxplore,tian2018deeptest, kim2019rvfuzzer, choi2020cyber,kim2020control,kimpgfuzz}.} However, none of the existing designs can be directly applied to our problem due to several unique design challenges specific to our problem definition: (C1) Lack of testing oracles to tell whether a change of \newparts{a} planning decision is \textit{overly} conservative or not. For example, directly putting obstacles ahead of the victim to cause DoS is not a vulnerability for behavioral planning; (C2) Need to systematically generate attacker-introduced physical objects following problem-specific physical constraints, e.g., avoiding road regions directly ahead of the victim as explained above; and (C3) Need to obtain fine-grained \newparts{code-level} feedback from the planning decision-making process to guide our vulnerability discovery, which is highly desired for us as the direct behavioral planning output is usually quite discrete (\S\ref{subsec:design_challenge}).




To achieve our goal, we design PlanFuzz, a novel dynamic testing approach that systematically addresses the aforementioned design challenges in an evolutionary testing framework. To address C1, we propose and identify \textit{Planning Invariant (PI)} as the problem-specific testing oracle, which defines a set of constraints for the attacker-introduced physical objects based on common driving norms such that if satisfied, the behavior planning should not give up the desired planning decision. To address C2, we design \textit{PI-aware physical-object generation}, which can systematically enforce the generated testing inputs to conform to both driving norms (e.g., traffic rules) and the PI constraints above, while maintaining diversity and inheritance properties desired for evolutionary testing. To address C3, we design \textit{behavioral planning vulnerability distance} to measure how close the current planning decision is to violate PI and thus trigger a vulnerability in the run time.




We implement a prototype of PlanFuzz and evaluate it on 3 different behavioral planning implementations from two open-source AD systems, Apollo~\cite{apollo} and Autoware~\cite{autoware}, which are both practical AD systems with full-stack implementations~\cite{AVtop4, apollo_news_taxi, carma-platform}. We use LGSVL, an industry-grade AD simulator, to generate diverse driving scenarios, which allows us to obtain 11,912 different initial testing seeds in total for 8 different driving scenarios. Using PlanFuzz with these seeds, all 3 behavioral planning implementations are found vulnerable, with 9 previously-unknown semantic DoS vulnerabilities discovered in total. Among them, 8 can prevent the victim from reaching the destination (7 can cause permanent stop), and the remaining 1 can cause emergency stop. \newparts{We also perform baseline comparisons by replacing different key components in our design, which shows statistically significant performance drops for almost all of the 9 vulnerabilities, leading to over 3.5$\times$ average slow-down or even failure in their discoveries.}
We also manually verified that no false positives are generated. 

To concretely understand the end-to-end impacts of the discovered vulnerabilities, we further perform 3 vulnerability exploitation case studies by constructing and evaluating real-world attack scenarios using simulation and real-vehicle sensor traces. The results show that the discovered vulnerabilities can cause the AD vehicle running Apollo or Autoware to (1) permanently stop in an empty road or in front of an empty intersection due to completely off-road cardboard boxes or parked bikes; or (2) give up necessary lane changing purely due to a following vehicle without any intention to change \newparts{lanes}. Demos are at our website \textbf{\url{https://sites.google.com/view/cav-sec/planfuzz}}~\cite{planfuzzwebsite}. We also discuss root causes and potential fixes. We also release our code at our website~\cite{planfuzzwebsite}.

In summary, this work makes the following contributions:
\begin{itemize}
\setlength{\itemsep}{0pt}
\setlength{\parskip}{0pt}
    \item To the best our knowledge, we are the first to perform AD planning-specific semantic vulnerability discovery. We focus on semantic DoS vulnerabilities, which can damage the availability of AD services. We formulate the problem with a domain-specific vulnerability definition and a practical threat model that only allows adding seemingly benign physical objects to the driving environment.
    
    \item To systematically discover the vulnerability, we design PlanFuzz, a novel dynamic testing approach that addresses various problem-specific design challenges. Specifically, we (1) propose and identify PIs as the testing oracle, (2) design a novel PI-aware physical object generation to systematically enforce problem-specific input constraints; and (3) design a behavioral planning vulnerability distance metric to effectively guide the discovery. 
    
    
    
    \item We evaluate PlanFuzz on 3 planning implementations from two practical open-source AD systems. We find that PlanFuzz can effectively discover DoS vulnerabilities in all 3 implementations without incurring false positives. In total, 9 previously-unknown semantic DoS vulnerabilities are discovered, which can all be exploited to either prevent the victim from reaching its destination or cause an emergency stop. We find all our main designs are necessary, \newparts{as without each design, statistically significant drops in performance are generally observed.}
    
    
    
    
    \item We further perform 3 vulnerability exploitation case studies using simulation. The results show that the discovered vulnerabilities can cause the AD vehicle to unnecessarily stop permanently or give up necessary driving decisions. We also discuss root causes and potential mitigations.

    
\end{itemize}

%% file: background.tex
\nsection{Background \& Problem Definition} \label{sec:background} 
\vspace{0.1in}
\nsubsection{Behavioral Planning (BP) in AD Systems}
\label{subsec:background_bp} 

\cut{

\begin{figure}
    \centering
    \includegraphics[width=0.99\linewidth]{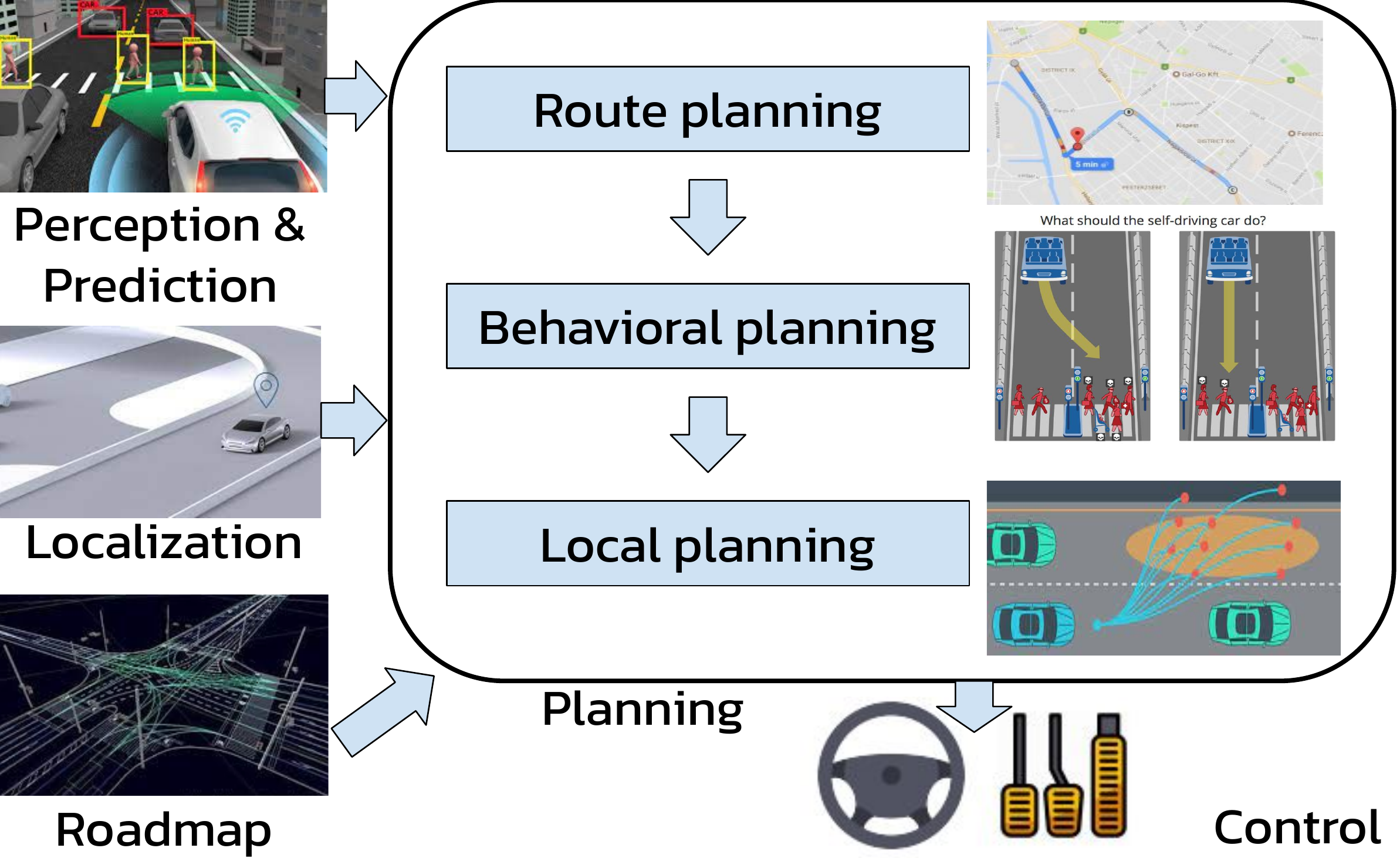}
    \caption{The framework of planning module.}
    \label{fig:planning_framework}
\end{figure}
}

\textbf{AD planning.} In high-level (e.g., Level-4~\cite{sae2018}) AD systems, \textit{planning} is a critical module designed to generate safe, efficient, and smooth driving trajectories to reach the destination. In industry-grade AD, such a module typically adopt a 3-layer design: \textit{route planning}, \textit{behavioral planning} (\textit{BP}), and \textit{local planning}~\cite{paden2016survey, ilievski2019design, urmson2008autonomous, buehler2009darpa, apollo, autoware,  gonzalez2015review, pendleton2017perception,schwarting2018planning}. Given the destination, \textit{route planning} selects a route from the map. To follow the selected route while ensuring safety and correctness (e.g., conform to traffic rules), \textit{BP} then makes high-level driving decisions such as cruising, \newparts{stopping}, lane changing, etc. based on the real-time driving environment. For example, when the AD vehicle needs to pass a signaled intersection, the BP layer needs to consider both the traffic light and the dynamic behaviors of surrounding vehicles and pedestrians to decide whether and when to proceed. Next, \textit{local planning} translates the high-level decisions into concrete low-level driving trajectories (e.g., waypoints), which will then be passed to the vehicle control module to actuate.

\textbf{Our focus: AD Behavior Planning (BP).} As described above, BP is at the core of AD decision making, and thus highly security/safety critical: any mistakes made in it can directly lead to undesired driving behaviors such as driving too \textit{aggressively}, which can cause collisions, or too \textit{conservatively}, which can cause unnecessary emergency stops and road blocking.
In BP, the decisions can be generated from \textit{programmed} or \textit{learned} logic. There are some recent works exploring learning-based planning~\cite{sun2018fast, chen2019learning,abbeel2008apprenticeship}, but so far they are all experimental (e.g., designed and evaluated only for simulated racing game setups~\cite{10.1145/2463372.2463509, wang2018deep, loiacono2010learning, capo2020short, sallab2017deep}) and generally lack the necessary capability and support for real-world driving (e.g., limited to only cruising without handling intersections, cross-walks, pedestrians~\cite{sallab2017deep, 9196730, kendall2019learning, kuefler2017imitating}, and no consideration of traffic rules~\cite{han2020neuro, bai2014integrated,brechtel2013solving, wray2017online,kuefler2017imitating, behbahani2019learning}). Such a learning-based approach is also generally known to suffer from difficulties with debugging and interpreting~\cite{chi2017deep}, and enforcing safety rules/measures~\cite{yurtsever2020survey}, while the latter is especially critical for production AD. Thus, the BP in today's industry-grade AD systems generally adopts programmed logic~\cite{udacity_planning, argo_safety_report, apollo, autoware}. Such program-based BP is thus also the main target of our design, which \newparts{raises} design challenges as detailed in~\S\ref{subsec:design_challenge}.

\nsubsection{Attack Goal and Incentives}\label{subsec:attack_goal}

\textbf{Attack goal: Semantic Denial-of-Service (DoS) of BP.} In this paper, we target an attack goal of causing \textit{semantic Denial-of-Service (DoS)} on BP, which we define as causing it to change a normal driving decision to an \textit{overly-conservative} one so that the victim AD vehicle will have a \textit{failed} or \textit{significantly-degraded} mission performance (e.g., never reach the destination). Specifically, we focus on 2 concrete types of such DoS in \newparts{an} AD context: (1) causing \newparts{an} emergency/permanent stop, and (2) causing the victim to give up a mission-critical driving decision, such as necessary left/right turns and lane changing on the route. To achieve this goal, in this paper we target \textit{physical-world attack vectors} in the AD context (\newparts{e.g.}, adding seemingly-benign static/dynamic physical road objects, detailed in~\S\ref{subsec:threat_model}) for high practicality and realism.

\begin{figure}
    \centering
    \includegraphics[width=0.99\linewidth]{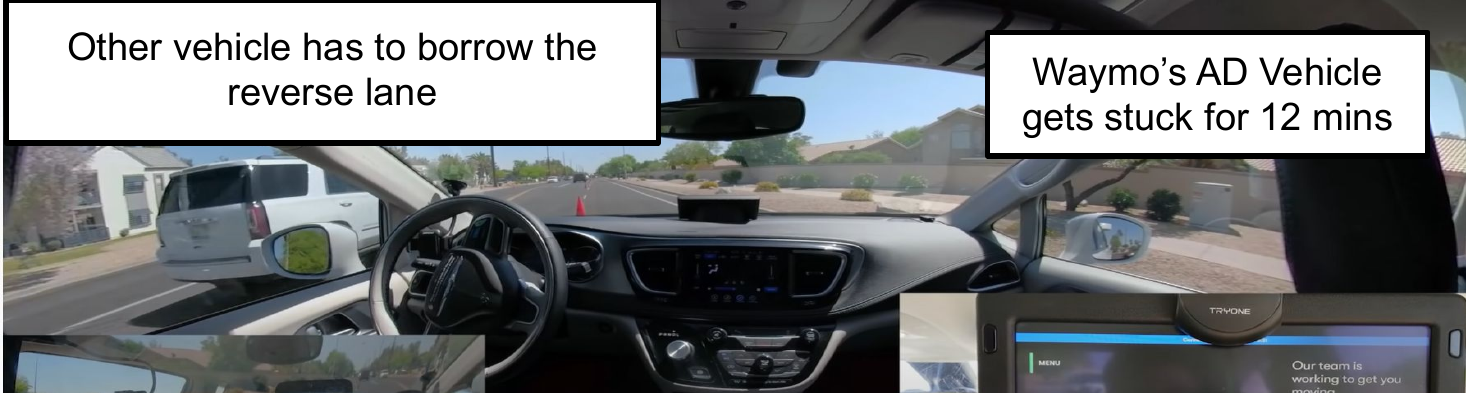}
    \vspace{-0.2in}
    \caption{Real-world overly-conservative driving behavior observed for a Waymo AD vehicle, which got stuck in the middle of the road and force other vehicles to borrow the reverse lane~\cite{youtube_stuck_3}.}
    \vspace{-0.15in}
    \label{fig:waymo_dos}
\end{figure}

We choose to focus on causing overly-conservative driving decisions instead of overly-aggressive ones because we hypothesize that real-world BP, especially those in production settings, will \textit{generally try to be as conservative as possible by design to avoid any possible safety incidents}. This is based on the fact that one single fatal crash (e.g., the Uber one~\cite{uber_crash} and increasingly more Tesla ones~\cite{tesla_crash_truck,tesla_crash_texas, tesla_mich}), no matter whether it is mainly due to AD system flaws or not, can cause great reputational damage, lawsuits, and business being paused or even sold~\cite{tesla_inves, uber_acquire}. 
In fact, such overly-conservative behaviors have \textit{already been observed in real-world production AD settings}, causing troubles for AD service and surrounding traffic flows.
For example, there are videos showing the AD vehicle from Waymo, a world-leading AD company, getting stuck by non-road-blocking traffic cones for $>$10 mins~\cite{youtube_stuck_3}, in the parking lot when no other moving objects are around~\cite{youtube_stuck_1}, and at an intersection \newparts{resulting in blocking} normal traffic~\cite{youtube_stuck_2}. A snapshot of such a problem is in Fig.~\ref{fig:waymo_dos}. \newparts{Since there is no official report on the root cause, we cannot be certain that this is caused by planning flaws/bugs; however, we suspect so since the environmental perception results are all correct according to the in-vehicle display~\cite{youtube_stuck_3}.} Furthermore, a few AD companies (e.g., Waymo, Volvo, Uber) have reported that their AD vehicles have been ``bullied'' by human drivers~\cite{mit_tech_review, uber_bully, volvo_bully}. For example, Waymo AD vehicle found it difficult to pull away from stop signs since they were too timid and were taken advantage \newparts{of} by passing human drivers~\cite{mit_tech_review}.

\textbf{Problem seriousness and incentives.} With the growing deployment of commercial AD services without safety drivers (e.g., by Waymo, Nuro, and Baidu~\cite{baidu_service, waymo_service_commerical, nuro_service}), such DoS problem can severely damage the availability of these services and thus ruin their user experience, reputation, and also revenues.
It can help if remote operators are available, but such helps are not necessarily effective. For example, in the Waymo video above (Fig.~\ref{fig:waymo_dos}), the AD vehicle called remote-operator help twice but it actually made the problem worse, and eventually called road assistance team to physically arrive, causing >10min trip delay in total. Such semantic DoS may also cause safety problems, e.g., by triggering emergency brakes in dangerous road segments like highway exit ramps, or blocking the road and thus forcing other vehicles to borrow the reverse lane like in Fig.~\ref{fig:waymo_dos}. Since such consequences can at least damage the reputation of the victim AD company, one potential attack incentive is for business competition (e.g., by a rival AD company to unfairly gain competitive advantages). Also, from the transportation system's perspective, such attacks can damage traffic mobility and city functions (e.g., by causing traffic jams), and also waste fuel and \newparts{an} individual's time~\cite{traffic_jam_cost}. Similar to DoS attacks on the Internet, such attacks may be politically or financially incentivized~\cite{chen2018exposing}.

\begin{figure}[t] 
      \centering
          \includegraphics[width=1.\linewidth]{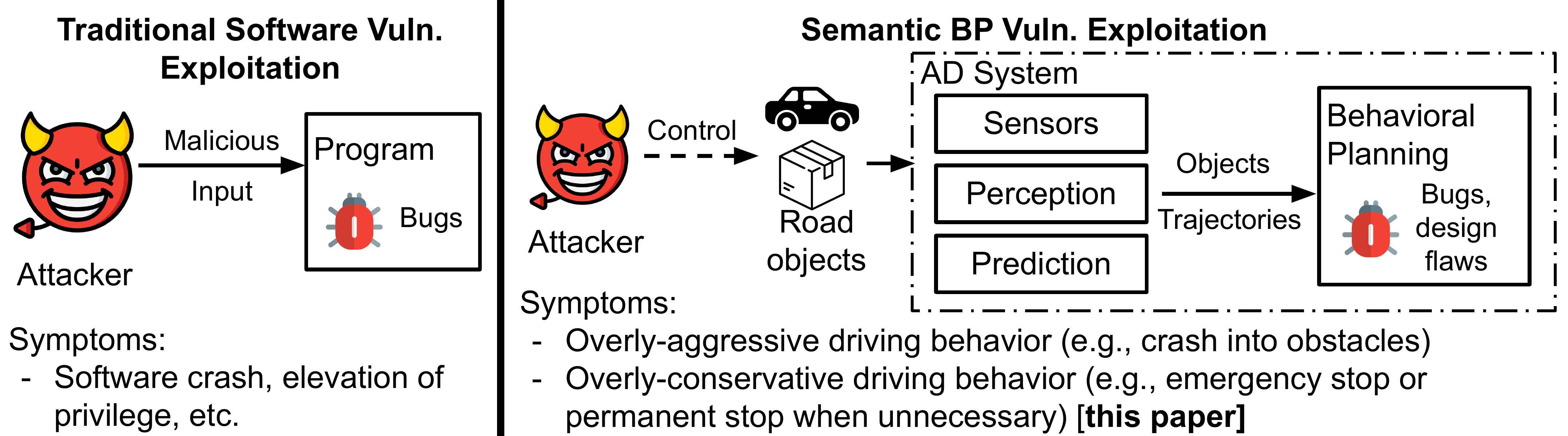}  
	\vspace{-0.2in}
	\caption{\newparts{Illustration of the domain-specific threat model and exploitation symptoms of the semantic BP DoS vulnerabilities targeted in this paper, along with their distinctions to those of traditional software vulnerabilities.}}
	\vspace{-0.15in}
		\label{fig:threat_model_demo}
\end{figure}

\newparts{
\textbf{Distinction to traditional software bugs.} As illustrated in Fig.~\ref{fig:threat_model_demo}, the semantic BP DoS vulnerability targeted in this paper is a type of semantic software vulnerability for BP, which can be caused by either software design flaws or implementation bugs. The key distinction of such semantic vulnerabilities to traditional software bugs is that their symptoms are erroneous behaviors at the BP decision logic level (e.g., keep driving or not, change lane or not) instead of at the generic computer program level (e.g., software crash, memory corruption, hang). In our problem setting, since we target physical-world attack vectors (\S\ref{subsec:threat_model}), the semantic BP vulnerabilities we target further differ in that the vulnerability triggering is via physical-world realizable perturbations (e.g., by adding attacker-controllable road objects) instead of generic program input changes (e.g., bit-level value changes of BP inputs), which is also illustrated in Fig.~\ref{fig:threat_model_demo}.
}



\cut{

observation: focus on dos -> convertaive -> hypothesize that dos will be more common -> focus on dos

}

\nsubsection{Threat Model}\label{subsec:threat_model}



\textbf{Attack vector: Attacker-controllable common physical-world road objects.}
\newparts{Fig.~\ref{fig:threat_model_demo} illustrates our threat model with a comparison to traditional software vulnerability exploitation. Instead of directly sending malicious inputs to the program}, we assume that the attacker can only exploit semantic BP DoS vulnerabilities via \textit{introducing common and easily-controllable external physical-world objects} to the driving environment, e.g., dumped cardboard boxes, parked bikes on the road side, vehicles driving on the roadway, or pedestrians walking on road pavements. We choose such a threat model because the attacker can more realistically launch such an attack in a practical setting since the attacker does not need to compromise or tamper with the internals of the victim AD system and the objects can pretend to be benign once they follow basic traffic laws and driving norms. \newparts{In~\S\ref{sec:limitation}, we also discuss the capabilities of a stronger threat model that might also be able to compromise the perceptional sensors.} Since this work aims at developing a systematic BP vulnerability discovery system for AD system developers, our system design assumes white-box access to the BP implementation.

\newparts{\textbf{Distinction to safety problems.} Under such a physical-world attack threat model, the attack-targeted unintended BP decision behaviors are also possible to naturally occur in non-adversarial settings, making them also in the scope of general safety or robustness problems. Here, the distinction is that we focus on the set of such unintended behaviors that are \textit{(more) triggerable} by attackers in the driving environment, e.g., easily-controllable road objects such as cardboard boxes, bikes, and attacker-driven vehicles, instead of weather conditions and road-side building locations/shapes. Such a security focus makes the discovered vulnerabilities arguably more severe, since with an adversary such unintended behaviors can be more frequently, controllably, and strategically triggered to cause more severe real-world consequences, e.g., causing emergency brakes in more dangerous road segments such as highway ramps, and causing traffic blocking in mission-critical roads such as in front of police stations or fire stations.
}

%% file: motivation.tex
\nsection{Motivation and Challenges}
In this section, we use a motivating example to concretely describe the BP DoS vulnerabilities targeted in this paper and the design challenges to systematically discover them.

\nsubsection{Motivating Example} \label{subsec:mov_example}

\begin{figure}[tbp]
\centering
\begin{minipage}{\linewidth}
\hrule
\begin{minted}[fontsize=\footnotesize,breaklines, linenos,escapeinside=||,mathescape=true,xleftmargin=8pt,numbersep=4pt]{python}
# Pre-defined lateral safety buffer, in meter
obstacle_lat_buffer |$\leftarrow$| 0.4f 
ADC_width |$\leftarrow$| 2.11f #Default AD veh width,in meter
#Initialize left/right boundaries of drivable space
{left_bound, right_bound} |$\leftarrow$| {left_lane_boundary, right_lane_boundary}
# Iterate over static obstacles at this longitudinal position  
for (each obs in obstacles_list) 
  if ((obs.min_l+obs.max_l)/2<0): # Left-side obstacles
    left_bound |$\leftarrow$| max(left_bound, obs.max_l + obstacle_lat_buffer)
  else: # Right-side obstacles
    right_bound |$\leftarrow$| min(right_bound, obs.min_l - obstacle_lat_buffer)
# Check whether exists drivable space laterally
if (right_bound - left_bound < ADC_width):
  path_blocked |$\leftarrow$| true #Conclude: lane is blocked
\end{minted}
\hrule
\vspace{0.02in}
  \caption{Simplified pseudo code for a semantic DoS vulnerability PlanFuzz discovered from BP in Apollo, an industry-grade AD system~\cite{apollo}.}
	\vspace{-0.2in}
  \label{fig:code_example}
\end{minipage}
\end{figure}

\begin{figure}[tbp]
\centering
\begin{minipage}{\linewidth}
    \centering
    \includegraphics[width=\linewidth]{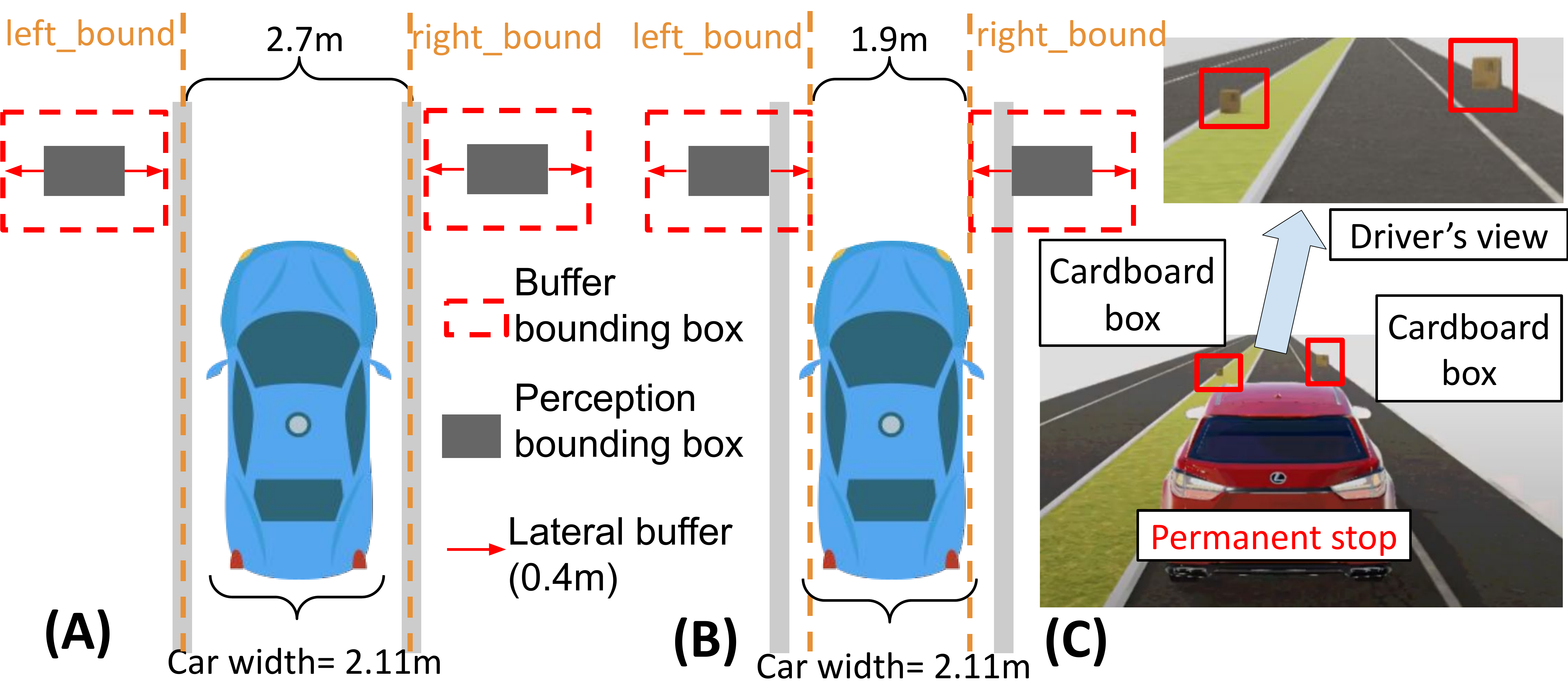}
	\vspace{-0.2in}
    \caption{Illustration of (A) BP decision logic in Fig.~\ref{fig:code_example}; (B) its semantic BP DoS vulnerability; (C) potential real-world exploitation of it from our end-to-end simulation case study (\S\ref{sec:case_study}).} 
	\vspace{-0.2in}
    \label{fig:path_bounx_demo}
\end{minipage}
\end{figure}

Fig.~\ref{fig:code_example} shows the simplified pseudo code for a BP DoS vulnerability our system discovered from version 5.0 of Apollo, an open-source industry-grade AD system~\cite{apollo} (also confirmed that such vulnerability also exist in 6.0, the latest version). This logic is from \texttt{path\_bounds\_decider}, one of the BP decision-making steps for the lane following scenario that checks whether the current lane has enough space in the lateral direction (left/right) for the AD vehicle to drive; if not, it considers the lane as blocked.






\textbf{Decision logic.} As shown in Fig.~\ref{fig:code_example} and illustrated in Fig.~\ref{fig:path_bounx_demo} (A), this code maintains 2 variables, \texttt{left\_bound} and \texttt{right\_bound}, to represent the leftmost and rightmost boundaries of the drivable space for a given longitudinal (forward/backward) position. If the space between these two boundaries is smaller than the AD vehicle width \texttt{ADC\_width} (line 13), it means no drivable space laterally and thus the current lane is blocked. The two variables are initialized with the left and right lane boundaries (line 5). Next, it iterates over the list of detected surrounding static obstacles, and use the rightmost lateral boundaries  (\texttt{obs.max\_l}) of the left-side obstacles to update \texttt{left\_bound} (line 8-9), and the leftmost lateral boundaries  (\texttt{obs.min\_l}) of the right-side obstacles to update \texttt{right\_bound} (line 10-11). Such logic can thus avoid causing the vehicle body to hit/touch the static obstacles. Here, a lateral \textit{obstacle safety buffer} (\texttt{obstacle\_lat\_buffer}, line 2) is conservatively applied to the leftmost and rightmost boundary calculation (line 9, 11) on top of the detected obstacle boundaries from the perception module, to ensure that the vehicle can keep enough distance from the obstacles.



\textbf{BP DoS vulnerability.} In Apollo 5.0, the obstacle safety buffer above is set to a constant value, 0.4m, regardless of obstacle positions in the driving environment. Meanwhile, the minimal lane width is 2.7m in urban areas~\cite{lane_width}, and thus static obstacles out of the boundaries of such lanes can in the extreme case reduce the lateral drivable space between \texttt{left\_bound} and \texttt{right\_bound} to 1.9m ($2.7 - 0.4\times2$). This is actually narrower than most vehicle models popularly used for AD today, e.g., 2.11-2.14m (with mirrors) for Lincoln MKZ, Lexus RX, Jaguar I-Pace, etc.~\cite{autonomous_product,ford_width,lincolnwidth,jager_width}. As shown in Fig.~\ref{fig:code_example}, the default value in Apollo is set to 2.11m (line 3). This means that with such logic a lane can be considered as blocked \textit{even when its designed drivable space is completely empty with no static obstacles invading its lane boundaries}. For a single-lane road, this means that the whole road is considered blocked and thus the AD vehicle will \textit{permanently} stop at the current lane, as illustrated in Fig.~\ref{fig:path_bounx_demo} (B).
This is a decision flaw for BP as such driving behavior is \textit{too conservative} to the extent that it \textit{directly} violates common driving norms. For example, the requirement that a vehicle ``should never block the normal and reasonable movement of traffic''~\cite{calihandbook}, which can \newparts{result in the vehicle being cited if this requirement} is violated, especially when \newparts{such a violation occurs} inside a tunnel or 15 ft within a fire station's driveway~\cite{cvc22500}.




Based on the code logic, the root cause for such an overly-conservative BP decision flaw is in the use and setting of the obstacle safety buffer. We found that such a flaw is not specific to Apollo: our system discovers that Autoware, another popular open-source full-stack AD system~\cite{autoware}, \original{has similar flawed BP logic and is actually even more conservative in such buffer settings (\S\ref{subsec:vuln_dis_res}).} \newparts{has similar flawed BP logic even though the concrete implementation is quite different from Apollo's. Also, Autoware is even more conservative in such buffer settings (\S\ref{subsec:vuln_dis_res}).} This thus conforms to our general design hypothesis in~\S\ref{subsec:attack_goal} that BP for practical AD systems tends to be as conservative as possible, leading to a general susceptibility to semantic DoS vulnerabilities.



\textbf{Exploitation.} To exploit this vulnerability in the real world, an attacker simply needs to prepare 2 easy-to-carry static objects that are not too uncommon in road regions, e.g., cardboard boxes, and place them close to but still off the lane boundaries on each side of a single-lane road, as illustrated in Fig.~\ref{fig:path_bounx_demo} (B). This is seemingly benign as these boxes are not blocking road and such randomly-dumped garbage on road side is not entirely uncommon. However, it can cause the AD vehicle with the vulnerable BP logic above to get permanently stuck at this road position and block traffic. Note that these 2 boxes do not have to be at exactly the same longitudinal position; from the code, they can be up to 5m (in \newparts{the} latest Apollo version) apart in longitudinal direction while still causing such a permanent stop decision, which can make such exploitation look more benign and thus stealthier. We did not include such longitudinal direction logic in Fig.~\ref{fig:code_example} for the ease to understand the key vulnerable logic. Fig.~\ref{fig:path_bounx_demo} (C) shows a snapshot of our end-to-end simulation case study of this exploit for Autoware (detailed in \S\ref{sec:case_study}). As shown, the two boxes are clearly off road and far from blocking the road, but the AD vehicle is forced to permanently stop there. 

\nsubsection{Design Challenges} \label{subsec:design_challenge}
Motivated by the concrete example above and similar problems observed in real-world production AD settings today (\S\ref{subsec:attack_goal}), it is highly desired to develop a systematic approach to discover such BP DoS vulnerabilities at the AD system development stage so that the developers can proactively find and fix them before deployment.


Recently, property-based testing have achieved great success in discovering safety violations in AD software~\cite{fremont2020formal,dreossi2019verifai,kimpgfuzz,fremont2020formal2, dreossi2019compositional,hekmatnejad2020search, tuncali2018simulation}. We follow the same general framework to develop a tool which can systematically discover DoS vulnerabilities in AD software. In the motivating example, the property we aim to \original{falsification}\newparts{falsify} can be formally expressed as: 
\begin{equation}
\footnotesize
\label{eq:overall_prop}
    IsSafetoDrive \rightarrow  \neg Stop 
\end{equation}
The above property indicates that when the current lane is safe to drive, the vehicle should be able to normally follow the lane instead of getting stuck by irrelevant physical objects. Even though the property seems to be simple at the first glance, none of previous works can be directly applied \original{to us}\newparts{to solve this problem} due to the following design challenges:


\textbf{C1. Lack of testing oracles for semantic DoS vulnerability in BP.}
For our vulnerability definition (\S\ref{subsec:attack_goal}), a key challenge is how to judge whether the current situation is safe to perform a certain planning behavior. For example, in our motivating example in~\S\ref{subsec:mov_example}, we need to decide the value of predicate \textit{IsSafetoDrive} in Eq.~\ref{eq:overall_prop}. Previous works~\cite{dreossi2019compositional, dreossi2019verifai, fremont2020formal, fremont2020formal2, kimpgfuzz} focus on the safety properties, especially collision, which can be directly extracted from official documentations or easy to define. But when it comes to studying DoS properties, there is no clear boundary of whether it is safe for the vehicle to drive due to the uncertainty and complexity of the environment. The first challenge is that we need to \textit{concretize} the predicate \textit{IsSafetoDrive} in Eq.~\ref{eq:overall_prop} into expressions which can be directly computed from planning inputs.

\textbf{C2. Need to systematically generate attack inputs following \newparts{complicated} problem-specific physical constraints.}
To discover our BP DoS vulnerability, the dynamic testing process needs to effectively generate attacker-introduced physical-object properties (e.g., positions) that can (1) follow basic traffic rules and driving norms as required in~\S\ref{subsec:threat_model} to achieve high attack practicality and stealthiness, e.g., a moving attack vehicle should drive in a lane following the road direction, instead of on pavements or in the wrong direction; and (2) make sure the value of predicate \textit{IsSafetoDrive} is true, since we can only find counterexamples when this predicate is true. Both constraints require to resolve complicated problem-specific geometry constraints; \newparts{the closest solution so far is from Scenic~\cite{fremont2019scenic}, which can generate test inputs within several generic geometric constraints in AD context (e.g., certain distances of a road object to curb). However, it still cannot address the more complicated ones specific to our problem context. For example, since we specifically want to generate road objects that should not affect the ego vehicle driving decision, their geometry constraints are inherently dependent on the planned trajectory of the ego vehicle, e.g., cannot be on or have any intention to move to any lanes that the ego vehicle plans to drive on (PI-C1, C4, C5 in Table~\ref{tab:phy_constraints}). However, Scenic’s current design does not consider such dependencies.
}

\original{\ziwenres{Original version}\textbf{C3.  Need to obtain fine-grained feedback from the planning decision-making program to guide our vulnerability discovery.} For automated software vulnerability discovery in general, existing dynamic testing methods popularly obtain code-level feedback to guide the discovery process, which shows superior effectiveness over treating the software as a black-box~\cite{10.1145/3243734.3243804,chen2018hawkeye,pham2019smart,bohme2017coverage, fioraldi2020weizz}. Also, recent sampling-based property falsifications~\cite{hekmatnejad2020search, kimpgfuzz} use the robustness metric as the guidance. However, their approach can only give a boolean guidance for finding DoS vulnerabilities since we can only know whether the AD vehicle stops or not\ziwenres{This sentense is misleading. Since robustness can provide quantitive feedback. The core problem here is that we need to get feedback even though the decision is unchanged}. The testing process cannot obtain any guidance on which test inputs are closer to triggering the vulnerability\ziwenres{This sentence is considered overclaim by RE.}. This thus makes it necessary to obtain \textit{finer-grained} feedback from the planning decision-making process, e.g., the distance between \texttt{left\_bound} and \texttt{right\_bound} in Fig.~\ref{fig:code_example}. So far, existing testing methods for semantic vulnerabilities/bugs in AD systems either do not consider obtaining such feedback (i.e., black-box testing)~\cite{li2020av}, or require the algorithm under test to be differentiable (e.g., DNN-based)~\cite{pei2017deepxplore, tian2018deeptest, zhang2018deeproad} and thus are not applicable to the representative program-based BP designs (\S\ref{sec:background_bp}) targeted in this work.}

\textbf{C3. Need to obtain more fine-grained feedback from the planning decision-making \newparts{code level (i.e., decision code branches)} to guide our vulnerability discovery.} For automated software vulnerability discovery in general, existing dynamic testing methods popularly obtain code-level feedback to guide the discovery process, which shows superior effectiveness over treating the software as a black-box~\cite{10.1145/3243734.3243804,chen2018hawkeye,pham2019smart,bohme2017coverage, fioraldi2020weizz}. \newparts{In the general CPS testing domain, prior works have used quantitative feedback such as the robustness metrics to guide testing~\cite{hekmatnejad2020search, kimpgfuzz}, but such a guidance still treats the code-level decision logic (i.e., decision-making code branches) as black-box, which thus cannot provide guidance once the overall planning output stays unchanged. In our problem setting, it is desired if we can improve this with more fine-grained code-level guidance such as the distance between the current inputs and the planning decision boundary at the code branch level. For example, in Fig.~\ref{fig:code_example}, the guidance can be more effective if we can leverage the value distance between \texttt{left\_bound} and \texttt{right\_bound} at the decision code branch at line 13.}

%% file: design.tex
\nsection{Design: PlanFuzz} \label{sec:design}
In this paper, we are the first to address the 3 challenges in \S\ref{subsec:design_challenge} by designing an automated approach to systematically discover BP DoS vulnerabilities (\S\ref{subsec:attack_goal}), called \textit{PlanFuzz}.

\newparts{
\textbf{Design goal.} The goal of PlanFuzz is to discover previously-unknown semantic DoS vulnerabilities defined at the BP decision code level (an example is in~\S\ref{subsec:mov_example}). Note that our current focus is not on the comprehensive identification of the triggering scenarios for the discovered vulnerabilities; nevertheless, a few concrete triggering scenarios will come with the discovery to provide the vulnerability triggerability since we adopt a dynamic testing framework (detailed below).
}



\nsubsection{Overview of Key Designs} \label{sec:design_overview}


At a high level, PlanFuzz follows an evolutionary testing framework, which is widely adopted by prior works on domain-specific vulnerability discovery with high generality, effectiveness, and efficiency~\cite{choi2020cyber,petsios2017nezha, rawat2017vuzzer,sparks2007automated,rawat2010evolutionary}. To address the challenges in \S\ref{subsec:design_challenge}, the following key designs are introduced: 

\begin{figure*}[t]
    \centering
    \includegraphics[width=\linewidth]{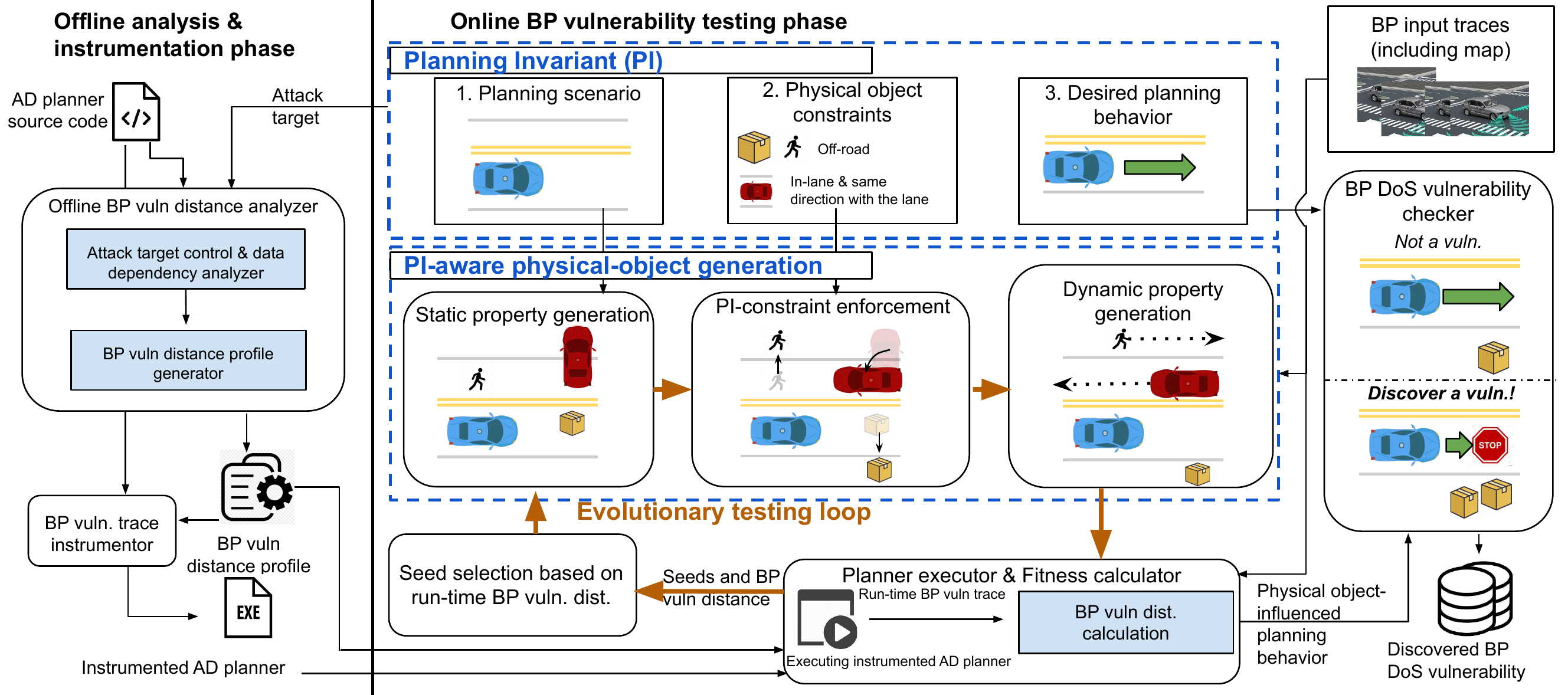}
    \vspace{-0.22in}
    \caption{Overview of PlanFuzz, our automated approach for systematically discovering BP DoS vulnerabilities in AD systems.}
    \label{fig:framework}
    \vspace{-0.1in}
\end{figure*}

\textbf{Planning Invariant (PI) as testing oracle.} To address $C1$, we propose \textit{Planning Invariant (PI)} as the problem-specific testing oracle for BP DoS vulnerabilities. PI has 3 components: \textit{planning scenario}, \textit{constraints for physical objects}, and \textit{desired planning behavior}, which together define an invariant property for BP: In a planning scenario, the BP output should always conform to the desired planning behavior as long as the PI constraints for surrounding physical objects are satisfied. For example, for our motivating example in~\S\ref{subsec:mov_example}, the planning scenario is lane following; the PI constraints for physical objects \newparts{are} that the surrounding physical objects are all static and located outside the lane boundaries; the desired planning behavior is that the AD vehicle should keep \newparts{driving forward}. Such invariant properties are derived from common driving norms (e.g., off-road static obstacles should not be considered as blocking the road, detailed in~\S\ref{subsec:PI}) so that its violations can be used to tell an overly-conservative decision. 


\textbf{PI-aware physical-object generation.} To address $C2$, we need to design physical-object generation methods that conform to both \newparts{the} driving norms (e.g., traffic rules) and PI constraints above given a planning scenario and desired planning behavior. A direct solution direction is to directly generate random physical-object properties within these constraints. However, generating them following a random distribution is very difficult since these constraints are quite irregular in the real world (e.g., curvy and zigzag road boundaries), not to mention the problem-specific constraints due to the dependence on the planned trajectory of the ego vehicle (\S\ref{subsec:design_challenge}). To address this, our strategy is to first generate objects without considering these constraints, and then add a PI constraint enforcement step afterwards to adjust its properties (e.g., positions) \newparts{for} conformation. Specifically, here different PI constraint enforcement operators are designed following the diversity and inheritance principles desired for the random input generation step in genetic algorithms~\cite{whitley1994genetic}.

\textbf{BP vulnerability distance.} To overcome $C3$, we design \textit{BP vulnerability distance} as the code-level feedback to measure how close the current BP execution is to violate a PI and thus trigger a BP DoS vulnerability. Specifically, such a distance is defined based on the control- and data-flow differences between the current executed code path and its closet path to a set of \textit{attack target positions}, which are code positions indicating violations of the PI (for example, the callsite of API \textit{BuildStopDecision}). To facilitate its run-time calculation during the dynamic testing, we first perform an off-line analysis of the BP code to (1) identify the key predicates that the attack target positions control- and data-dependent on, and compute information that can be pre-calculated about their run-time control- and data-flow distances to the target positions, which forms a \textit{BP vulnerability distance profile}; and (2) instrument these predicates to collect their execution status, called \textit{BP vulnerability trace}, in the run time. During the testing, based on the execution status of these key predicates collected in BP vulnerability trace, the control- and data-flow distances of the executed key predicates are calculated with the pre-computed information from BP vulnerability distance profile, which are then combined to calculate the final vulnerability distance. 

Next, we describe the whole PlanFuzz system design incorporating these key designs, and then provide details for each.

\nsubsection{PlanFuzz System Design}
\label{sec:bpfuzz_design}

\textbf{System input and output.} Fig.~\ref{fig:framework} shows an overview of PlanFuzz system with the key designs above. As shown, the whole PlanFuzz system requires 3 inputs: (1) BP source code (we assume \newparts{it is avaiable} since PlanFuzz is designed for AD developers); (2) BP input traces (including map) for the planning scenarios of interest; and (3) a set of PIs for these scenarios. \newparts{The output is a set of test inputs that can trigger a BP DoS vulnerability. The AD developers can then utilize them to identify the vulnerable code logic and analyze root causes, with the goal of developing vulnerability fixes.}


\newparts{\textbf{Manual efforts.} Three types of manual efforts are required for using PlanFuzz: (1) Collecting BP input traces for the targeted scenarios. Since PlanFuzz is designed for AD developers, we assume they have access to such traces from their AD system testing or operations; (2) identifying PIs for these targeted scenarios. Note that such identification is an one-time effort and there are general ones applicable across multiple scenarios identified later in~\S\ref{tab:summary_planning_invariant}; and (3) identifying and annotating the attack target positions in the source code. For AD developers, we assume they can have high-level knowledge of the BP code (e.g., BP decision APIs and state variable class) to identify these based on the desired planning behavior of PI and the state variables associated with the scenarios. In our experiments, it took less than 50 lines in total and less than 1 hour of manual efforts for an author who is experienced with the AD planning code.
}




\textbf{System framework.} With the inputs above, the vulnerability discovery process has 2 phases as \newparts{shown} in Fig.~\ref{fig:framework}: (1) \textit{offline analysis and instrumentation}, and (2) \textit{online BP vulnerability testing}. On the left side of Fig.~\ref{fig:framework}, the offline phase takes the BP source code and attack target positions based on PIs, and generate the BP vulnerability profile and instrumented BP code. On the right side of Fig.~\ref{fig:framework}, the online phase follows the evolutionary testing framework in which each single physical object is considered as a ``gene''. It customizes each component of a genetic algorithm: (1) \textit{fitness}, for which we use \newparts{the} run-time BP vulnerability distance to measure how close a BP execution is to trigger a BP DoS vulnerability. It is calculated by executing the instrumented BP code in the planner executor and \newparts{the} fitness calculator in Fig.~\ref{fig:framework}; (2) \textit{mutation and crossover}, for which we use the PI-aware physical-object generation to mutate and exchange physical objects; (3) \textit{seed selection}, for which we select test cases with smaller BP vulnerability distances \newparts{as the new seeds}.


To get started, the planner executor \newparts{extracts the input frames corresponding to each BP decision from the BP input trace}, and \newparts{then feeds} these frames to the instrumented BP code. To \newparts{maintain the consistency of} internal BP states, we record a snapshot of the internal state variables beforehand for the initial seed, and recover it before feeding the testing inputs later. During the testing, \newparts{the} BP DoS vulnerability checker \newparts{determines} whether a certain generated testing case violates the PI; if so, it outputs the discovered vulnerability.

\cut{
Besides the novel design components, we customize the genetic algorithm framework to fit into our testing system to discover BP DoS vulnerability effectively and efficiently. Each single physical object is considered as a ``gene'' in the context of genetic algorithm. The details of how we customize each component of genetic algorithm are listed below:
\begin{itemize}
    \item \textbf{Fitness.}  The final goal of our testing system is to discover BP DoS vulnerability. And the fitness represents how close a planning execution is to reaching the BP DoS vulnerability. The ``gene'', physical objects need to be concatenated with the planning scenario to form the input of behavior planner to calculate the BP vulnerability distance at runtime. 
    \item \textbf{Mutation.} We use the PI-aware physical object mutation component as the mutation function. The PI-aware physical object mutation can mutate a single physical object while maintaining the constraints of planning invariant. 
    \item \textbf{Crossover.} Following the the original concept of crossover in genetic algorithm, we exchange the ``gene'', physical objects between 2 seeds. 
    \item \textbf{Seed selection.} Following the common design of genetic algorithm, seed selection component tends to select test cases seeds with smaller BP vulnerability distance such that the testing process is moving towards to a DoS vulnerability. 
\end{itemize}
Besides the evolutionary testing loop, a BP DoS vulnerability checker is designed to check whether a certain generated testing case violates the planning invariant.
}

\nsubsection{Planning Invariant} \label{subsec:PI}

\begin{table}[tbp!]
\footnotesize
\centering
\caption{General PI (Planning Invariant) constraints across different driving scenarios from the full list of PIs identified and used in this paper (Table~\ref{tab:summary_planning_invariant}).}
\label{tab:phy_constraints}
\setlength{\tabcolsep}{3.2pt}
\begin{tabular}{c|l}
\hline
\textbf{Physical object type} & \multicolumn{1}{c}{\textbf{PI constraints for physical objects}}\\
\hline
\begin{tabular}{c} Static obstacle (cardboard \\boxes, parked bikes, etc.) \end{tabular} &\begin{tabular}{l} PI-C1. $StaticOffRoad(x)$\end{tabular}\\
\hline
\multirow{2}{*}{Vehicle} & \begin{tabular}{l}PI-C2. $IsFollowingVehicle(x)$ \end{tabular}\\
 & \begin{tabular}{l}PI-C3. $IrrevalentVehicle(x)$ \end{tabular}\\
 \hline
\multirow{2}{*}{Pedestrian} & \begin{tabular}{l}PI-C4. $StaticOffRoad(x)$\end{tabular}\\
 &\begin{tabular}{l}PI-C5. $DynamicOffRoad(x)$  \end{tabular}\\

 \hline
\end{tabular}
\vspace{-0.2in}

\end{table}
\cut{
\begin{figure}
    \centering
    \includegraphics[width=0.95\linewidth]{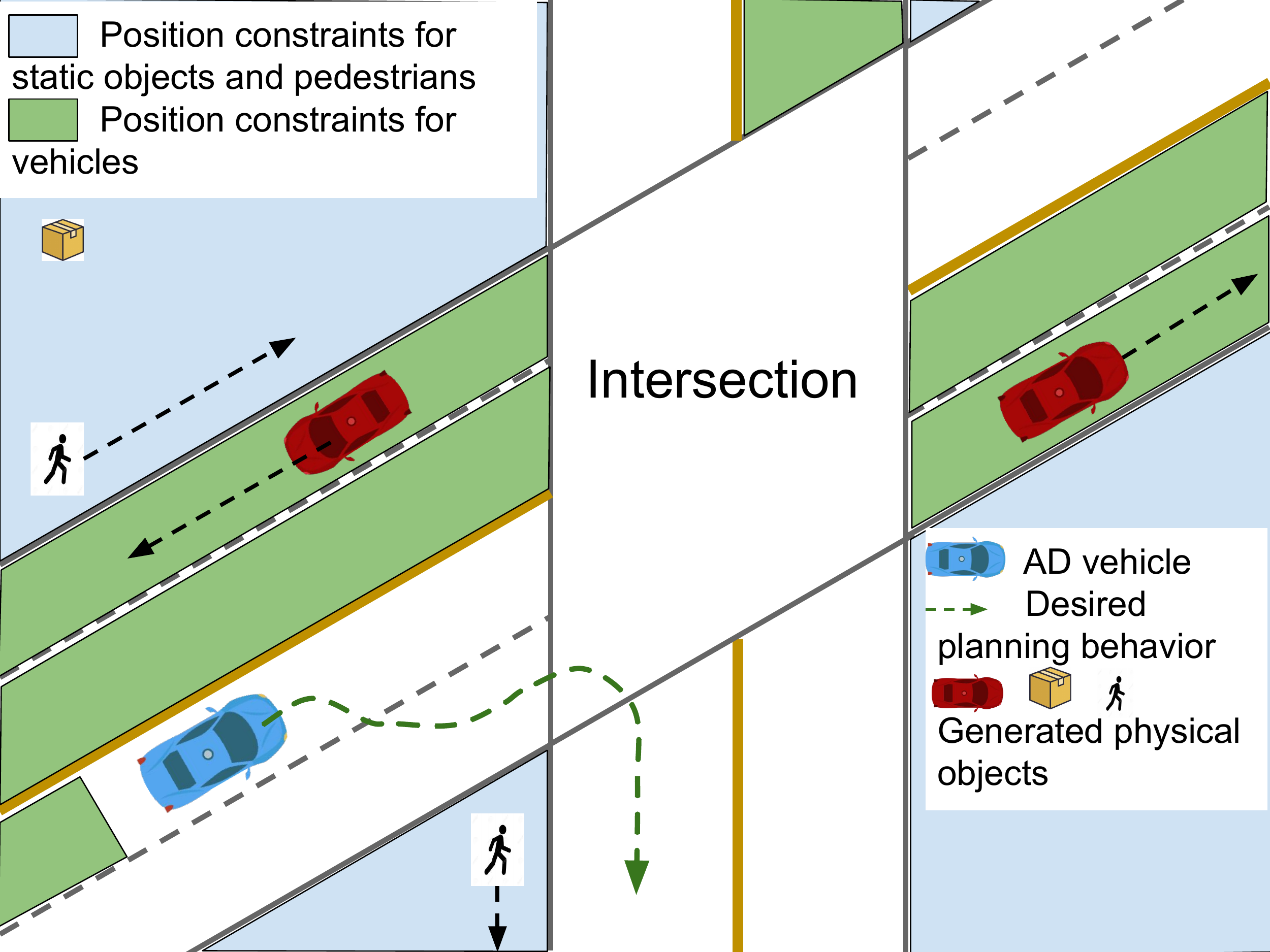}
    \caption{Illustration of general PI constraints for physical objects in Table~\ref{tab:phy_constraints} for common driving scenarios.} 
    \label{fig:mutation_demo_1}
\end{figure}}

As introduced in~\S\ref{sec:design_overview} and shown in Fig.~\ref{fig:framework}, each PI has 3 components: planning scenario, constraints for physical objects, and desired planning behavior, which together form \newparts{the} BP invariant properties that concretely define \newparts{the} overly-conservative BP decisions. In this paper, we consider 8 different planning scenarios \newparts{commonly} supported by industry-grade AD systems~\cite{apollo, autoware,gm_safety_report}, covering various basic \newparts{real-world driving} scenarios, e.g., lane following, lane changing, intersections with stop signs and traffic signals, lane borrowing, bare intersection, and parking. The full list is in Table~\ref{tab:summary_planning_invariant}. Since we target semantic DoS vulnerabilities, the desired planning behavior for each scenario \newparts{is} usually to just keep the intended driving behavior, e.g., keep cruising for lane following, keep moving to pass the intersection, and finish the lane-change/borrow/parking actions.

Given a planning scenario and the desired driving behavior, the next is to identify the physical-world constraints of the attacker-introduced physical objects such that if satisfied, the BP logic should not give up the desired driving behavior. We derive such constraints conservatively based on common driving norms, e.g., descriptions from the driver's handbook~\cite{calihandbook} such as those quoted in~\S\ref{subsec:mov_example}. Specifically, in this paper we focus on the general constraints that are applicable to multiple driving scenarios, which are summarized in Table~\ref{tab:phy_constraints} and denoted as \textit{PI-C}. For example, for static obstacles such as cardboard boxes and parked bikes, the ones that are completely off-road and without any violation of the boundaries of the lanes, which the AD vehicle plans to drive on, should generally not cause the BP logic to give up lane following, changing, borrowing, or passing an intersection. For pedestrians and vehicles, the ones moving in their commonly-designated regions (e.g., off-road pavement for pedestrians, and \newparts{traffic} lanes for vehicles) without showing any intention to move towards the AD vehicle or the lane regions the AD vehicles plans to drive one should not give up those desired driving decisions. Here for the simplicity, we define a list of functions $StaticOffRoad(x), DynamicOffroad(x), FollowingVeh$ $icle(x), IrrevalentVehicle(x)$ to express the geometry relationship between the road network, the trajectory of a certain physical object $x$, and the trajectory of the AD vehicle. The complete set of PI-Cs for all planning scenarios and the formal \newparts{definitions} of the functions are in Table~\ref{tab:summary_planning_invariant} in Appendix. With the defined PI-Cs, we are able to concrete the high-level property in Eq~\ref{eq:overall_prop} into the following format to describe the property in lane following scenario:
\begin{equation}
\footnotesize
\begin{aligned}
StaticCons(x) &:= (x.type == Static)\rightarrow (StaticOffRoad(x))\\
VehicleCons(x) &:= (x.type == Vehicle) \rightarrow\\ 
& (FollowVehicle(x) \vee IrrelevantVehicle(x)\\ 
PedestrianCons(x) &:= (x.type == Pedestrian) \rightarrow \\
  &(DynamicOffroad(x)\vee StaticOffRoad(x))\\
\wedge_{x\in O}&((StaticCons(x))\wedge (VehicleCons(x))\\
&\wedge (PedestrianCons(x)) \rightarrow \neg Stop
\end{aligned}
\end{equation}
Here $O$ is the set of physical objects and $x$ is a physical object in the set. We define the constraints for each type of objects and merge them in the end to define the availability property.

\cut{
We define the planning invariant as the testing oracle for our testing framework. The planning invariant can be formulated as three components:
\begin{itemize}
    \item \textbf{Planning scenario.} The planning scenario is the collection of basic information required for the planner to make a behavior decision. The planning scenario includes the routing path, the road map, the current position of AV, and current traffic participants. 
    \item \textbf{Physical objects constraints.} We assign constraints to the physical objects so that no matter how the physical objects are changed under the constraints, the semantic meaning of the planning scenario is not changed. The constraints come from two sources. First, the physical objects should obey traffic rules, e.g., the attacker can not put a trash can in the middle of the lane. Second, the constraints also come from the traffic interaction with the AV, i.e., the generated physical objects should not affect the planning behavior of the AV. Specifically, for the static objects, the requirement is that each object has no intersection with the current lane. For the pedestrians, they should not have any moving tendency towards to the current lane and will not go across the current lane. For vehicle objects, we expect the object will not share the same lane with AV in any future moment unless the object is following the AV.
    \item \textbf{Desired planning behavior.} The desired planning behavior is the expected behavior generated by the planner given the current planning scenario.
\end{itemize}
The principle behind our planning invariant design is that the AD planner's behavior decision should remain unchanged given the certain planning scenario and constraints to the physical objects. For example, as shown in Figure~\ref{fig:framework}, the car is driving on a single lane road and there are no other traffic participants around it. This forms the planning scenario and it is the basic information for a behavior planning. The desired planning behavior is to move forward since clearly the AV can move forward safely. The constrains of physical objects can be described based on the type of physical objects. For a static object such as a cardboard box, it should be off the road. The same off-the-road rule applies for the pedestrian and also a pedestrian should not have any tendency to move towards the current lane of AV. As for a moving vehicle, it can either follow the AV on the current lane or drive on the reverse lane with the reverse direction. It is obvious that any physical object fits to the constraint here should not affect the AV's normal driving. The specific planning invariant is that the AV should always move forward on this single lane road and ignore the off-road static objects, off-road pedestrians with no intention to move to the current lane, following vehicles, and vehicles normally driving on the reverse lane. 

Another concept related to the planning invariant is the \textit{attack target} in the AD planner source code. As we introduced before, the attacker wants to trigger a different planner decision to achieve the attack goal. Here we define the attack position as a set of positions in the source code where the attacker-wanted decision will be generated. Currently, we generate such attack positions with manual efforts. The efforts of identifying such position in the code is small since there will be very clear sign in both the function call signatures and also the log information. For example, we can simply associate the decision to stop in front of a stop sign with the function call ``BuildStopDecision'' with a "STOPSIGN" stop reason in the function arguments.
}
\nsubsection{PI-aware Physical-Object Generation} \label{subsec:design_mutation}

After concreting the property, the next step is to generate the inputs that can always satisfy the constraints in planning invariant. In \newparts{other words}, we want to make sure that the left side of Eq.~\ref{eq:overall_prop} is always true. We design \newparts{PI}-aware physical object generation to satisfy this requirement. Due to the page limit, we leave most of the details in Appendix~\ref{sec:app_pienforcer}. The input generation contains three main steps:

\textbf{Static property initialization and mutation.} In this step, we first randomly generate the static properties of the objects, e.g., position, type, and size, without considering PI-Cs during the testing input initialization and mutation processes. For each generated physical object, we will assign \newparts{an} appropriate size given the randomly generated object type.

\cut{
\begin{table}[]
\footnotesize
\centering
\caption{Data fields and the corresponding initialization/mutation strategies of a physical object.}
\label{tab:obj_defination}
\setlength{\tabcolsep}{2.5pt}
\begin{tabular}{c|c}
\hline
Physical object data field & Initialization/mutation strategy\\
\hline
 Position & randomly generated \\
 \hline
 Type & randomly picked \\
 \hline
Size & randomly generate for static obstacle only\\ 
\hline
 Speed & processed by further steps \\
\hline
 Trajectory & processed by further steps\\
\hline
\end{tabular}
\end{table}
}

\textbf{PI-constraints enforcement.} The second step is to change the position and heading of each physical object to enforce the position correctness for each object. The high-level idea of this enforcement step is to adjust the property value to the closest one that does not violate PI-C. For example, moving \newparts{a} static obstacle, which violates the lane boundaries, to the closest off-lane position. We also change the lateral position and longitudinal position separately to keep the diversity and inheritance during the evolutionary algorithm.  


\cut{
Since the type and the position of physical objects are generated in object position initialization/mutation step without considering any PI constraints, the goal of PI constraints enforcer is to adjust the position and generate the heading of a physical object given its type, the road map, and also driving norm.

When generating the attacker-controllable physical objects for a certain scenario, we should consider two requirements so that the attacker can behave normally:
\begin{itemize}
    \item The physical objects should obey the traffic rules.
    \item The physical objects do not have any possibility to affect the desired driving behavior according to the driving norm. \todo{write in a better way...}
\end{itemize}
Following this high-level requirements, we define concrete generation constraints for each specific type of physical object in our mutation function:
\begin{itemize}
    \item \textbf{Static obstacles.} We directly constrain the static obstacles to be off the driving lane of the AV. As a result, the off-lane static obstacle will not have any intersection with the current driving path of the AV and should not affect its desired driving behavior under any circumstances.  
    \item \textbf{Vehicles.} A vehicle object should be at least driving inside the lane to behave normally. Besides that, we need to take the planning trajectory of the AV and the traffic rules into consideration. We enforce that the vehicle will not affect the behavior of AV by forcing the vehicle to either follow the AV or drive in a lane which will not be used by the AV in the future. If the AV is going to pass an intersection with stop sign, we have to further make sure that the vehicle will not enter the same intersection. An exception of this is that when testing lane borrowing scenario, the vehicle can park stationary in the current lane so that the victim AV has to borrow the reverse lane. 
    \item \textbf{Pedestrians.} The pedestrian is walking or standing off the road. Since the pedestrian can be walking, we also have to make sure that the pedestrian does not have any tendency to move close to the desired planning trajectory. 
\end{itemize}

Figure~\ref{fig:mutation_demo_1} shows an example of position constraints added to the physical objects. For computation convenience, we transfer all the positions into world coordinate into the Frenet-Serret coordinate~\cite{slcoordinate} based on the current lane. The static objects and the pedestrians should be off-lane and the corresponding area is marked with light blue in the figure. The places where the mutation function can put vehicle objects are marked with orange in the figure.

}
\textbf{Dynamic property generation.}
The last step is to add dynamic properties (e.g., speed and moving trajectory) for dynamic physical objects such as driving vehicles and walking pedestrians. Here, the generated properties need to ensure (1) satisfying the driving norms (e.g., traffic rules) as in~\S\ref{subsec:threat_model}, and (2) conforming to the related PI-Cs such as not showing any intention to move towards the AD vehicle or the lanes it plans to drive on (PI-C3,5 in Table~\ref{tab:phy_constraints}).

\nsubsection{BP Vulnerability Distance} \label{subsec:vuln_dist}
\begin{figure}[t]
      \centering
          \includegraphics[width=1.\linewidth]{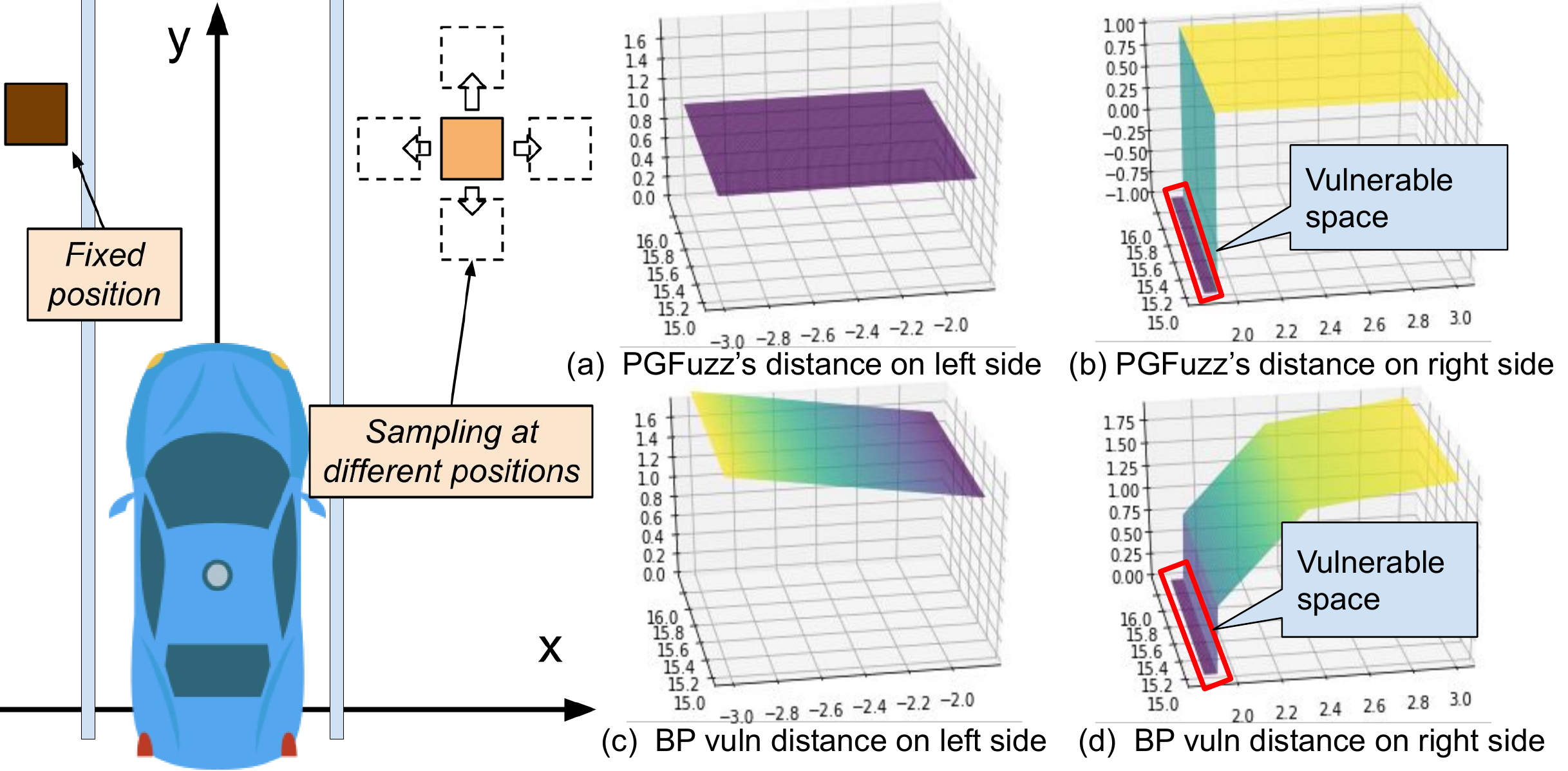}  
    \vspace{-0.1in}
	\caption{\newparts{Illustration of the benefit of our BP vulnerability distance design compared to other fitness function choices.}}
		\label{fig:distance_demo}
		\vspace{-0.1in}
\end{figure}



The BP vulnerability distance is calculated based on information from both offline BP vulnerability distance profile and runtime BP vulnerability trace. If we recall the example code in Fig.~\ref{fig:code_example}, our goal here is to find a set of physical objects, which satisfy the constraints of PI, and make the program execution reach line 14. Inspired by directed greybox fuzzing~\cite{bohme2017directed}, we design a distance metric to quantify the distance between current execution trace and the attack target positions.
The directed greybox fuzzing techniques so far generally only consider control flow distance defined in~\cite{bohme2017directed}. In this paper, we further add a data flow distance into our distance metric design. This is motivated by our observation: (1) The BP \newparts{decision-making} is usually based on a list of predicates with floating-point number comparisons. Thus, measuring the difference between floating points operands can help guide the testing in a more fine-grained way.
(2) The control-flow changes in our problem setting is not very significant. For example, in the path bounds case shown in \S\ref{subsec:mov_example}, changing the position of a certain obstacle will not change the overall control flow unless a different decision is made.


Specifically, in the offline analysis and instrumentation phase as shown in Fig.~\ref{fig:framework}, we first build a Program Dependence Graph (PDG) of the BP code and then use it to identify all the predicates that the attack target positions control- and data-dependent on. We further divide the predicates into two parts: critical predicates and non-critical predicates. Critical predicates refer to the predicates that connect themselves with the attack target position with control dependence edges only. On the other side, non-critical predicates refer to the predicates that must connect with the attack target via data dependence edges. Examples for these two types of predicates based on our motivating example are included in Appendix~\ref{sec:app_bp_dist}. 

For each of these predicates, we calculate their individual control- and data-flow distances, and use the sum as the overall BP vulnerability distance. Our control-flow distance design is similar to that in prior directed grey-box fuzzing methods~\cite{bohme2017directed}, and thus next we focus on explaining our data-flow distance design using run-time collected BP vulnerability traces. For critical predicates, our offline phase determines which branch is reachable or closer to the target positions. In the runtime, we calculate how close the execution is to triggering such a branch using the differences between the operands and the execution number of each branch. For non-critical predicates, since we do not know which branch is the closer branch towards the target positions, our strategy is to minimize the difference between operands to get more diverse execution traces.  


We use the motivating example in~\S\ref{subsec:mov_example} to demonstrate the benefit of this design. As illustrated in Fig.~\ref{fig:distance_demo}, we fix the position of one static object right next to the left lane line and 15m in front of the AD vehicle, and \newparts{at} the same time, we arbitrarily \newparts{move} the other static object and measure the robustness distance metric used in~\cite{kimpgfuzz, tuncali2018simulation, hekmatnejad2020search} and BP vulnerability distance proposed by our paper. In Fig.~\ref{fig:distance_demo} (a)(b), the robustness metric can only give a boolean guidance \newparts{on} whether the decision is changed. However, our distance metric in Fig.~\ref{fig:distance_demo} (c)(d) can guide the position of the second object into \newparts{the} vulnerable area due to that \newparts{the} operand distance of predicate on Line 8 and Line 13 in Fig.~\ref{fig:code_example} becomes smaller when the object is approaching to the vulnerable area. 


%% file: evaluation.tex
\nsection{Evaluation} \label{sec:evaluation}
\nsubsection{Evaluation Setup} \label{subsec:evaluation_setup}
\textbf{Subject BP implementations.} 
We evaluate PlanFuzz on the BP code from 2 open-source AD systems, Baidu Apollo~\cite{apollo} and Autoware~\cite{autoware}. Both are practical full-stack AD systems that can be readily installed on real vehicles for driving on public roads~\cite{autoware-car,Apollo-lincon}, and also have representativeness for industry-grade AD systems as Baidu has been recently ranked among the top 4 leading industrial AD developers with Waymo, Ford, and Cruise~\cite{AVtop4} and has been providing self-driving taxi services in China for months~\cite{apollo_news_taxi}, while Autoware is adopted by the USDOT in their AD vehicle fleet~\cite{carma-platform}.

Specifically, we select the BP implementations in 2 Apollo versions, 3.0 and 5.0, as their design and implementations are significantly different based on the release log~\cite{apollorelease}. We did not evaluate on 6.0, the latest Apollo version, as it only made minor changes to 5.0 in BP~\cite{apollorelease}. We have confirmed that all the discovered vulnerabilities from 5.0 also exist in 6.0. The Autoware BP implementation we evaluate on is from Autoware.AI 1.14.0~\cite{autoware, autoware_software}, the latest version with an implementation of the open planner~\cite{darweesh2017open}. Both Autoware and Apollo use rule-based logic to make decisions. 
\newparts{Their main difference is that Autoware's BP adopts a single finite state machine-based design where all planning behavior changes are modeled as state transitions, while Apollo's is more modular, which first decomposes the whole driving decision making into independent primitive tasks (e.g., obstacle avoidance, lane changing, intersection passing, velocity selection, etc.) and then selects the appropriate ones based on different driving scenarios (e.g., lane following, intersection, stop sign).}


\begin{table*}[tbp]
\footnotesize
\centering
\caption{Discovered BP DoS vulnerabilities from Apollo (3.0, 5.0) and Autoware. PI identifiers refer to PIs in Table~\ref{tab:summary_planning_invariant} in Appendix. All the vulnerabilities discovered in Apollo 5.0 are confirmed to also exist in the latest version Apollo 6.0.}
\label{tab:vuln_discovered}
\setlength{\tabcolsep}{2.5pt}
\begin{tabular}{@{}cccccc@{}}
\toprule
Vuln \# & Driving Scenario & Software & \begin{tabular}{c} Violated PI \#\end{tabular} & \begin{tabular}{c} Attack-influenced Planning Behavior\end{tabular} & \begin{tabular}{c} Triggering Objects \end{tabular} \\
\midrule
V1 & Lane following & Apollo 3.0/5.0 & PI1 (PI-C1) & Permanent stop & Static obstacle \\
\midrule
V2 & Lane changing & Apollo 3.0/5.0 & PI3 (PI-C2, 3)& Fail to change lane and never reach destination & Vehicle \\
\midrule
V3 & Lane borrow & Apollo 3.0/5.0 & \begin{tabular}{c}PI4 (SP-PI-C1, 2) \end{tabular} & Fail to borrow the lane and permanent stop & \begin{tabular}{c}Vehicle or static obstacles \\ in front of blocking vehicle \end{tabular}  \\
\midrule
V4 & Lane borrow & Apollo 3.0/5.0 & \begin{tabular}{c}PI4 (PI-C1, 4, 5) \end{tabular} & Fail to borrow the lane and permanent stop & Off-road static obstacle \\
\midrule
V5 & Intersection w/ traffic signal & Apollo 3.0/5.0 & PI6 (PI-C4) & Fail to pass intersection and permanent stop & Static pedestrian \\
\midrule
V6 & Intersection w/ traffic signal & Apollo 3.0/5.0 & PI6 (PI-C5) & \begin{tabular}{c}Emergency stop; possible to cause\\ permanent stop and thus fail to pass intersection\end{tabular} & \begin{tabular}{c}Pedestrian who is\\ leaving the intersection \end{tabular}\\
\midrule
V7 & Intersection w/ stop sign & Apollo 3.0/5.0  & \begin{tabular}{c}PI5 (PI-C1)\end{tabular} & Fail to pass intersection and permanent stop & \begin{tabular}{c}Static bicycle off the road \end{tabular}\\
\midrule

V8 & Lane following & Autoware  & PI1 (PI-C1) & Permanent stop & \begin{tabular}{c}Static obstacle off the lane \end{tabular}\\
\midrule

V9 & Lane following & Autoware  & PI1 (PI-C3) & Emergency stop & \begin{tabular}{c}Moving vehicle off the lane \end{tabular}\\
\bottomrule
\end{tabular}
\vspace{-0.1in}
\end{table*}

\cut{
\begin{table}[tbp!]
\footnotesize
\centering
\caption{Subject AD software tested in our study.}
\label{tab:factor_importance}
\setlength{\tabcolsep}{3.2pt}
\begin{tabular}{c|c}
\hline
Name and Version of AD system & Robotics Middleware \\
\hline
Baidu Apollo 3.0 & ROS \\
\hline
Baidu Apollo 5.0 & Cyber RT \\
\hline
Autoware.AI 1.14.0 & ROS \\
\hline
\end{tabular}
\end{table}
}

\textbf{BP input trace collection.}
We use LGSVL, an production-grade AD simulator~\cite{lgsvl}, to collect the BP input traces as the initial testing seed (\S\ref{sec:bpfuzz_design}). The benefit of using a simulator to generate seeds is that it is easier to (1) create different planning scenarios to increase testing diversity, and (2) control the scenario to prevent any irrelevant physical objects from affecting the generation of desired planning behavior. Note that such simulation-based testing is widely used in the AD industry for flexibility, scenario coverage, and safety~\cite{arg_sim_1, arg_sim_2,arg_sim_3}. LGSVL itself is also designed for performance and safety testing of production AD systems~\cite{lgsvl}.

\newparts{In total, 40 traces are collected under 8 different driving scenarios (5 per scenario). For each scenario, the 5 traces have diversity in driving tasks (e.g., drive straight or make turns) and road layouts (e.g., width of the local road's lane width or the highway). Each trace spans up to 47sec with 100-2400 BP decisions, and the input frame for each BP decision is used as an individual testing seed. These traces lead to 28,789 (9,676 for Apollo, 19,113 for Autoware) different initial testing seeds in total (3,598 per scenario on average) used in our evaluation. More details are in Table~\ref{tab:seed_collection} in Appendix. In these traces, both the AD vehicle itself and other traffic participants are behaving normally/correctly (e.g., follow traffic rules and driving norms, and can correctly execute the designed driving maneuvers).}


\newparts{\textbf{Test input generation.} For each initial testing seed, PlanFuzz generates the attack's physical objects as described in~\S\ref{subsec:design_mutation} and injects them into the planning input. Here, we directly use their ground-truth physical properties (e.g., type, bounding-box size and shape), which can thus avoid finding violation cases due to errors in upstream modules (e.g., perception) instead of BP. As described in~\S\ref{subsec:design_mutation}, for each attack object, PlanFuzz initializes and mutates their properties within common feasible ranges according to their types. For example, the positions of generated objects is within 80m to the ego vehicle (a common range of AD perception~\cite{teslasensorcoverage}); the sizes for static objects are 0.5-2m each dimension, and those for pedestrians and vehicles are the same as the default ones used in the simulator. Details are in Table~\ref{tab:valid_input} in Appendix. Note that consistent with our threat model (\S\ref{subsec:threat_model}), we do not mutate the non-attacker-controllable planning inputs such as weather conditions and the ego vehicle's driving speed; all of them just inherit the valid values from original BP input traces.}


\textbf{Evaluation metrics and setup.} 
We consider a vulnerability as discovered when any attack target position is triggered (\S\ref{sec:bpfuzz_design}). \newparts{We consider a discovered BP DoS vulnerability is \textit{unique} if its code-level decision logic (branches) that causes such vulnerability (e.g., those identified in~\S\ref{subsec:mov_example}) is different from the others, which is similar to unique crashes in traditional fuzzers~\cite{afl, yun2018qsym, chen2018angora, stephens2016driller}).} For each initial seed, we run PlanFuzz multiple times to increase the chance of finding unique vulnerabilities. For each run, the testing terminates when either a vulnerability is found, or the optimal fitness value is unchanged for 100 generations.
We manually verified each discovered vulnerability and did not find any false positives. All experiments are run on AMD EPYC 7551 CPUs.

\nsubsection{Vulnerability Discovery Effectiveness}\label{subsec:vuln_dis_res}



Throughout our experiments, PlanFuzz discovered 9 unique BP DoS vulnerabilities in Apollo and Autoware as shown in Table~\ref{tab:vuln_discovered}. All these vulnerabilities can be exploited to adversely delay the progress of the AD vehicle from reaching its destination; some of them can cause the AD vehicle to permanently stop on the road. We classify the vulnerabilities into three types based on the attack scenarios. In this section, we provide a summary of the attack scenarios, including (1) the driving scenarios and symptoms of the relevant vulnerabilities, (2) the violated PIs, (3) the root causes, and (4) the potential real-world exploitations. Pseudo code and detailed analysis of the vulnerabilities can be found in Appendix~\ref{sec:appendix_PIsummary}.

\textbf{Attack scenario 1: lane following DoS attack.}\label{subsec:evaluation_result_lane_following} 
In this scenario, the AD vehicle keeps cruising in the current lane while static or dynamic obstacles located outside of the current lane boundaries. Leveraging V1, V8, or V9, the attacker can cause the AD vehicle to decelerate or permanently stop in the current lane, which effectively prevents the AD vehicle from reaching the destination. In this scenario, since the designed drivable space (i.e., the current lane) is completely empty with no obstacles invading the lane boundaries, such BP decisions thus violate PI1 or PI2 depending on the road structure (i.e., single-lane or multiple-lane road). As discussed in \S\ref{subsec:mov_example}, the root cause of such vulnerabilities is the setting and usage of the lateral obstacle safety buffer, which leads to the overly-conservative BP decisions.

To exploit V1 or V8, the attacker needs to find a narrow urban road (e.g., 2.7m in width~\cite{lane_width}) and place static obstacles close to the lane boundaries. The AD vehicle will then permanently stop in front of the static obstacles. To exploit V9, multiple attackers can coordinate to drive two vehicles in front of the AD vehicle on the lanes other than AD vehicle's current lane to trigger a deceleration decision.

\textbf{Attack scenario 2: intersection passing DoS attack.}\label{subsec:evaluation_result_intersection}
The third type of attack happens when the AD vehicle is approaching an intersection. V5--7 belongs to this type and enable the attacker to cause the AD vehicle to stop in front of the crosswalk or even permanently stop before the stop sign. Since the static objects are all off-road and the dynamic objects' movement will not affect AD vehicle's planning behavior, the stopping decisions produced by BP thus violate PI5 and PI6. When passing the crosswalk, the AD vehicle needs to make sure there is no pedestrian inside the crosswalk or with the intention to move into the crosswalk. However, due to the overly-conservative distance checking between the AD vehicle's driving path and a standing pedestrian (V5) or trajectory "collision checking between AD vehicle and a moving pedestrian (V6), the BP decides to stop before the crosswalk despite its driving path is in fact clear. For the intersection with stop signs, the BP maintains a watch list for the objects that arrive earlier such that it can proceed following a first-come first-serve convention. However, due to the overly-conservative distance threshold to the closest lanes when considering which objects it should wait for, the BP mistakenly includes parked bikes off the road into the watch list. Since these bikes are static, the AD vehicle keeps waiting for them and thus permanently stops before the stop line.


By exploiting V5 and V7, the attacker can cause the AD vehicle to permanently stop at the intersection. To achieve that, the attacker only needs to place \newparts{a} standing pedestrian or parked bikes around the intersection. For V6, since the pedestrian must be moving and will eventually leave the intersection, the attacker can thus carefully control the movement of the pedestrian such that the AD vehicle continuously applies a large deceleration, which may pose safety threats to the passengers and other vehicles (\S\ref{subsec:attack_goal}).


\textbf{Attack scenario 3: lane changing DoS attack.}\label{subsec:evaluation_result_lane_change}
This type of attack happens when the AD vehicle is about to change lanes or borrow the reverse lane due to the routing requirement or a blocking static obstacle. By exploiting V2--4, the attacker is able to use non-blocking static obstacles or following vehicles to prevent the AD vehicle from performing the desired lane changing or borrowing behaviors. As the changing and borrowing lanes are clear in such scenarios, the vulnerabilities thus make the BP violate PI3 and PI4 for the lane changing and borrowing scenarios. Specifically, V2 and V3 are caused by overly-conservative design when checking if the AD vehicle's future driving path overlaps with other vehicles during the lane changing or borrowing. Although PlanFuzz mainly aims \newparts{to} find BP DoS vulnerabilities introduced by overly-conservative planning decisions, V4 is in fact due to an implementation bug in the BP when it checks whether the perception range is blocked by any obstacles before performing lane borrowing. More details of the vulnerabilities can be found in Appendix~\ref{sec:appendix_v3_lane_borrow_off}.

The attacker can exploit V2 by driving a vehicle tailgating the AD vehicle in the same lane. As long as the attacker's vehicle is close to the left lane line (but without touching the lane line), the AD vehicle will mistakenly interpret that the changing lane is blocked and give up the lane changing attempt, which in the worst case causes significant delays for the AD vehicle to reach its destination if the attacker keeps performing such \newparts{an} attack. To exploit V3 and V4, 
the attacker first needs to find a lane borrowing condition where the road is blocked by a static obstacle (e.g., a truck which is unloading the cargo). Second, the attacker can park another vehicle in front of the truck (V3) or simply place an off-road cardboard box 5m away from the AD vehicle (V4).

\cut{ 
\textbf{Experiment Results.}
Table~\ref{tab:vuln_discovered} shows the details of each BP DoS vulnerability discovered by our testing tool. We provide details of the vulnerable planner, how to exploit the vulnerability, the desired planning behavior, and the attack-influenced planning behavior. All these vulnerabilities can be exploited to adversely delay the progress of the AV from reaching its destination, and some of them can even cause the AV to permanently stop on the road. Vulnerabilities V1--4 exist in both Apollo 3.0 and 5.0. The root cause of V1 is the over-approximation of danger level of static obstacles off the road. V2 is caused by an overly-large check range when planner checks whether the lane changing path is clear or not. V3 happens since Apollo's planner mistakenly thinks the blocking vehicle is waiting for other driving vehicle despite that vehicle is static. V4 is caused by a wrong calculation of angle such that planner comes to the conclusion that perception is largely blocked even though the blocking obstacle is off the road and far away from the AV.
Both V5 and V6 are caused by improper filtering conditions. Apollo considers a pedestrian standing on the roadside or moving away from the intersection as a pedestrian it should be waiting for in front of the crosswalk. 
V7 is caused by a wrong obstacle selection logic. Apollo's planner even thinks a parked static bicycle off the road is waiting in front of a stop sign and the victim AV should wait until the static bicycle has left the intersection. Both V8 and V9 are caused by overly-conservative estimation of static or dynamic obstacles in Autoware so that the victim AV stops to avoid collision even though there is no chance to crash into other obstacles if driving normally. 

We further classify the vulnerabilities into three types based on the attack scenarios and introduce more details for each type of attacks.

\nsubsubsection{Single Lane Following Attack}\label{subsec:evaluation_result_lane_following} V1, V8, and V9 can be exploited when the victim AV is driving on a single lane. V1 and V8 share the same symptom, the victim AV will permanently stop in the current lane when the attacker put static objects off the lane. This vulnerability can be discovered in both Apollo and Autoware. V9 is caused by dynamic physical objects. The victim AV will decelerate when there are parallel dynamic objects driving on the adjacent lanes. 

\textbf{Planning invariant.} We define the single lane following attack as the attack which happens when the victim is driving inside a single lane road. The attacker can exploit the vulnerability by using the off-lane static or dynamic physical objects to force the victim AV permanently stop or decelerate. By giving constraints to the position of physical objects, we already make sure that the current lane is clear to drive, but the victim AV still fails to drive normally. The width of the traffic lane is designed to be enough for the vehicle to safely pass as long as the velocity is under the speed limit of the lane. The violated driving norm behind this is that the victim should never block the normal and reasonable movement of traffic. Based on California driver handbook~\cite{calihandbook}, the driver may be cited if violating this. Or in a more extreme scenario, if the victim AV is attacked inside a tunnel or 15 feet within a fire station's driveway, the victim AV is explicitly breaking the traffic rule (California Vehicle Code 22500~\cite{cvc22500}) and may cause more severe consequences in real world. \todo{i like the extreme cases here. maybe also consider ambush attacks? Think about personal gain attack}

\textbf{Root cause.}
The root cause of this type of scenario is the improper setup of the lateral buffer when deciding whether the current lane is blocked or not. As we described in the motivating example (\S\ref{subsec:mov_example}), Apollo maintains a 0.4m lateral buffer between the boundary of the car and the boundary of the static obstacle regardless of the current lane width. If the remaining space inside the lane is not enough for the lateral buffers, a stop decision is required to avoid collision in Apollo's view. When preventing future collision with static obstacles, Autoware will add a safety border around the AV with a width 2.4m plus the car width. As a result, two static objects within 2.4m plus car width can successfully mark all the potential planned trajectories as blocked and thus block the victim AV permanently. Autoware uses a similar design which enforces that dynamic obstacles should always be at least 1.2m plus half car width away from the potential planned trajectory. Thus two dynamic obstacles off the current lane of the victim AV can also cause the AV to make deceleration decision to avoid future collision. 

\textbf{Potential exploits in real world.}
To exploit V1 or V8, the attacker needs to use innocent static obstacles, such as cardboard boxes or trash cans to avoid suspicion of the traffic participants.
Another requirement of the attack is to find a road segment where the attacker can put static obstacles in the side of the lane without raising any suspicion. This type of road is very common in the metropolitan area \todo{give examples: single-lane one-way roads}. Once the attacker is aware of the vulnerability, the attacker just need to make sure the lateral distance between the static obstacles on the different side is within the attack range, the victim AV then will permanently stop in front of the static objects. To exploit V9, the attacker needs to carefully control two vehicles on the other lanes to trigger deceleration decision. There is less limit to the road segment since the only requirement here is there are at least 3 lanes regardless of directions. The attacker can control the timing of two dynamic vehicles such that two dynamic vehicles appear at the same time in front of the victim AV. The attack consequence is that the victim AV will decelerate at the attacked moment. If the attack happens when the victim AV is driving at a high speed, the victim will decelerate with the maximal deceleration allowed to avoid collision, which could be dangerous especially on the highway. At the same time, sudden deceleration will greatly lower the passengers' trip experience. 

We further create a concrete end-to-end case study with simulator for V8 in the next section.

\nsubsubsection{Changing Lane Attack}\label{subsec:evaluation_result_lane_change} This type of attack happens when the planning behavior of the victim AV involves multiple lanes, such as changing a lane and borrowing the reverse lane to pass the blocking obstacles. Exploiting V2--4 can be classified into this type. 

\textbf{Planning invariant.} By enforcing the constraints of objects of planning invariant, the lane in front of the victim AV is clear and the lane changing behavior should not be impacted. In real world driving, if the victim AV has requirements to change to the other lane or borrow the reverse lane to bypass the obstacles, the AV should change the lane to avoid blocking normal traffic.

\textbf{Root cause.} The root cause behind V3 is that Apollo recognizes a blocking static object as a driving vehicle waiting in a queue. Apollo mistakenly recognize two static vehicles on the same lane as two cars waiting in a line. As a result, Apollo should wait for the two previous vehicles until the lane is clear. The root cause behind V4 is that when making a lane changing decision, Apollo will first check whether the perception is blocked. Due to the wrong checking condition, even an off-road static obstacle is considered as blocking the perception and Apollo will not change the lane out of safety consideration. V2 is caused by the overestimated danger of surrounding vehicles. When checking whether the changing lane is clear, Apollo will set a lateral and longitudinal threshold around the AV and make sure that the area within this threshold is clear. Due to the large threshold set in Apollo, a car following the victim AV in the current lane will be recognized as a danger vehicle for the lane changing. 

\textbf{Potential exploit in real world.} Attacker may want to exploit V3 and V4 since the two vulnerabilities can block the normal traffic flow and cause traffic congestion. The attacker first needs to create a lane borrow scenario for the AV. The attacker can park a vehicle on a single lane road with emergency light. The AV by design should borrow the reverse lane to bypass the blocking vehicle when it is safe to do so. However, the attacker can exploit V3 by parking another vehicle in front of the previous vehicle or exploit V4 by putting a static object off the road. Our testing result shows that one 1m cube box can successfully trigger V4 even it is 5m away from the victim AV, which makes the attack easy to do without raising any 
suspicion.
As a result, when an victim AV is driving on this lane, it will trigger a permanent stop in front of the attacker's car. To exploit V5, the attacker can drive another car and tailgate the victim AV. As a result, the victim AV will always fail to change the lane. This will greatly damage the user experience since the traveling time will increase and the victim AV may fail to reach the destination in a timely manner. 

We further create a concrete end-to-end case study with simulator for V2 in next section.

\nsubsubsection{Intersection Attack}\label{subsec:evaluation_result_intersection} The third type of attack happens at intersection. The planning logic tends to be more complicated since the AV need to interact with the pedestrian and the vehicles from other direction to safely pass the intersection. V5--7 belong to this type of vulnerability. 

\textbf{Planning invariant.} By adding the constraints that the static objects are all off-road and the dynamic objects' movement will not affect AV's planning behavior, we make sure that the intersection is already clear for the victim AV to pass as long as the traffic rule allows it. Under such circumstance, the AV should always leave the intersection, or otherwise it will violate the traffic rule due to illegal parking in the intersection~\cite{cvc22500}.

\textbf{Root cause.}
V5 and V6 are caused by the improper handling for the crosswalk. When passing the crosswalk, Apollo needs to make sure there is no pedestrian is inside the crosswalk or has an intention to get into the crosswalk. V5 mistakenly considers a standing pedestrian off the road as on road due to a wrong configuration. V6 happens when Apollo checks whether the pedestrian is moving towards to the victim AV. This moving tendency is calculated by the inner-product between the velocity vector of the pedestrian and the relative position of the pedestrian. Due to the improper threshold of the inner-product, the victim AV will consider a pedestrian is moving towards it as long as the distance between it and the pedestrian is getting smaller. This can happen even when the pedestrian is normally walking away from the intersection. V7 happens when Apollo handles when to start passing the intersection with stop signs. Due to the traffic rule, AD software has to maintain a waiting list when approaching a stop line. Whenever a vehicle stops earlier than the AV, AV needs to put it into the waiting list and remove it when the vehicle has left the current intersection. The AV can start only if this list is clear. However, Apollo does not check which objects should be put into the waiting list in a clear way, and as a result, a static bicycle off the road is put into the waiting list. Since the bicycles are static, the bicycles will always be on the list and the car will stop in front of the stop line forever. An interesting fact is that Apollo has a time out design to handle the situation that one object is always in the list and the AV is stuck forever.  When the time out is reached and there is only one object in the watch list, Apollo will ignore that object. However, the time out design can not defend against exploiting this DoS vulnerability since the attacker can control more than one object to bully the AV. 

\textbf{Potential exploit in real world.} By exploiting the intersection vulnerabilities, the attacker can achieve two attack goals: force the AV improperly stop at an intersection or trigger an emergency brake. The attacker can exploit V5 and V7 to trigger a permanent stop at the intersection. The attacker needs to place a static pedestrian or static bicycles around the intersection. Due to the pedestrian must move in V6, it might be hard to trigger a permanent stop for V6. But the attacker can carefully control the movement of the pedestrian so that the victim AV decelerate in front of the crosswalk. This may lead to severe danger problem.
}  

\nsubsection{Baseline Comparison} \label{subsec:evaluation_ablation}
\newparts{Since there is no existing alternative fuzzer that directly performs BP DoS vulnerability discovery, we evaluate the benefits of our designs by replacing important design components in PlanFuzz with possible baseline designs. Specifically, through such an evaluation we aim at answering the following methodology-level research questions (RQ):



\textbf{RQ1.} Can our BP vulnerability distance design (\S\ref{subsec:vuln_dist}) provide effective guidance to benefit the vulnerability discovery?

\textbf{RQ2.} Can our PI-aware physical-object generation design (\S\ref{subsec:design_mutation}) benefit the vulnerability discovery?

\textbf{RQ3.} Can traditional fuzzing techniques without using any of our problem-specific fuzzing designs also effectively discover BP DoS vulnerabilities?}

\newparts{\textbf{Baseline setups.} To answer RQ1, we create a baseline setup, \textit{PlanFuzz}$^{-\textrm{\textit{guide}}}$, that replaces the BP vulnerability distance-guided genetic algorithm with random sampling (while still using all other PlanFuzz components such as PI-aware physical-object generation). For RQ2, we create another baseline setup, \textit{PlanFuzz}$^{-\textrm{\textit{PI}}}$, that keeps the BP vulnerability distance design but remove the steps that enforce PI constraints in the generated attack objects' static and dynamic properties, which are the key problem-specific designs in PI-aware physical-object generation (\S\ref{subsec:design_mutation}). For RQ3, we remove both BP vulnerability distance and PI-aware physical-object generation designs; we directly use Protobuf-mutator for the entire test input generation process (denoted as \textit{PB-M}). Protobuf-mutator is a readily-available fuzzer designed specifically for the data structure of protobuffers~\cite{protobufmutator}, which is directly compatible with the BP input data structure in both Apollo and Autoware.}






\newparts{\textbf{Evaluation setup and metrics.} We use the initial testing seeds that allow PlanFuzz to discover the 9 vulnerabilities in Table~\ref{tab:vuln_discovered} to perform this baseline evaluation. For each seed, we run each of the 4 setups (full PlanFuzz and the 3 baselines) for 10 times, each time for 24 hours, to avoid variance as suggested by prior work~\cite{klees2018evaluating}. For each fuzzer setup, we measure the discovered unique vulnerabilities (defined in~\S\ref{subsec:evaluation_setup}), and  their average Time-to-Exposure ($\mu$TTE), i.e., the time taken to discover them. When comparing a given baseline setup with the full PlanFuzz, we also use statistical testing following the suggestions in~\cite{arcuri2011practical, klees2018evaluating, bohme2017directed}. Specifically, we use the \textit{Vargha-Delaney statistic $\Hat{A}_{12}$} to measure the effect size, and use \textit{Mann-Whitney U}~\cite{arcuri2011practical} to measure the statistical significance of the $\mu$TTE performance drop, which is the recommended measure for assessing randomized algorithms like fuzzers~\cite{arcuri2011practical, bohme2017directed}.}

\newparts{\textbf{Results.} As shown in the last column of Table~\ref{tab:ablation_study}, PB-M, the setup without using any of our main problem-specific designs, fails to detect \textit{any} of the 9 BP DoS vulnerabilities within 24h, which is thus at least $57\times$ less efficient/effective. This concretely shows that our two main problem-specific fuzzer designs, BP vulnerability distance (\S\ref{subsec:vuln_dist}) and PI-aware physical-object generation (\S\ref{subsec:design_mutation}), are \textit{necessary} for effectively discovering BP DoS vulnerabilities (RQ3).


For PlanFuzz$^{-\textrm{guide}}$ and PlanFuzz$^{-\textrm{PI}}$, the setups that each still retains one of these two problem-specific designs, the same set of unique vulnerabilities (as the full PlanFuzz) are still discoverable within 24h. However, their discovery efficiency degrades substantially compared to the full PlanFuzz. Specifically, for almost all vulnerabilities (7/9 for PlanFuzz$^{-\textrm{guide}}$, and 8/9 for PlanFuzz$^{-\textrm{PI}}$), the $\mu$TTE performance drops are \textit{statistically significant} (bold $\Hat{A}_{12}$ values in Table~\ref{tab:ablation_study}). From the $\mu$TTE values, PlanFuzz$^{-\textrm{guide}}$ and PlanFuzz$^{-\textrm{PI}}$ are on average \textit{$>$4.5$\times$} and \textit{$>$3.5$\times$} slower respectively; for complicated cases such as V1, such degradation can be even \textit{over 7.7$\times$} for PlanFuzz$^{-\textrm{guide}}$. Among the 9 vulnerabilities, V4 is the only one that does not show significant $\mu$TTE differences for both PlanFuzz$^{-\textrm{guide}}$ and PlanFuzz$^{-\textrm{PI}}$; this is because it is relatively easier to trigger by nature due to a likely range-checking bug (Appendix~\S\ref{sec:appendix_v4_lane_borrow_in}), which can be seen by its much lower $\mu$TTE values (1 sec). However, even so, without \textit{both} of these problem-specific designs (BP vulnerability distance and PI-aware physical-object generation), they still \textit{cannot} be discovered by traditional fuzzers like PB-M even given 24h as shown in Table~\ref{tab:ablation_study}.
}

\begin{table}[tbp]
\footnotesize
\centering
\caption{\newparts{Results of baseline comparison. PlanFuzz$^{-\textrm{guide}}$: PlanFuzz without BP vulnerability distance as guidance. PlanFuzz$^{-\textrm{PI}}$: PlanFuzz without PI-aware physical-object generation. PB-M: Protobuf-mutator (a directly-compatible traditional fuzzer). Each setup is run 10 times, 24h each time. $\mu$TTE is average Time-To-Exposure. NF: Vulnerability not found in 24h. \textbf{Bold} $\Hat{A}_{12}$ values denote statistically significant $\mu$TTE performance drops.}}
\label{tab:ablation_study}

\setlength{\tabcolsep}{2.5pt}

\begin{tabular}{ccccccccc}
\midrule
          \multirow{2}{*}{\begin{tabular}{c}Seed\\\end{tabular}}        &  \multirow{2}{*}{\begin{tabular}{c}Uniq.\\ vuln\end{tabular}}  & PlanFuzz & \multicolumn{2}{c}{\begin{tabular}{c} PlanFuzz$^{-\textrm{guide}}$ \end{tabular}}  & 
          \multicolumn{2}{c}{\begin{tabular}{c} PlanFuzz$^{-\textrm{PI}}$\end{tabular}} & \multicolumn{1}{c}{\begin{tabular}{c} PB-M \end{tabular}} \\
                  &    &$\mu$TTE    & $\mu$TTE        & $\Hat{A}_{12}$ & $\mu$TTE       & $\Hat{A}_{12}$   & $\mu$TTE    \\
\midrule
          1        & V1 &  165s   & 1278s (7.74$\times$)    &   \textbf{0.97}    & 276s (1.67$\times$)& \textbf{0.88} & NF        \\
                  \midrule
           2       & V2 &   19s   &  21s (1.11$\times$)     &  0.55              & 117s (6.15$\times$) & \textbf{0.98}       & NF        \\
                  \midrule
\multirow{2}{*}{3}    & V3 & 16s  &  34s (2.12$\times$)     &  \textbf{0.88}     & 53s (3.31$\times$) &     \textbf{0.88}   & NF     \\ 
                    & V4 &  1s    &   1s (1.00$\times$)     &  0.57              & 2s (2.00$\times$)  &   0.58     & NF        \\
                
                  \midrule

\multirow{2}{*}{4} & V5 &     47s  &  92s (1.95$\times$)    &  \textbf{0.89}     & 167s (3.55$\times$) &    \textbf{0.98}    & NF        \\
                  & V6 &     35s   &  78s (2.22$\times$)    &  \textbf{0.83}     & 148s (4.22$\times$) & \textbf{0.93} &NF        \\
                  \midrule
          5        & V7 &      45s & 119s (2.64$\times$)    &  \textbf{0.97}     & 208s (4.62$\times$) & \textbf{0.96} &NF         \\
                  \midrule
          6        & V8 &    53s   & 193s (3.64$\times$)    &  \textbf{0.96}     & 327s (6.16$\times$) & \textbf{0.90} &NF       \\
                 \midrule 
          7        & V9 &     37s  & 57s (1.54$\times$)     &  \textbf{0.93}     & 188s (5.08$\times$) & \textbf{0.96} &NF     \\
          \midrule
    \multicolumn{2}{c}{Average}& 46s  & 208s ($4.52\times$)    &  \textbf{0.83}     & 165s (3.58$\times$)   &  \textbf{0.89} &    NF  \\
                 \bottomrule
\end{tabular}
\vspace{-0.1in}

\end{table}

\cut{
\begin{table}[tbp]
\footnotesize
\centering
\caption{\newparts{Results of baseline comparison. PlanFuzz$^{-\textrm{data}}$: PlanFuzz without data flow distance in the guidance. PlanFuzz$^{-\textrm{guide}}$: PlanFuzz without the whole BP vulnerability distance as guidance. PB-M: Protobuf-mutator (a directly-compatible traditional fuzzer). Each setup is run 10 times, 24h each time. $\mu$TTE is average Time-To-Exposure. NF: Vulnerability not found in 24h. }}
\label{tab:ablation_study}
\begin{threeparttable}

\setlength{\tabcolsep}{2.5pt}
\begin{tabular}{ccccccccc}
\midrule
          \multirow{2}{*}{\begin{tabular}{c}Seed\\\end{tabular}}        &  \multirow{2}{*}{\begin{tabular}{c}Uniq.\\ vuln\end{tabular}}  & PlanFuzz & \multicolumn{2}{l}{\begin{tabular}{c} PlanFuzz$^{-\textrm{data}}$\end{tabular}}  & \multicolumn{2}{l}{\begin{tabular}{c} PlanFuzz$^{-\textrm{guide}}$\end{tabular}} & \multicolumn{1}{l}{\begin{tabular}{c} PB-M \end{tabular}} \\
                  &    &$\mu$TTE    & $\mu$TTE        & $\Hat{A}_{12}$ & $\mu$TTE       & $\Hat{A}_{12}$   & $\mu$TTE    \\
\midrule
          1        & V1 &    165s    &    374s (2.26$\times$)   &    \textbf{0.89}  & 1278s (7.74$\times$)                &   \textbf{0.97}    & NF        \\
                  \midrule
           2       & V2 &      19s  &  18s (0.94$\times$)      &   0.53    &   21s (1.11$\times$)               &  0.55     & NF        \\
                  \midrule
\multirow{2}{*}{3}    & V3 &   16s     &   21s (1.31$\times$)          &0.75       &    34s (2.12$\times$)             &  \textbf{0.88}     & NF     \\ 
                    & V4 &    1s    &   2s (2.00$\times$)       &  0.59     &    1s (1.00$\times$)          &   0.57    & NF        \\
                
                  \midrule

\multirow{2}{*}{4} & V5 &     47s   &    62s (1.32$\times$)      &\textbf{0.79}       &    92s (1.95$\times$)             & \textbf{0.89}      & NF        \\
                  & V6 &     35s   &    39s (1.11$\times$ )  &  0.67     &   78s (2.22$\times$)              &  \textbf{0.83}     & NF        \\
                  \midrule
          5        & V7 &      45s  &   73s (1.62$\times$)  &    \textbf{0.90}   &  119s (2.64$\times$)                &  \textbf{0.97}     & NF         \\
                  \midrule
          6        & V8 &    53s    &    87s (1.64$\times$)     &  \textbf{0.83}     &  193s (3.64$\times$)      &          \textbf{0.96}    & NF       \\
                 \midrule
          7        & V9 &     37s   &    39s (1.05$\times$)     &    0.52   & 57s (1.54$\times$)       &             \textbf{0.93}    & NF     \\
                 \bottomrule
\end{tabular}
\end{threeparttable}

\end{table}
}

%% file: case_study.tex
\nsection{Exploitation Case Studies} \label{sec:case_study}
In this section, we provide three case studies on the BP DoS vulnerabilities discovered by our testing framework and demonstrate how an attacker can exploit the vulnerabilities to disrupt the normal driving behavior of the AD vehicle without raising suspicion. Specifically, we select one vulnerability from each of the attack scenarios categorized in \S\ref{subsec:vuln_dis_res}.
The case studies are conducted in an end-to-end way, where we create the concrete driving environments with the attack obstacles in an AD simulator, LGSVL~\cite{lgsvl}, and simulate with the complete AD stack (Apollo or Autoware) in the loop more than 10 times, which also include other AD system components such as localization, perception, and control. We create attack demos for all the case studies listed in this section. Demo videos are available at our project website \textbf{\url{https://sites.google.com/view/cav-sec/planfuzz}}~\cite{planfuzzwebsite}.


\nsubsection{Lane Following DoS Attack on Autoware} 
\label{sec:case_study_lane-following}
\textbf{Vulnerable decision logic.} 
The lane following DoS attack on Autoware (V8) is able to change the AD vehicle's lane following decision into an overly-conservative decision to permanently stop on a clear road, by exploiting the vulnerable decision logic in \texttt{trajectory\_evaluator}~\cite{autoware_traj_eval} (pseudo code can be found in Fig.~\ref{fig:app_v8_code} in the Appendix). This evaluator is designed to decide whether there is a candidate trajectory for AD vehicle to safely pass without crashing into a static obstacle. Similar to Apollo's design illustrated in \S\ref{subsec:mov_example}, Autoware predefined an overly-conservative lateral safety buffer of 1.2 meters between the AD vehicle and any obstacles.
As described in \S\ref{subsec:mov_example}, since the minimal urban lane width is 2.7m and typical width of AD vehicle (with mirrors) is larger than 2m, two static obstacles out of lane boundaries can block all the candidate trajectories for the AD vehicle.

\cut{
\textbf{Vulnerable decision logic.} We presented the pseudo code of the vulnerable planner logic in Autoware in Figure~\ref{fig:lane_follow_vuln}. This part of code is designed to handle static objects. Autoware generates a list of parallel candidate planning trajectories for the AV and check whether each trajectory is blocked by the static obstacle one by one.
Specifically, Line 9 iterates all the contour points of a physical objects to get precise relationship between the candidate trajectory and the physical object. $lateraldist$ and $longitudinaldist$ are calculated based on the current candidate trajectory and represent the lateral and longitudinal distances between the AV center and the contour point. On line 15, if the lateral distance is within the critical lateral distance range and it is ahead of the AV within the following distance, the object is considered as a blocking static object on line 18. 
After that, it then determines whether the AV is fully blocked by merging the decisions on the candidate trajectories (line 20--23). A fully blocking decision is made if all candidate trajectories are blocked by the static object.
We can exploit this logic by putting static objects off the road and block all the candidate trajectories since the $critical\_lateral\_distance$ is much larger than the half lane width~\cite{lane_width}. }

\newparts{\textbf{Exploitation method.} To exploit this vulnerability, the attacker can find a single-lane road up to 4.35m wide (applicable to both common local and highway roads~\cite{lane_width}), and put two easy-to-carry objects, e.g., cardboard boxes, on each side of the road. There is no strict relative longitudinal position requirement for them, as long as they are in the same planning range (35m for Autoware). Since Autoware uses a more conservative safety buffer size than Apollo, for a narrow road width such as 2.7m~\cite{lane_width}, each of such object can be $>$80cm to the road boundary to make them look more stealthy. An example setup in the simulator is shown in Fig.~\ref{fig:path_bounx_demo}~(C). This can cause an emergency stop and/or permanent stop of the victim AD vehicle and thus damage the AD service, block traffic, and also potentially damage road safety (from emergency stop).}

\textbf{End-to-end attack results.} We set up the above exploitation scenario in the simulator. From the simulation, after detecting the cardboard boxes, the AD vehicle immediately starts to decelerate and finally comes to a complete stop at $\sim$10m in front of the boxes. After the full stop, we keep the simulator and AD system running to observe if the AD vehicle will move forward. After waiting for $\sim$30min, the vehicle still stops at the original position. We manually examined the code and confirmed that this will be a permanent stop. A demo video is at our website~\cite{planfuzzwebsite} and the snapshot when the AD vehicle stops in front of the boxes is in Fig.~\ref{fig:path_bounx_demo}~(C). As shown, the cardboard boxes are located very far from the lane boundaries and the lane is completely clear to drive from the driver's view.

Such a vulnerability can lead to severe congestion and even rear-end collisions if the following vehicle is not able to react in time; video demos are at our website~\cite{planfuzzwebsite}. The attack consequence can be especially severe if launched at critical road segments such as highway exit ramps, or in front of police or fire stations (e.g., to block emergency responder actions). 



\begin{figure*}[t]
\centering
\begin{minipage}{0.185\linewidth}
  \includegraphics[width=\columnwidth]{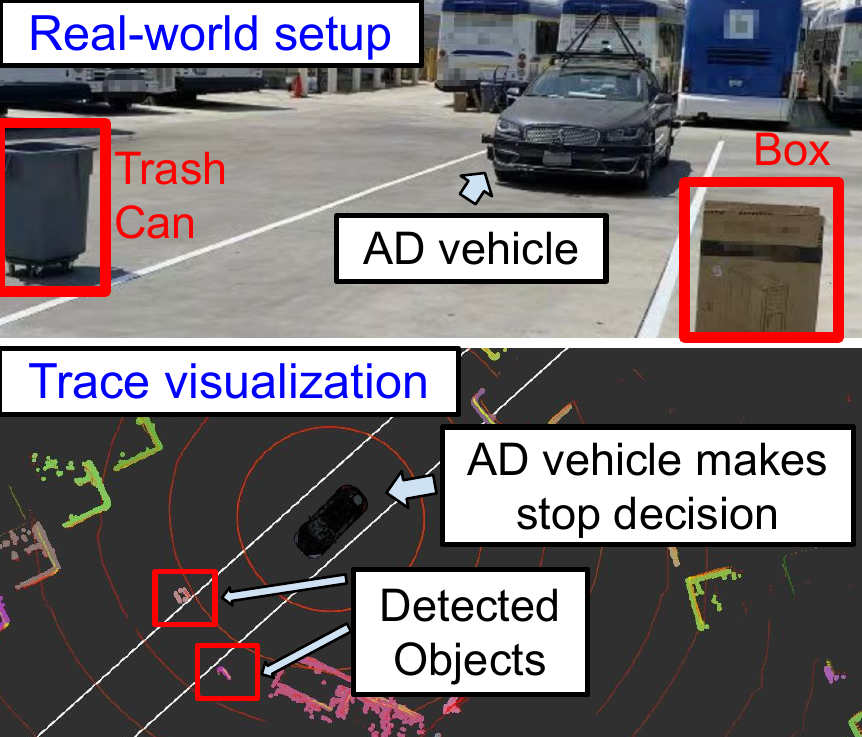}
  \vspace{-0.2in}
  \caption{Real-world experiment setup (top) and sensor trace visualization (bottom).} 
  \label{fig:case_study_lane_follow_real_world}
\end{minipage} \hfill
\begin{minipage}{0.395\linewidth}
  \includegraphics[width=\columnwidth]{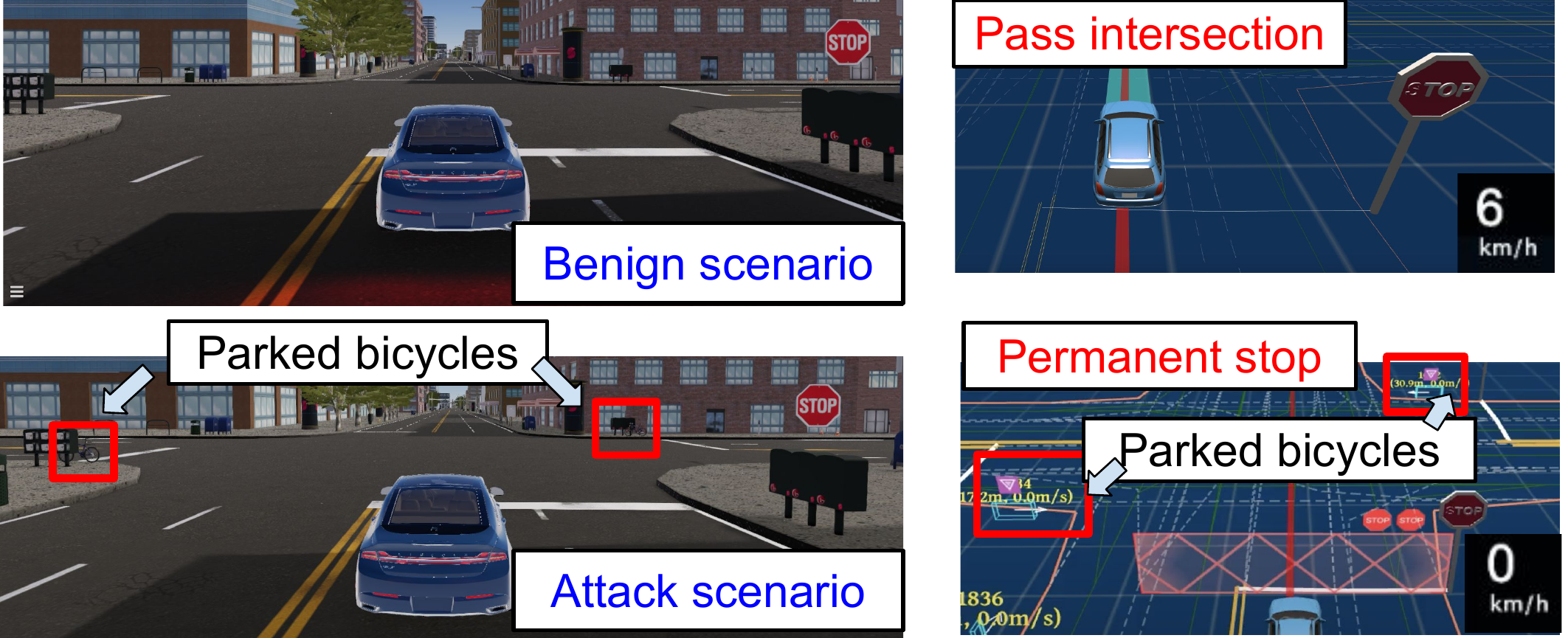}
  \vspace{-0.2in}
  \caption{Benign (top): AD vehicle passes the intersection. Attack (bottom): AD vehicle permanently stops because of the parked bicycles placed by the attacker.} 
  \label{fig:case_study_stop_sign}
\end{minipage} \hfill
\begin{minipage}{0.395\linewidth}
\includegraphics[width=\columnwidth]{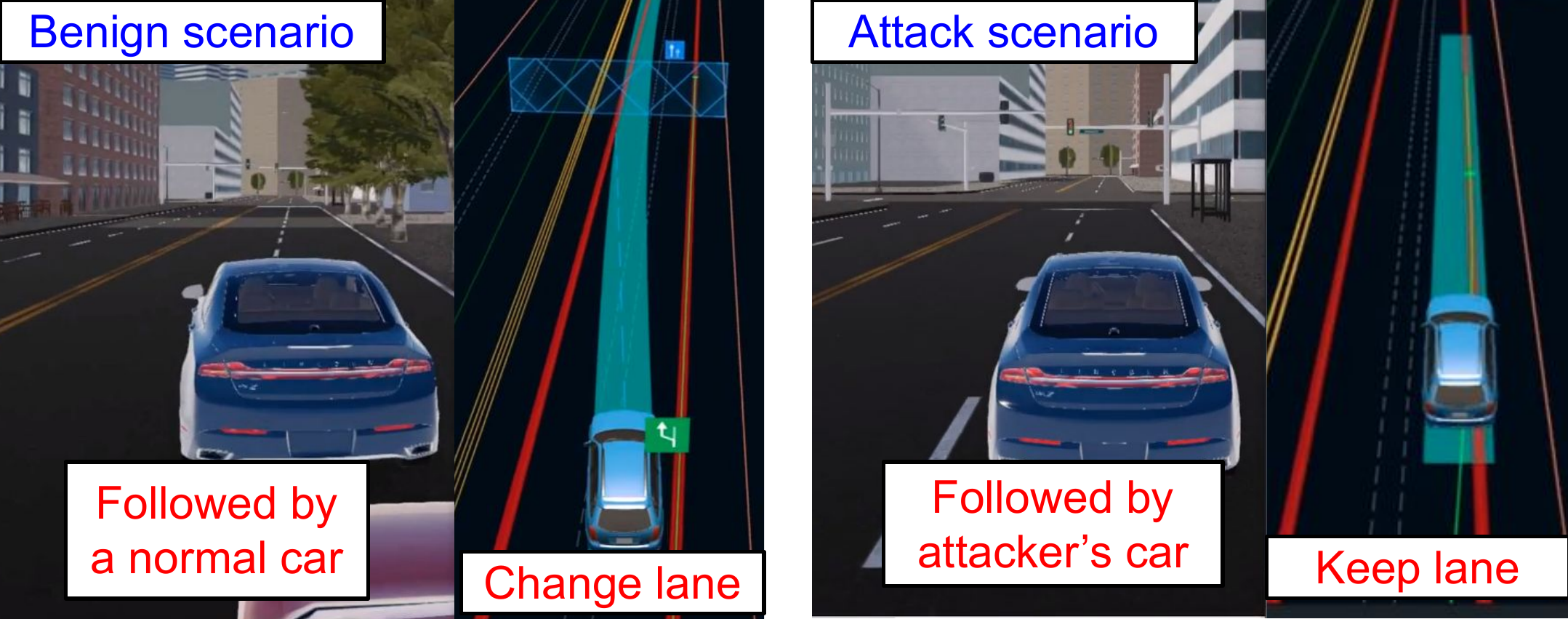}
  \vspace{-0.2in}
\caption{Benign (left): Successfully changes lanes even if it is followed by another vehicle. Attack (right): Fails to change lanes due to the attacker's vehicle.} 
\label{fig:case_study_lane_change}
\end{minipage}
\vspace{-0.05in}
\end{figure*}

\textbf{Physical-world experiment.} We further collect attack traces in real world to justify the attack realism. We conduct the experiment with a Lincoln MKZ~\cite{lincolnmkzdatasheet} equipped with Velodyne VLP-32C LiDAR~\cite{lidardatasheet} and NovAtel Positioning Kits~\cite{gnssdatasheet}. We mark a traffic lane with 3.5m width inside the parking lot and use a cardboard box and a trash can to construct the attack scenario. Due to the safety concern, we manually drive the car following the traffic lane to collect the attack trace. We run the Autoware's LiDAR clustering and tracking nodes~\cite{autowarelidarcluster, autowarelidartrack} to get the object detection results and launch the op\_planner~\cite{darweesh2017open} nodes to get the planning decisions. For all the frames in our collected traces, the two static obstacles can be detected correctly and the stop decision is made for every single frame in the traces. Thus, we come to the conclusion that the Autoware's lane following vulnerability can be reconstructed in real world. Fig.~\ref{fig:case_study_lane_follow_real_world} shows our experiment setup and trace visualization result.

\nsubsection{Intersection Passing DoS Attack on Apollo} \label{sec:case_study_stop_sign}

\textbf{Vulnerable decision logic.}
This DoS vulnerability (V7) appears in Apollo, where the attacker can force an AD vehicle to stop permanently in front of an intersection with a 4-way stop sign. As shown in Fig.~\ref{fig:app_v7_code}~\cite{apollo_stop_sign}, Apollo BP follows the ``first come, first serve'' traffic principle when the AD vehicle arrives at a 4-way stop sign. The vehicle maintains a watch list containing all vehicles and bicycles which reaches the intersection earlier than itself and waits until all vehicles and bicycles on that list have left the intersection. However, due to an overly-conservative distance threshold (5m), which is much larger than even the typical \textit{highway} lane width (3.6m~\cite{lane_width}), when judging if a vehicle or bicycle is on a lane and waiting for a stop sign, a parked bicycle that is not on the road will be mistakenly added to the watch list, which causes the AD vehicle to unnecessarily wait for off-road objects that are irrelevant to the stop sign-based intersection passing norms.





\newparts{\textbf{Exploitation method.} To exploit this, the attacker can find any stop sign-based intersections, and place road-side parked bikes as long as they are within 5m from the lane center of any lanes in the intersection (illustrated in Fig.~\ref{fig:app_v7_scenario}). Here, 2 parked bicycles are enough to bypass all timeout mechanisms in Apollo BP (detailed in Appendix~\ref{sec:appendix_v7_stop_sign}). This can then force the AD vehicles to stop in front of the stop line forever.}



\textbf{End-to-end attack result.} We set up the exploitation scenario above in the simulator. Fig.~\ref{fig:case_study_stop_sign} shows the snapshots of the benign and attack demos with and without the 2 parked bicycles. In the benign scenario, the AD vehicle can smoothly proceed and pass the intersection with stop signs. However, in the attack scenario, after the AD vehicle arrives \newparts{at} the intersection, it became stuck in front of the stop sign due to the two roadside parked bicycles despite the intersection \newparts{being} completely empty without any other vehicles.

\cut{
\textbf{Vulnerable decision logic.} We presented the pseudo code of the vulnerability in Figure~\ref{fig:lane_change_vuln}. The whole program shown in the figure is used to decide whether it is safe to perform lane changing given the positions of surrounding vehicle objects. All the position variables in the code are in the Frenet coordinate based on the target lane of the lane changing. The first check on line 5 is used to check whether the vehicle object is located on the target lane. If not, this vehicle will be ignored. The second check on line 8 is used to check the longitudinal distance between the AV and the vehicle. If the distance is smaller than a backward safety distance, then Apollo will come to the conclusion that it is not a good time to change the lane since another vehicle is driving on the target lane and the longitudinal distance between them is less than a safety distance. The vulnerable code is on line 5. Since the lateral buffer is too large, a vehicle following the victim AV is also considered as a blocking vehicle (usually the half lane width is smaller than 2m even on the highway~\cite{lane_width}).}


\cut{
to demonstrate the \nsubsection{Lane Changing DoS Attack on Apollo} \label{sec:case_study_lane_changing}}

\nsubsection{Lane Changing DoS Attack on Apollo}
\label{sec:case_study_lane_changing}
\textbf{Vulnerable decision logic.} The lane-changing DoS attack on Apollo (V2) is able to force the AD vehicle to give up a lane-changing decision by exploiting decision logic in \texttt{lane\_change\_decider}~\cite{apollo_lane_change_decicder}.
As shown by the pseudo code in Fig.~\ref{fig:app_v2_code}, the decision logic determines whether the target lane is clear for performing lane changing by checking if any vehicle occupies the target lane and is close to the AD vehicle. In the code, all the position variables (e.g., \texttt{start\_l}, \texttt{start\_s}) are in the Frenet coordinate relative to the target lane.
Due to an overly-conservative lateral distance threshold (2.5m in line 5), a nearby vehicle following the AD vehicle or driving on the other adjacent lane will be considered as occupying the target lane and force the AD vehicle to give up the lane-changing decision.



\begin{figure}[tb]
\centering
\hrule
\begin{minted}[fontsize=\footnotesize,numbersep=5pt,xleftmargin=10pt,linenos,escapeinside=||,mathescape=true]{python}
# Iterate over the obstacle list
for (each obs : obstacle_list):
  # Judge based on the lateral position of a veh.
  # start_l, end_l are veh's left/right boundaries
  if (obs.end_l < -2.5 or obs.start_l > 2.5):
    continue
  # The backward safe buffer
  BackwardSafeBuffer |$\leftarrow$| 4.0f
  # Check whether the veh is close 
  if (ego_start_s-obs.end_s<BackwardSafeBuffer):
    IsClearChangeLane |$\leftarrow$| false
\end{minted}
\hrule
\vspace{0.05in}
\caption{Simplified vulnerable pseudo code of V2 (lane changing).}
\vspace{-0.2in}
\label{fig:app_v2_code}
\end{figure}

\newparts{\textbf{Exploitation method.} To exploit this, for lanes with $\leq$3.55m width (applicable to almost all common local and highway lanes (up to 3.6m wide)~\cite{lane_width}), the attacker just needs to drive a normal-size car (e.g., 2.11m as in~\S\ref{subsec:mov_example}) to follow a victim AD vehicle with close following distance (up to 8m). This can then prevent the AD vehicle from making lane changes forever, which can make it fail to arrive at the destination (especially critical for AD services such as robo-taxi/delivery). For lanes wider than 3.55m, the attacker just needs to drive with a slight deviation to the lane center (e.g., 5cm for a 3.6m-wide lane, which is far from touching the lane line ($>$70cm)) towards the direction she wants to prevent the victim from changing lane to.}

\textbf{End-to-end attack results.} We set up the exploitation scenario above in the simulator. Fig.~\ref{fig:case_study_lane_change} shows the snapshots of the benign and attack demo videos. In the benign scenario, the AD vehicle can successfully change lanes since the following vehicle's lateral position does not satisfy the vulnerability condition. However, in the attack scenario, although the changing lane is completely empty without any vehicles in the front or back, the AD vehicle still gives up the lane-changing decision to stay on the current lane, which causes it to miss the optimal route to the destination. For AD vehicles, such a BP decision often entails a re-routing step to recalculate a new route to reach the destination. However, since the attacker can simply keep following the AD vehicle and perform such attacks on every new route, it is possible for the AD vehicle to never reach the destination in the worst case.



%% file: discussion.tex
\nsection{Discussion and Future works}
\vspace{0.1in}
\nsubsection{Root Cause and Solution Discussions} \label{sec:root_cause_countermeasure}

From the vulnerability discussions in \S\ref{subsec:vuln_dis_res} and Appendix~\ref{sec:appendix_PIsummary}), only 1 (V4) is likely an implementation bug, while the remaining 8 are due to the overly-conservative planning parameters/logic in judging safety. Specifically, V1, V8 are due to overly-conservative safety buffer configuration to road-side static objects; V9 is due to overly-conservative safe buffer configuration to moving vehicle trajectories in other lanes; V2, V3, V5, V6, and V7 are due to the overly-conservative logic in judging the intention of surrounding vehicles (V2, V3), pedestrians (V5, V6), and parked bikes (V7).


\newparts{
\textbf{Solution discussions.} Based on the causes, V4 can be potentially fixed by a bug patch; the other 8 are harder to fundamentally fix as it is non-trivial to effectively and practically balance the trade-off between safety and availability in an AD context. For example, considering the vulnerability causes are overly-conservative parameters/logic, a direct idea is to just make the design/implementation more aggressive, e.g., reducing the safety buffer configuration to road-side static objects. However, since such existing configurations may already be sufficiently tuned to ensure safety, changing them may compromise safety. One potential design direction for better addressing this trade-off is to consider dynamic configurations instead of fixed ones, e.g., dynamically adjusting the safety margin based on the velocity, since the required safety distance will decrease with a smaller velocity~\cite{shalev2017formal} (which is one reason why highway lanes are wider than local ones~\cite{lane_width_speed_relationship}). However, how to design such dynamic adjustment is also non-trivial since various internal and external driving factors need to be systematically considered (e.g., internal ones such as vehicle size/speed, external ones such as the static/dynamic road conditions), which we thus leave as future work. Besides improving the decision parameter/logic, another promising direction for future work is to detect such attacks that exploit conservative designs, e.g., detecting the placement patterns of road-side static obstacles in V1. 
}

\cut{
To eliminate these vulnerabilities, countermeasures can be deployed at different stages of the software development: (1) \textit{development stage}. For example, as V1, V8, and V9 are caused by a overly-conservative pre-defined safety buffer, the developers may need to re-consider its configurations, e.g., setting it up in a more dynamic way considering the road semantics

The developer should design the threshold in a dynamic and careful way to consider all the surrounding traffic environments \junjie{should include an example of how to address the V1, V8, or V9 vuln based on this suggestion}. Also, the developers should carefully consider which object should be filtered out when making a decision \junjie{should include the vuln. that can be eliminated by this suggestion}.

\begin{itemize}
    \item \textbf{Development.} Note that in V1, V8, and V9 are caused by a large pre-defined buffer. The developer should design the threshold in a dynamic and careful way to consider all the surrounding traffic environments \junjie{should include an example of how to address the V1, V8, or V9 vuln based on this suggestion}. Also, the developers should carefully consider which object should be filtered out when making a decision \junjie{should include the vuln. that can be eliminated by this suggestion}. 
    \item \textbf{Runtime.} During the runtime of AD software, we suggest adding emergency handling mechanisms such that a permanent stop decision can be handled without blocking the normal traffic operation. For example, when a stop decision is unchanged longer than a timeout threshold due to surrounding physical objects, the AV should be aware that itself is blocking the traffic and take a proactive strategy, \add{such as re-plan the path or call the human assistance.}  
    \junjie{i like the timeout idea, you should list it here as potential mitigation.}
\end{itemize}
}
\nsubsection{Limitations and Future Work}\label{sec:limitation}
\cut{\textbf{Tool effectiveness.} In the design and implementation of PlanFuzz, all the discovered vulnerabilities has been confirmed by the vulnerability checker and proven to violate the planning invariant. As a result, the analysis results do not have false positives. On the other side, as a dynamic analysis tool, we cannot give any guarantee that there is no false negative. The false negatives may come from the following sources: (1) The evolutionary testing fails to find the vulnerability because it terminates after finding a shallow one or gets stuck at a local minimum. (2) Since the driving trace collection relies on manual efforts and is limited to the maps provided by LGSVL in our current evaluation setting, the collected driving traces can only cover a subset of all the possible planning scenarios. (3) To ensure that the mutated physical objects can fit into the PI constraints, we design the physical object mutation function in a very conservative way. For example, we force the vehicle object to have the same heading with the road. If there is a BP DoS vulnerability triggered by a vehicle object with a different heading, then our tool can not discover it at the current stage. This may limit the tool's vulnerability discovery ability. Considering all the sources of false negatives, we plan to extend PlanFuzz by improving the evolutionary algorithm, automating the trace collection, and extending the range of mutation function.}

\textbf{Tool effectiveness.} PlanFuzz's results do not have false positives since all the discovered vulnerabilities \newparts{have} been confirmed by the vulnerability checker. However, similar to all dynamic testing approaches, PlanFuzz cannot give any guarantee on the non-existence of the vulnerability and thus can have false negatives (FNs). Specifically, FNs can come from: (1) the evolutionary algorithm gets stuck at a local minimum or terminate too early. We plan to try other generic optimization algorithms in the future; and (2) the vulnerability only exists in a specific road condition (e.g., a specific road layout and/or surrounding traffic pattern) that is not included in our BP input traces. These vulnerabilities may be arguably less important though as their triggering conditions are narrower. To also capture these, one potential future direction is to also mutate the road conditions. However, how to ensure that the PI still \newparts{holds} after such mutations is a challenge.

\newparts{
In addition, similar to other dynamic testing approaches, the output of our current design is a set of concrete test inputs that can trigger a discovered vulnerability. However, this can only provide several example triggering placements of the physical objects, not the complete \textit{triggering ranges} of such placements, which can be more informative. To achieve the latter, more sophisticated formal methods such as symbolic execution~\cite{cadar2008klee} can be leveraged, but such methods are known to generally suffer from scalability limitations when applied to real-world code bases~\cite{ramos2015under}, not to mention that in our problem settings one needs to further overcome the semantic gaps by systematically locating and extracting the physical object-related code-level constraints. We thus leave a systematic exploration of this direction to future work.}

\cut{
\textbf{Manual efforts.} Manual efforts are still required even though most of the components in PlanFuzz are automated. For each driving scenario, the user first needs to collect a driving trace in the clean environment as the seed for dynamic testing. A possible improvement is to automated the trace collection process. \junjie{moved this sentence here. should give some suggestion on how to automate the trace collection.} Second, the testing system should be aware which part of the code could lead to decisions expected by the attacker. The user of PlanFuzz needs to specify the attack target positions in the code by adding one line of annotation in the code at each attack position. However, this process only takes one hour to annotate all the attack targets\todo{Needs to be more clear.} for a person who is familiar with the code base. \red{Thus, the BP developers are the best candidates to perform this step since they are naturally familiar with the code.}
}


\cut{
\textbf{Lack of real world experiments.} Although we tried our best \jhl{as an academic work} \josh{Do we need to emphasize that this work is academic? That seems awkward to me.} to demonstrate the exploits of the discovered vulnerabilities in simulation environments (\S\ref{sec:case_study}), there is still some gap between the simulated world and the real world due to the limitation of \st{equipment} \red{inadequate modeling fidelity in the simulator and the perception noises in the AD software} \junjie{here you should explain why the gap exist rather than why we didn't do real world experiment}. This makes us unclear how the imperfection of real-world sensors \red{and perception algorithms} will \st{damage} \red{affect} the exploits. Nevertheless, we tried our best by \st{simulating the attack scenario} \red{leveraging the industry-grade AD simulator, LGSVL,} and \st{launching} \red{executing} the full AD software stack \red{in our simulations}. Note that similar to us, companies such as Waymo and Uber also heavily rely on \st{trace-based and} simulation\st{-based evaluations} \junjie{try not the emphasize ``evaluation'' since this limitation is only on the ``demonstration'' of the discovered bugs.} when developing and testing their driving automation systems for safety and budget considerations~\cite{arg_sim_1, arg_sim_2, arg_sim_3}. In the future, we plan to conduct miniature-scale experiments used by other works~\cite{ningfei-msf-adv} \junjie{cite DRP attack paper as well} to demonstrate the exploits of the discovered vulnerabilities.}


\textbf{Applicability and generality.} Due to the limitation of available code bases, we only evaluate the applicability and generality of PlanFuzz on open-source \newparts{AD systems}. Nevertheless, the design of PlanFuzz is general at both design and implementation levels. The design of PI and PI-aware physical object mutation \newparts{does} not make any assumption about the \newparts{AD system designs and implementations under test} as long as they are developed for driving on public roads. For the code instrumentation part, we use LLVM~\cite{llvm_project}, \newparts{which} can support a variety of programming languages. Besides, our system should be able to use \newparts{the gradient} to replace BP vulnerability distance if learning-based planners becomes mainstream and easier to interpret, debug, and enforce safety measures in the future.

\newparts{\textbf{Stronger threat models.} To trigger the semantic DoS vulnerabilities in this paper, adding road objects (e.g., carbon boxes, bikes, vehicles) is not the only possible threat model. For example, attackers may also attack the perceptional sensors, such as by sensor spoofing~\cite{cao2019adversarial} or compromising internal AD system components~\cite{jha2020ml}, to introduce malicious inputs to BP. Since these attack vectors may have stronger BP input perturbation capabilities/flexibility, more BP DoS vulnerabilities may be discovered. However, the downside is that such attack vectors may also introduce new attack requirements/costs, e.g., sensor spoofing equipment~\cite{cao2019adversarial} and access to internals or the supply chain of AD systems~\cite{jha2020ml}, making the vulnerability exploitation potentially less realistic/practical. We leave the systematic exploration of such a direction to future work.
}

\cut{
\textbf{Applicability and generality.} Although our results have shown that our tool can \st{detect} \red{discover} DoS vulnerabilities in behavior planner in both Apollo and Autoware. Ideally, it is better to evaluate on \st{production-grade ones} \red{more full-stack (ideally industry-grade similar to Apollo) AD systems}\ziwenres{Should we aim high at this sentence?}, but very unfortunately,
Apollo and Autoware are the only AD systems that are publicly available so far and it is unlikely for other AV companies to publicly release their implementations in the near future. Thus, due to the lack of information, it is unclear whether \red{the PlanFuzz proposed in this work can be applied to} \red{AD systems developed by} other leading \red{AV} companies, e.g., Waymo and GM\st{, are vulnerable to the BP DoS vulnerability}. Nevertheless, the design of PlanFuzz is general both at the design and implementation levels. The design of planning invariant and PI-aware physical object mutation function do not have any assumption on the implementation of behavior planner and they should be generally applicable to \st{the} \red{other} behavior planners \red{as well}. When analyzing the source code and calculating the BP vulnerability distance, we use \red{the widely-adopted} LLVM \red{framework} \ziwenres{add ciation here} \junjie{cite} to achieve the analysis and instrumentation, which fully supports C and C++ programming languages, the \red{most} widely-used languages in development of AD software~\cite{garcia2020comprehensive}. Nevertheless, for the potential emergence of machine learning based behavior planners \red{proposed in many academic works} \ziwenres{Also need to add ciation}, the design of planning invariant and PI-aware mutation physical object mutation can still be applied. Even though the BP vulnerability distance can not be directly applied, getting the guidance for testing could be easier since we can directly \st{extract} \red{calculate} the gradient from the machine learning models to guide the testing. }

%% file: related_work.tex
\nsection{Related Work} \label{sec:related_work}
\textbf{Autonomous Driving (AD) systems security.} 
Since AD systems heavily rely on sensors, prior works have studied \textit{sensor attacks} in AD context such as sensor spoofing/jamming~\cite{yan2016can, nassi2020phantom,ningfei-msf-adv,shin2017illusion, cao2019adversarial, tu2018injected}. Besides sensor-level attacks, prior works also studied attacks and defenses of AD system components related to environmental sensing, such as object detection and tracking, localization, and lane detection~\cite{sato2021dirty, sato2021wip, ningfei2021msfadv, liang2021wip,kanglan2021,junjie:usenix:2020,jia2019fooling,eykholt2018physical, chen2018shapeshifter, zhao2018seeing, cao2019adversarial, povolny2020adas, sun2020towards, quinonez2020savior, choi2018detecting}. However, so far none of them considered security problems specific to downstream modules such as BP like \newparts{in} this paper.

\textbf{Vulnerability discovery and property falsification in AD/RV (Robot Vehicle) software.}
Recently, increasingly more works consider software vulnerabilities/bugs in AD context~\cite{garcia2020avbug, hong2020avguardian}. Some developed methods to discover semantic vulnerabilities in the DNN models in AD\cite{eykholt2018physical, chen2018shapeshifter, zhao2018seeing, cao2019adversarial, sato2021dirty, ningfei2021msfadv, sato2021wip, liang2021wip, pei2017deepxplore, tian2018deeptest, zhou2020deepbillboard, zhang2018deeproad}. These methods assume differentiability of the test subject, which thus cannot be applied to the more industry-representative program-based BP targeted in this paper (\S\ref{subsec:background_bp}). Previous works also falsify safety properties on AD/RV software~\cite{li2020av,fremont2020formal, fremont2020formal2, tuncali2018simulation, hekmatnejad2020search, dreossi2019compositional, dreossi2019verifai, kimpgfuzz}. They can not be directly applied since the problem scopes are different and the guidance is only limited to black-box guidance. 



Vulnerability discovery in \newparts{RVs, such as drones or rovers}, is a closely-related research domain. Compared to AD systems, RVs typically follow the control commands sent by a base station without the need to make planning decisions by itself based on the surrounding environment. Thus, existing works concentrate on designs to find control-specific vulnerabilities~\cite{kim2019rvfuzzer,choi2020cyber, kimpgfuzz} or highly rely on control-specific knowledge~\cite{choi2018detecting, kim2020control}, which are thus orthogonal to the design challenges we need to address for discovering semantic vulnerabilities for BP in \newparts{the AD context.}

\newparts{\textbf{Traditional fuzzers.} There are various traditional fuzzers~\cite{afl,libfuzzer} that can automatically discover generic software vulnerabilities such as software crash and memory corruptions. However, they cannot readily be used for our problem setting since they need to at least change the testing oracle from generic software symptoms (e.g., crashes) to the semantic ones (overly-conservative planning decisions). For the latter, a valid and systematic definition is the very first design challenge that needs to be overcome (\S\ref{subsec:design_challenge}). However, that is still not enough since to achieve effective discovery, they further need to add new problem-specific input generation and guidance designs such as those we designed (\S\ref{subsec:vuln_dist}, \S\ref{subsec:design_mutation}); otherwise, they still cannot discover any such vulnerabilities even if given 24h as shown in~\S\ref{subsec:evaluation_ablation}. 
}

%% file: conclusion.tex
   \nsection{Conclusion}

In this paper, we design PlanFuzz, a novel dynamic testing approach to systematically discover BP DoS vulnerabilities under physical-world attacks. We propose and identify PIs as novel testing oracles, and design novel problem-specific fuzzing designs such as PI-aware physical-object generation and BP vulnerability distance. We evaluate PlanFuzz on 3 BP implementations from practical AD systems, and find that it can effectively discover 9 previously-unknown semantic DoS vulnerabilities without false positives. We further perform exploitation case studies using simulation and real-vehicle traces to validate their realism and impacts. We also discuss root causes and potential vulnerability solution directions. We hope that our findings and insights can inspire effective solution designs in future works.

%% file: acknowledge.tex
\section*{Acknowledgment}
We would like to thank Ningfei Wang, Yunpeng Luo, Takami Sato, Junze Liu, and the anonymous reviewers for valuable feedback on our work. This research was supported in part by the NSF under grants CNS-1850533, CNS-1929771, CNS-2145493, CNS-1823262, and CMMI-2054710.

%% file: appendix.tex
\begin{appendices}

\section{Summary of Planning Invariant}\label{appendix:summary_planning_invariant}
\vspace{-0.05in}
We introduce the formal definition of constraints on physical objects commonly used among different driving scenarios. Note that more customized properties are needed for specific scenarios. When processing the geometry relationship among the objects, AD vehicle, and the road, a necessary step is to transform the position of a physical object from a unified coordinate system (UTM coordinate system in Apollo and Autoware) into a coordinate system relative to a certain lane. We define such functionality as the following function:

\begin{equation}
\footnotesize
    (s, l, laneID) = transform(pos, m)
\end{equation}

We define the function with input position $pos$ in UTM coordinate system and the map ($m$). The outputs of this function are longitudinal position ($s$) and lateral position ($l$) relative to the closest lane, and the ID of the closest lane ($laneID$). We do not expand this function in a formal way since this involves more than one thousand lines of source code in Apollo.

We now introduce the constraints of physical objects in detail. For the static objects, the most common constraint is that the static objects should be off the road and should not intersect with any lane boundary. This can be formally defined based on a specific static physical object $x$ and a set of polygon points of its boundary $x.polygon$:
\begin{equation}
\footnotesize
\begin{aligned}
    & StaticOffRoad(x) := \\
    & \wedge_{p\in x.polygon} | transform(p, m).l | > halfLaneWidth
\end{aligned}
\end{equation}
This means every polygon point of the object boundary is not within the range of any lane. The \textit{halfLaneWidth} also needs to be queried from the map. For simplicity, we just describe it as a constant.
For a pedestrian $x$, we want to make sure that it will not touch any lane in its future moving trajectory and does not show any intention to enter the traffic lane or intersection. In this case, for a pedestrian $x$, the waypoints set $x.W$ to describe its moving trajectory, $x.polygons(w)$ for the polygon points at waypoint $w$, the constraints for the pedestrian can be defined as:
\begin{equation}
\footnotesize
\begin{aligned}
    & DynamicOffRoad(x) := \\
    & \left( \wedge_{w\in x.W} ( \wedge_{p\in x.polygon(w)} | transform(p, m).l | > halfLaneWidth) \right) \\
    &\wedge \left(innerProduct(x.heading, directionTowardsLane) < 0 \right)
\end{aligned}
\end{equation}
The property $DynamicOffRoad$ will check all the waypoints in the future moving trajectory of a physical object $x$ and make sure that the any points on the boundary shall not touch the lane boundary. Besides, it also checks the heading of the moving trajectory is not towards to the lane.

There are two possible constraints for a vehicle. First, a vehicle can simply following the AD vehicle and it should not affect the planning behavior. For simplicity, we also use a property $driveInLane(x, laneID)$ to describe that the vehicle $x$ keeps driving within the lane denoted by the lane ID ($laneID$). This can be modeled as:
\begin{equation}
\footnotesize
\begin{aligned}
    & FollowVehicle(x) := \\
    & (transform(x.pos, m).laneID == ego.laneID)\\
    & \wedge (transform(x.pos, m).s + safetyFollowingDistance < ego.s) \\
    & \wedge (x.velocity \le ego.velocity)\\
    & \wedge driveInLane(x, transform(x.pos, m).laneID) \\
\end{aligned}
\end{equation}
Another possibility is that the vehicle is driving normally on another lane. We formally define it as:
\begin{equation}
\footnotesize
\begin{aligned}
    & IrrevalentVehicle(x):= \\
    & (transform(x.pos, m).laneID \neq ego.laneID)\\
    & \wedge driveInLane(x, transform(x.pos, m).laneID)
\end{aligned}
\end{equation}
\begin{table*}[tbp]
\footnotesize
\centering
\caption{Summary of Planning Invariants (PI) identified and used in the paper. }
\label{tab:summary_planning_invariant}
\vspace{-0.05in}
\setlength{\tabcolsep}{2.5pt}
\begin{tabular}{@{}lllll@{}}
\toprule
PI Index & Planning Scenario & Object Type & Constraints on Physical Objects & Desired Planning Behavior \\ \midrule
\multirow{4}{*}{PI1} & \multirow{4}{*}{\begin{tabular}[c]{@{}l@{}}Lane following\\ (single-lane road)\end{tabular}} & Static obstacles & \begin{tabular}[c]{@{}l@{}}\textbf{PI-C1.} Off-road and w/o any violation of the boundaries\\ of the lanes the AD vehicle plans to drive on\end{tabular} & \multirow{4}{*}{Keep cruising in the current lane} \\
 &  & \multirow{2}{*}{Vehicles} & \textbf{PI-C2.} Follow the AD vehicle &  \\
 &  &  & \textbf{PI-C3.} Drive on reverse lane &  \\
 &  & Pedestrians & \begin{tabular}[c]{@{}l@{}}\textbf{PI-C4+5.} Off-road and w/o any intention to move towards to\\ the AD vehicle or the lanes the AD vehicle plans to drive on\end{tabular} &  \\ \midrule
\multirow{4}{*}{PI2} & \multirow{4}{*}{\begin{tabular}[c]{@{}l@{}}Lane following\\ (multiple-lane road)\end{tabular}} & Static obstacles & \begin{tabular}[c]{@{}l@{}}\textbf{PI-C1.} Off-road and w/o any violation of the boundaries\\ of the lanes the AD vehicle plans to drive on\end{tabular} & \multirow{4}{*}{Keep cruising in the current lane} \\
 &  & \multirow{2}{*}{Vehicles} & \textbf{PI-C2.} Follow the AD vehicle &  \\
 &  &  & \textbf{PI-C3.} Drive on other lanes &  \\
 &  & Pedestrians & \begin{tabular}[c]{@{}l@{}}\textbf{PI-C4+5.} Off-road and w/o any intention to move towards to\\ the AD vehicle or the lanes the AD vehicle plans to drive on\end{tabular} &  \\ \hline
\multirow{4}{*}{PI3} & \multirow{4}{*}{Lane change} & Static obstacles & \begin{tabular}[c]{@{}l@{}}\textbf{PI-C1.} Off-road and w/o any violation of the boundaries\\ of the lanes the AD vehicle plans to drive on\end{tabular} & Finish changing to the targeted lane \\
 &  & \multirow{2}{*}{Vehicles} & \textbf{PI-C2.} Follow the AD vehicle &  \\
 &  &  & \textbf{PI-C3.} Drive on other lanes except current and targeted lanes &  \\
 &  & Pedestrians & \begin{tabular}[c]{@{}l@{}}\textbf{PI-C4+5.} Off-road and w/o any intention to move towards to\\ the AD vehicle or the lanes the AD vehicle plans to drive on\end{tabular} &  \\ \midrule
\multirow{6}{*}{PI4} & \multirow{6}{*}{\begin{tabular}[c]{@{}l@{}}Lane borrow\\ (due to a blocking obstacle)\end{tabular}} & \multirow{2}{*}{Static obstacles} & \begin{tabular}[c]{@{}l@{}}\textbf{PI-C1.} Off-road and w/o any violation of the boundaries\\ of the lanes the AD vehicle plans to drive on\end{tabular} & \multirow{6}{*}{\begin{tabular}[c]{@{}l@{}}Finish borrowing the reverse lane\\ and pass blocking vehicle\end{tabular}} \\
 &  &  & \textbf{SP-PI-C1.} On-lane and in front of the blocking obstacle &  \\
 &  & \multirow{3}{*}{Vehicles} & \textbf{PI-C2.} Follow the AD vehicle &  \\
 &  &  & \textbf{PI-C3.} Drive on other lanes except current and targeted lanes &  \\
 &  &  & \textbf{SP-PI-C2.} On-lane and park in front of the blocking obstacle &  \\
 &  & Pedestrians & \begin{tabular}[c]{@{}l@{}}\textbf{PI-C4+5.} Off-road and w/o any intention to move towards to\\ the AD vehicle or the lanes the AD vehicle plans to drive on\end{tabular} &  \\ \midrule
\multirow{4}{*}{PI5} & \multirow{4}{*}{Intersection w/ stop sign} & Static obstacles & \begin{tabular}[c]{@{}l@{}}\textbf{PI-C1.} Off-road and w/o any violation of the boundaries\\ of the lanes the AD vehicle plans to drive on\\ and the intersection the AD vehicle is going to pass\end{tabular} & \multirow{4}{*}{\begin{tabular}[c]{@{}l@{}}Pass intersection w/ stop sign\\ following the traffic rule\end{tabular}} \\
 &  & \multirow{2}{*}{Vehicles} & \textbf{PI-C2.} Follow the AD vehicle &  \\
 &  &  & \textbf{PI-C3.} Drive on other lanes except current and targeted lanes &  \\
 &  & Pedestrians & \begin{tabular}[c]{@{}l@{}}\textbf{PI-C4+5.} Off-road and w/o any intention to move towards to\\ the AD vehicle or the lanes the AD vehicle plans to drive on\end{tabular} &  \\ \midrule
\multirow{4}{*}{PI6} & \multirow{4}{*}{Intersection w/ traffic signal} & Static obstacles & \begin{tabular}[c]{@{}l@{}}\textbf{PI-C1.} Off-road and w/o any violation of the boundaries\\ of the lanes the AD vehicle plans to drive on\\ and the intersection the AD vehicle is going to pass\end{tabular} & \multirow{4}{*}{\begin{tabular}[c]{@{}l@{}}Pass intersection w/ traffic signal\\ following the traffic rule\end{tabular}} \\
 &  & \multirow{2}{*}{Vehicles} & \textbf{PI-C2.} Follow the AD vehicle &  \\
 &  &  & \textbf{PI-C3.} Drive on other lanes except current and targeted lanes &  \\
 &  & Pedestrians & \begin{tabular}[c]{@{}l@{}}\textbf{PI-C4+5.} Off-road and w/o any intention to move towards to\\ the AD vehicle or the lanes the AD vehicle plans to drive on\end{tabular} &  \\ \midrule
\multirow{4}{*}{PI7} & \multirow{4}{*}{Bare intersection} & Static obstacles & \begin{tabular}[c]{@{}l@{}}\textbf{PI-C1.} Off-road and w/o any violation of the boundaries\\ of the lanes the AD vehicle plans to drive on\\ and the intersection the AD vehicle is going to pass\end{tabular} & \multirow{4}{*}{Pass the bare intersection} \\
 &  & \multirow{2}{*}{Vehicles} & \textbf{PI-C2.} Follow the AD vehicle &  \\
 &  &  & \textbf{PI-C3.} Drive on other lanes except current and targeted lanes &  \\
 &  & Pedestrians & \begin{tabular}[c]{@{}l@{}}\textbf{PI-C4+5.} Off-road and w/o any intention to move towards to\\ the AD vehicle or the lanes the AD vehicle plans to drive on\end{tabular} &  \\ \midrule
\multirow{3}{*}{PI8} & \multirow{3}{*}{Parking} & Static obstacles & \textbf{SP-PI-C3.} Placed on other parking spots & \multirow{3}{*}{\begin{tabular}[c]{@{}l@{}}Park into an empty\\ targeted parking spot\end{tabular}} \\
 &  & Vehicles & \textbf{SP-PI-C4.} Parked on other parking spots &  \\
 &  & Pedestrians & \textbf{SP-PI-C5.} Walking pedestrians moving away from AD vehicle &  \\ \bottomrule
\end{tabular}
\vspace{-0.2in}
\end{table*}

\cut{
\begin{table*}[tbp]
\footnotesize
\centering
\caption{Summary of Planning Invariants (PI) identified and used in the paper. }
\label{tab:summary_planning_invariant}
\vspace{-0.12in}
\setlength{\tabcolsep}{4pt}
\begin{tabular}{@{}lllll@{}}
\toprule
PI Index & Planning Scenario &Object Type & Constraints on Physical Objects & Desired Planning Behavior\\
\midrule
\multirow{4}{*}{PI1} & \multirow{4}{*}{\begin{tabular}{c}Lane following\\ (single-lane road)\end{tabular}} & Static obstacles & \begin{tabular}{c}\textbf{PI-C1.}Off-road and w/o any violation of the boundaries\\ of the lanes the AD vehicle plans to drive on\end{tabular}
  & \multirow{4}{*}{Keep cruising in the current lane}\\
      
   &   & \multirow{2}{*}{Vehicles} & \begin{tabular}{c}\textbf{PI-C2.}Follow the AD vehicle\end{tabular} &  \\
&  &  & \begin{tabular}{c}\textbf{PI-C3.}Drive on reverse lane \end{tabular}& \\    
    &  & Pedestrians & \begin{tabular}{c}\textbf{PI-C4+5.}Off-road and w/o any intention to move towards to\\ the AD vehicle or the lanes the AD vehicle plans to drive on \end{tabular}& \\
 \midrule
 
 \multirow{4}{*}{PI2} & \multirow{4}{*}{\begin{tabular}{c}Lane following\\ (multiple-lane road)\end{tabular}} & Static obstacles & \begin{tabular}{c}\textbf{PI-C1.}Off-road and w/o any violation of the boundaries\\ of the lanes the AD vehicle plans to drive on\end{tabular}
  & \multirow{4}{*}{Keep cruising in the current lane}\\
      
   &   & \multirow{2}{*}{Vehicles} & \begin{tabular}{c}\textbf{PI-C2.}Follow the AD vehicle\end{tabular} &  \\
&  &  & \begin{tabular}{c}\textbf{PI-C3.}Drive on other lanes \end{tabular}& \\    
    &  & Pedestrians & \begin{tabular}{c}\textbf{PI-C4+5.}Off-road and w/o any intention to move towards to\\ the AD vehicle or the lanes the AD vehicle plans to drive on \end{tabular}& \\
 \midrule
 
 \multirow{4}{*}{PI3} & \multirow{4}{*}{\begin{tabular}{ll}Lane change\end{tabular}} & Static obstacles & \begin{tabular}{c}\textbf{PI-C1.}Off-road and w/o any violation of the boundaries\\ of the lanes the AD vehicle plans to drive on\end{tabular}
  & \multirow{4}{*}{Finish changing to the targeted lane}\\
      
   &   & \multirow{2}{*}{Vehicles} & \begin{tabular}{c}\textbf{PI-C2.}Follow the AD vehicle\end{tabular} &  \\
&  &  & \begin{tabular}{c}\textbf{PI-C3.}Drive on other lanes except current and targeted lanes \end{tabular}& \\    
    &  & Pedestrians & \begin{tabular}{c}\textbf{PI-C4+5.}Off-road and w/o any intention to move towards to\\ the AD vehicle or the lanes the AD vehicle plans to drive on \end{tabular}& \\
 \midrule
 
  \multirow{6}{*}{PI4} & \multirow{6}{*}{\begin{tabular}{ll}Lane borrow\\ (due to a blocking obstacle)\end{tabular}} & \multirow{2}{*}{Static obstacles} & \begin{tabular}{c}\textbf{PI-C1.}Off-road and w/o any violation of the boundaries\\ of the lanes the AD vehicle plans to drive on\end{tabular}
  & \multirow{6}{*}{\begin{tabular}{c}Finish borrowing the reverse lane\\ and pass blocking vehicle\end{tabular}}\\
      &  &  & \begin{tabular}{c}\textbf{SP-PI-C1.}On-lane and in front of the blocking obstacle \end{tabular}& \\    
   &   & \multirow{3}{*}{Vehicles} & \begin{tabular}{c}\textbf{PI-C2.}Follow the AD vehicle\end{tabular} &  \\
&  &  & \begin{tabular}{c}\textbf{PI-C3.}Drive on other lanes except current and targeted lanes \end{tabular}& \\    
&  &  & \begin{tabular}{c}\textbf{SP-PI-C2.}On-lane and park in front of the blocking obstacle \end{tabular}& \\    

    &  & Pedestrians & \begin{tabular}{c}\textbf{PI-C4+5.}Off-road and w/o any intention to move towards to\\ the AD vehicle or the lanes the AD vehicle plans to drive on \end{tabular}& \\
 \midrule
 
  \multirow{4}{*}{PI5} & \multirow{4}{*}{\begin{tabular}{c}Intersection w/ stop sign\end{tabular}} & Static obstacles & \begin{tabular}{c}\textbf{PI-C1.}Off-road and w/o any violation of the boundaries\\ of the lanes the AD vehicle plans to drive on\\ and the intersection the AD vehicle is going to pass\end{tabular}
  & \multirow{4}{*}{\begin{tabular}{c}Pass the intersection with stop sign\\ following the traffic rule\end{tabular}}\\
      
   &   & \multirow{2}{*}{Vehicles} & \begin{tabular}{c}\textbf{PI-C2.}Follow the AD vehicle\end{tabular} &  \\
&  &  & \begin{tabular}{c}\textbf{PI-C3.}Drive on other lanes except current and targeted lanes \end{tabular}& \\    
    &  & Pedestrians & \begin{tabular}{c}\textbf{PI-C4+5.}Off-road and w/o any intention to move towards to\\ the AD vehicle or the lanes the AD vehicle plans to drive on\\
    \end{tabular}& \\
 \midrule
 
   \multirow{4}{*}{PI6} & \multirow{4}{*}{\begin{tabular}{c}Intersection w/ traffic signal\end{tabular}} & Static obstacles & \begin{tabular}{c}\textbf{PI-C1.}Off-road and w/o any violation of the boundaries\\ of the lanes the AD vehicle plans to drive on\\ and the intersection the AD vehicle is going to pass\end{tabular}
  & \multirow{4}{*}{\begin{tabular}{c}Pass the intersection with traffic signal\\ following the traffic rule\end{tabular}}\\
      
   &   & \multirow{2}{*}{Vehicles} & \begin{tabular}{c}\textbf{PI-C2.}Follow the AD vehicle\end{tabular} &  \\
&  &  & \begin{tabular}{c}\textbf{PI-C3.}Drive on other lanes except current and targeted lanes \end{tabular}& \\    
    &  & Pedestrians & \begin{tabular}{c}\textbf{PI-C4+5.}Off-road and w/o any intention to move towards to\\ the AD vehicle or the lanes the AD vehicle plans to drive on\\
    \end{tabular}& \\
 \midrule
 
   \multirow{4}{*}{PI7} & \multirow{4}{*}{\begin{tabular}{c}Bare intersection\end{tabular}} & Static obstacles & \begin{tabular}{c}\textbf{PI-C1.}Off-road and w/o any violation of the boundaries\\ of the lanes the AD vehicle plans to drive on\\ and the intersection the AD vehicle is going to pass\end{tabular}
  & \multirow{4}{*}{\begin{tabular}{c}Pass the bare intersection\end{tabular}}\\
      
   &   & \multirow{2}{*}{Vehicles} & \begin{tabular}{c}\textbf{PI-C2.}Follow the AD vehicle\end{tabular} &  \\
&  &  & \begin{tabular}{c}\textbf{PI-C3.}Drive on other lanes except current and targeted lanes \end{tabular}& \\    
    &  & Pedestrians & \begin{tabular}{c}\textbf{PI-C4+5.}Off-road and w/o any intention to move towards to\\ the AD vehicle or the lanes the AD vehicle plans to drive on\\
    \end{tabular}& \\
 \midrule
 
    \multirow{4}{*}{PI8} & \multirow{4}{*}{\begin{tabular}{c}Parking\end{tabular}} & Static obstacles & \begin{tabular}{c}\textbf{SP-PI-C3.}Placed on other parking spots\end{tabular}
  & \multirow{4}{*}{\begin{tabular}{c}Park into an empty\\ targeted parking spot\end{tabular}}\\
      
   &   & \multirow{1}{*}{Vehicles} & \begin{tabular}{c}\textbf{SP-PI-C4.}Parked on other parking spots\end{tabular} &  \\
    &  & Pedestrians & \begin{tabular}{c}\textbf{SP-PI-C5.}Walking pedestrians moving away from the AD vehicle\\
    \end{tabular}& \\
\bottomrule
\end{tabular}
\end{table*}
}

\begin{table*}[tbp]
\footnotesize
\centering
\caption{\newparts{The 40 collected BP input traces from LGSVL simulator for our evaluation (\S\ref{sec:evaluation}). \textit{Lexus} and \textit{Lincoln} are short for Lexus RX Hybrid 2016 and Lincoln MKZ 2017. More details (e.g., pictures) of the map, vehicle, and driving behaviors in each trace are on our website~\cite{planfuzzwebsite}.}}
\label{tab:seed_collection}
\setlength{\tabcolsep}{3pt}
\begin{tabular}{@{}ccccc@{}}
\toprule
Scenario & Driving Behavior & Map  & Vehicle & Duration (\# of Planing Decisions) \\
  \midrule
  \multirow{10}{*}{\makecell[c]{Lane Follow \\(Single lane road)}} 
    & \makecell[c]{Follow a 1-lane straight narrow road\\ (2.7m lane width)}  & Single-lane road & \makecell{Apollo: Lincoln\\Autoware: Lexus} & \makecell[c]{15.0s (133)\\25.4s (2394)} \\ \cline{4-5}
    & \makecell[c]{Follow a 1-lane straight medium road\\ (3.0m lane width)} & Single-lane road & \makecell{Apollo: Lincoln\\Autoware: Lexus} & \makecell[c]{14.3s (121)\\23.8s (2241)} \\ \cline{4-5}
    & \makecell[c]{Follow a 1-lane straight wide road \\(3.5m lane width)} & Single-lane road & \makecell{Apollo: Lincoln\\Autoware: Lexus} & \makecell[c]{18.6s (157)\\ 24.6s (2037)} \\ \cline{4-5}
    & \makecell[c]{Follow a 1-lane left-curved road} & CubeTown & \makecell{Apollo: Lincoln\\Autoware: Lexus} & \makecell[c]{21.3s (209)\\18.3s (1749)} \\ \cline{4-5}
    & \makecell[c]{Follow a 1-lane right-curved road} & CubeTown & \makecell{Apollo: Lincoln\\Autoware: Lexus} &  \makecell[c]{17.6s (172)\\21.3s (1978)} \\
\midrule
  \multirow{10}{*}{\makecell[c]{Lane Follow \\(Multiple lane road)}} 
    & \makecell[c]{Follow a 2-lane straight road}  & San Francisco & \makecell{Apollo: Lincoln\\Autoware: Lexus} & \makecell[c]{18.7s (177)\\15.4s (1379)} \\ \cline{4-5}
    & \makecell[c]{Follow a 3-lane straight road} & Modern City & \makecell{Apollo: Lincoln\\Autoware: Lexus} & \makecell[c]{14.3s (121)\\21.3s (1840)} \\ \cline{4-5}
    & \makecell[c]{Follow a 4-lane left-curved road} & San Francisco & \makecell{Apollo: Lincoln\\Autoware: Lexus} & \makecell[c]{18.7s (181)\\ 19.8s (1679)} \\ \cline{4-5}
    & \makecell[c]{Follow a 4-lane right-curved road} & San Francisco  & \makecell{Apollo: Lincoln\\Autoware: Lexus} &\makecell[c]{21.5s (208)\\25.9s (2379)} \\ \cline{4-5}
    & \makecell[c]{Follow a 4-lane straight road} & San Francisco & \makecell{Apollo: Lincoln\\Autoware: Lexus} & \makecell[c]{13.4s (129)\\19.5s (1437)} \\
\midrule

  \multirow{5}{*}{\makecell[c]{Lane Change}} 
    & \makecell[c]{Right change on a straight road}  & San Francisco & Apollo: Lincoln& 21.2s (203) \\ 
    & \makecell[c]{Left change on a straight road} & San Francisco &  Apollo: Lincoln& 15.7s (138) \\ 
    & \makecell[c]{Left change on a left-curved road} & San Francisco & Apollo: Lincoln& 13.4s (130) \\ 
    & \makecell[c]{Right change on a left-curved road} & San Francisco & Apollo: Lincoln& 18.7s (172) \\ 
    & \makecell[c]{Left change on a right-curved road} & San Francisco & Apollo: Lincoln& 16.4s (159) \\
\midrule
    \multirow{8}{*}{\makecell[c]{Lane Borrow}} 
    & \makecell[c]{Borrow lane on a straight narrow road\\ (2.7m lane width)} & \makecell{Single-lane road} & Apollo: Lincoln& 25.9s (238) \\ 
    & \makecell[c]{Borrow lane on a straight medium road\\ (3.0m lane width)} & \makecell{Single-lane road} & Apollo: Lincoln& 28.7s (279) \\ 
    & \makecell[c]{Borrow lane on a straight wide road\\ (3.5m lane width)} & \makecell{Single-lane road} & Apollo: Lincoln& 30.5s (317) \\ 
    & \makecell[c]{Borrow lane on a left-curved road} & CubeTown & Apollo: Lincoln& 27.3s (262) \\ 
    & \makecell[c]{Borrow lane on a right-curved road} & CubeTown & Apollo: Lincoln& 33.2s (329) \\
\midrule
    
\multirow{5}{*}{\makecell[c]{Traffic Signal\\ Intersection}} 
    & \makecell[c]{Turn left at a 4-way intersection}  & San Francisco & Apollo: Lincoln& 47.1s (453) \\ 
    & \makecell[c]{Turn right at a 4-way intersection} & San Francisco & Apollo: Lincoln& 36.8s (329) \\ 
    & \makecell[c]{Go straight at a 4-way intersection} & San Francisco & Apollo: Lincoln& 27.9s (288) \\ 
    & \makecell[c]{Turn right at a 3-way intersection} & San Francisco & Apollo: Lincoln&  26.4s (233) \\ 
    & \makecell[c]{Go straight at a 3-way intersection} & San Francisco & Apollo: Lincoln& 31.9s (308) \\
\midrule

\multirow{5}{*}{\makecell[c]{Stop sign\\ Intersection}} 
    & \makecell[c]{Turn left at a 4-way intersection}  & Shalun & Apollo: Lincoln& 32.3s (334) \\ 
    & \makecell[c]{Turn right at a 4-way intersection} & Shalun & Apollo: Lincoln& 27.9s (255) \\ 
    & \makecell[c]{Go straight at a 4-way intersection} & Shalun & Apollo: Lincoln& 23.8s (220) \\ 
    & \makecell[c]{Turn right at a 3-way intersection} & Shalun & Apollo: Lincoln& 33.2s (329) \\ 
    & \makecell[c]{Go straight at a 3-way intersection} & Shalun & Apollo: Lincoln& 29.7s (283) \\
\midrule
\multirow{5}{*}{\makecell[c]{Bare Intersection\\ Intersection}} 
    & \makecell[c]{Turn left at a 4-way intersection}  & GoMentum Station & Apollo: Lincoln& 37.9s (361) \\ 
    & \makecell[c]{Turn right at a 4-way intersection} & GoMentum Station & Apollo: Lincoln& 42.3s (391) \\ 
    & \makecell[c]{Go straight at a 4-way intersection} & GoMentum Station & Apollo: Lincoln&  30.1s (287) \\ 
    & \makecell[c]{Turn right at a 3-way intersection} & GoMentum Station & Apollo: Lincoln&  29.2s (288) \\ 
    & \makecell[c]{Go straight at a 3-way intersection} & GoMentum Station &  Apollo: Lincoln& 38.5s (379) \\
\midrule
    
\multirow{5}{*}{\makecell[c]{Parking}} 
    & \makecell[c]{Park to a front parking spot}  & GoMentum Station & Apollo: Lincoln&  23.4s (228) \\ 
    & \makecell[c]{Park to a left close parking spot} & GoMentum Station & Apollo: Lincoln& 30.5s (309) \\ 
    & \makecell[c]{Park to a right close parking spot} & GoMentum Station & Apollo: Lincoln& 27.6s (263) \\ 
    & \makecell[c]{Park to a left far parking spot} & GoMentum Station & Apollo: Lincoln& 24.3s (231) \\ 
    & \makecell[c]{Park to a right far parking spot} & GoMentum Station & Apollo: Lincoln& 17.9s (163) \\

   \bottomrule
\end{tabular}
\end{table*}

\begin{table*}
\caption{\newparts{The physical-property initialization and mutation ranges for the attack-introduced physical road objects used in our experiments (\S\ref{sec:evaluation}).}}
\label{tab:valid_input}
\begin{tabular}{ccccc}

\toprule
Input                 & Parameters                             & Description                                                                            & \multicolumn{2}{c}{Valid ranges}                                                                                   \\ \hline
    Type                  & type                                   & An enumeration that represents type of object                                          & \multicolumn{2}{c}{\makecell{\{Pedestrian, Vehicle, Bicycle, Static object\}}}                                                \\ \hline
Position              & x, y                                   & Coordinate of the object center in 2D space                                       & \multicolumn{2}{c}{\makecell{{[}ego.position.x - 80, ego.position.x + 80{]},\\ {[}ego.position.y - 80, ego.position.y + 80{]}}} \\ 
\hline
\multirow{4}{*}{Size} & \multirow{4}{*}{\makecell{length (m),\\ width (m),\\ height (m)}} & \multicolumn{1}{c}{\multirow{4}{*}{Length, width, height of the object bounding box}} & \multicolumn{1}{c}{Pedestrian} & 0.50, 0.50, 1.80\\ 
                      & & & Vehicle & 4.70, 2.06, 2.05 \\
                      & & & Parked Bicycle  & 1.80, 0.50, 1.00  \\
                      & & & Static object  & (0.5, 2.0), (0.5, 2.0), (0.5, 2.0)  \\
\hline                                   
\multirow{4}{*}{Velocity} & \multirow{4}{*}{\makecell{velocity (m/s)}} & \multicolumn{1}{c}{\multirow{4}{*}{\makecell[c]{The absolute velocity. It will be transformed into\\ a vector based on the heading when feeding into planning.}}} & \multicolumn{1}{c}{Pedestrian} & (0, 1.4)\\ 
                      & & & Vehicle & Ego velocity \\
                      & & & Parked Bicycle  & 0  \\
                      & & & Static object  & 0  \\
\hline                                   
\multirow{4}{*}{Heading} & \multirow{4}{*}{\makecell{heading (angle)}} & \multicolumn{1}{c}{\multirow{4}{*}{\makecell[c]{The heading of the physical object.}}} & \multicolumn{1}{c}{Pedestrian} & \makecell[c]{Parallel to the current road or\\ leave the road vertically}\\ 
                      & & & Vehicle & Current road heading \\
                      & & & Parked Bicycle  & [0, $\pi$)  \\
                      & & & Static object  & [0, $\pi$)  \\
\hline                                   
\multirow{4}{*}{Trajectory} & \multirow{4}{*}{\makecell{a list of waypoints}} & \multicolumn{1}{c}{\multirow{4}{*}{\makecell[c]{A list of the waypoints whch represents the predicted\\ trajectory of each object.}}} & \multicolumn{1}{c}{Pedestrian} & \makecell[c]{Sampling on a straight line\\ based on the heading} \\ 
                      & & & Vehicle & Sampling based on road network \\
                      & & & Parked Bicycle  & Empty  \\
                      & & & Static object  & Empty  \\
\bottomrule
\end{tabular}
\end{table*}

\cut{
\begin{table*}[tbp]
\footnotesize
\centering
\caption{\newparts{Detailed information of the maps for BP's input traces and initial testing seeds collection.}}
\label{tab:map_setup}
\setlength{\tabcolsep}{3pt}
\begin{tabular}{@{}ccc@{}}
\toprule
Map name & Common lane width & Map description\\
\midrule
Single lane road (narrow) & 2.7m & A single lane road with lane width 2.7m\\
\midrule
Single lane road (medium) & 3.0m & A single lane road with lane width 3.0m\\
\midrule
Single lane road (wide) & 3.5m & A single lane road with lane width 3.5m\\
\midrule
San Francisco & 3.5m & \makecell[c]{A re-creation of a section of SOMA San Francisco around Market Street.\\ It features many traffic light intersections and multi-lane streets.}\\
\midrule
GoMentum Station & 3.5m & A testing facility for autonomous driving vehicles.\\
\midrule
Shalun & 3.4m & A testing facility for autonomous driving vehicles.\\
\midrule
CubeTown & 3.4m & A virtual environment in simulation.\\
\bottomrule
\end{tabular}
\end{table*}
}

\section{Details of PI Constraint Enforcement in~\S\ref{subsec:design_mutation}}\label{sec:app_pienforcer}

As described in \S\ref{subsec:design_mutation}, in the PI-constraints enforcement we need to enforce the randomly-generated static properties to conform to PI constraints. As it is used in the mutation step of a genetic algorithm, such enforcement step needs to have the following proprieties:
\begin{itemize}
    \item \textbf{Diversity.} A good mutation function in genetic algorithm should introduce randomness and diversity to the next generation such that the genetic algorithm can explore different descent directions and avoid local minima.
    \item \textbf{Inheritance.} An other important property for the mutation function in genetic algorithm is inheritance. The problem-solving gene should be propagated to the next generation such that the whole genetic algorithm is approaching to the best solution.
\end{itemize}

\begin{figure}
    \centering
    \includegraphics[width=0.95\linewidth]{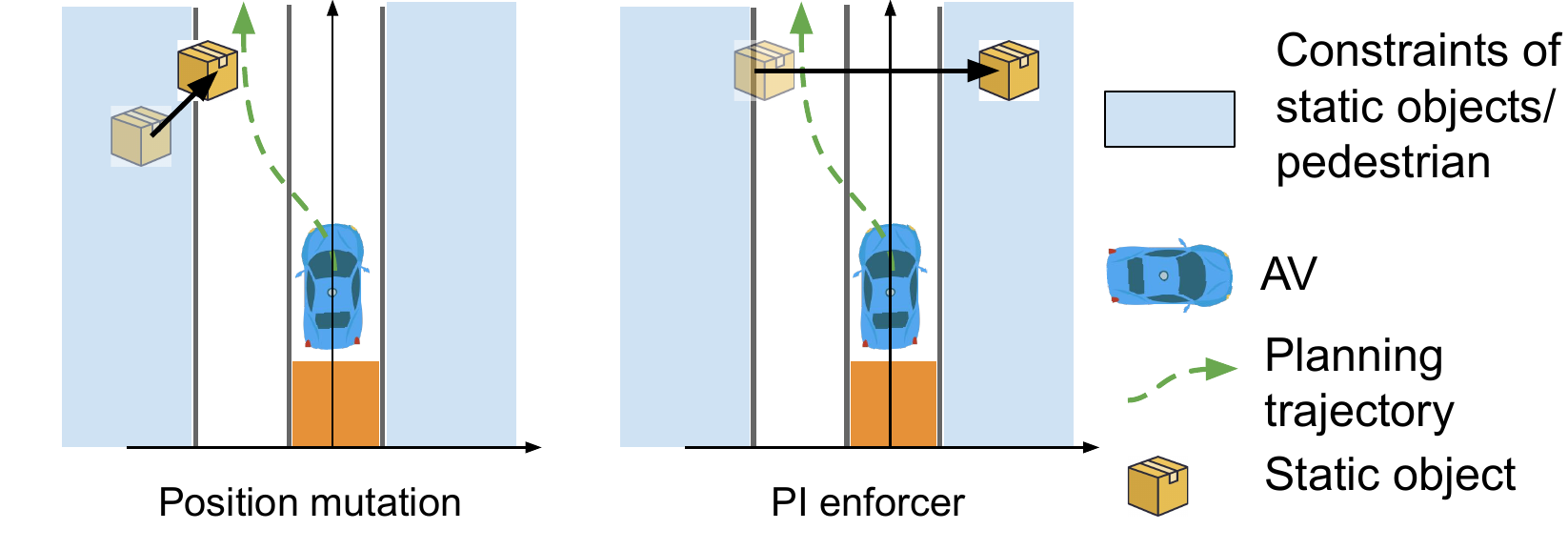}
    \caption{Off-road operation.} 
    \label{fig:mutation_off_lane}
\end{figure}

\begin{figure}
    \centering
    \includegraphics[width=0.95\linewidth]{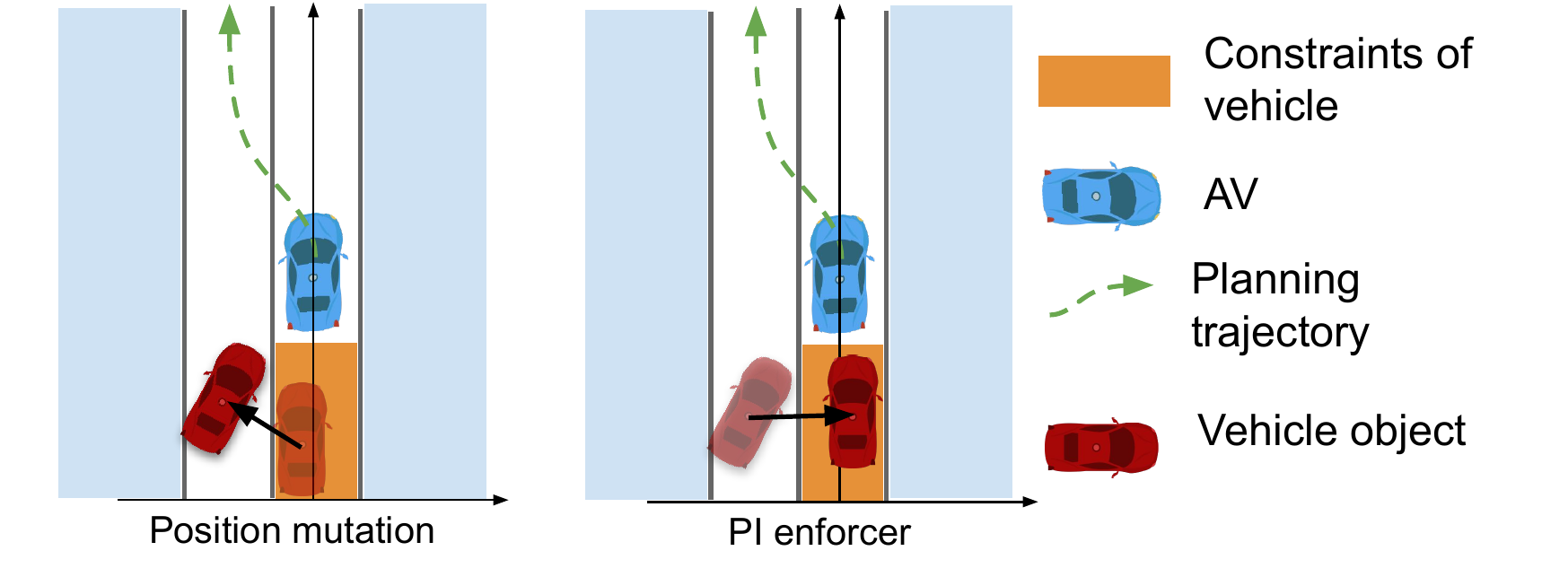}
    \caption{On-lane operation
    } 
    \label{fig:mutation_on_lane}
\end{figure}

Following these principles, we define the following enforcers for the PIs in this paper:
\begin{itemize}
    \item \textbf{Off-road constraint enforcer.} Off-road constraint is the most frequent constraint in the planning invariant. If the violation of off-lane rule is generated by an initialization process, we directly move the position to the closest point where the whole object is off-road. We noticed that if a violation is introduced by a mutation, we can easily infer the direction of this violation by comparing the position of the points and the boundary of constraints. We still want to keep the direction of mutation due to the diversity. As shown in Fig.~\ref{fig:mutation_off_lane}, suppose the object positions before and after the mutation are $(s, l)$ and $(s_m, l_m)$, and we can infer that this mutation violates the constraints in the lateral direction with an intention to increase the $l$. The enforcer will thus apply the following transformation:
    \begin{equation}
    \footnotesize
        \begin{aligned}
        s_e & = s_m\\
        l_e & = l_m + (road\_width + obstacle\_width)
        \end{aligned}
    \end{equation}
    We keep the position on the direction where no violation happens and shift the position with the length of the constraint on the other direction.In this particular case, we keep the longitudinal direction of the object position unchanged since it does not violate the constraint. However since the lateral direction violates the constraint, the enforcer then apply additional offset to the lateral direction to make it satisfy the constraint. The offset we apply here is the width of the road plus the width of the AD vehicle, which is the whole width of the area where the static object can not be placed. By applying this offset, we transmit the object to another feasible area. As a result, we can maintain the randomness from the mutation function and still enforce the constraints at the same time. 
     
    \item \textbf{On-lane constraint enforcer.} On-lane constraint is used to force the vehicle driving in a normal way. Same as the off-road constraints handling approach, if the violation is generated during the initialization, we use the nearest lane way point to replace the original point. Following the same idea used in off-road constraints, we use the same transformation operation to enforce the constraints. As shown in Fig.~\ref{fig:mutation_on_lane}, we still maintain the same mutation direction. When the boundary of the object has touched the constraint, we transmit the object to the other side of the lane. In the particular example in Fig.~\ref{fig:mutation_on_lane}, we can notice that the position generated by the position mutation already touches the target lane of AD vehicle's lane changing, which violates the constraint of PI. During the enforcer step, we divide the leftwards lateral mutation into two parts: the first part drags the object to the boundary of the constraints. At this point, we initialize the position of object to the right side of the feasible area. By doing that, we can make sure that the leftwards position mutation will still mutate the object inside the constraints. Then we can start with the ramaining leftwards mutation.
\end{itemize}

\section{Details of BP Vulnerability Distance Calculation in~\S\ref{subsec:vuln_dist}}\label{sec:app_bp_dist}

\begin{figure}
    \centering
    \includegraphics[width=0.99\linewidth]{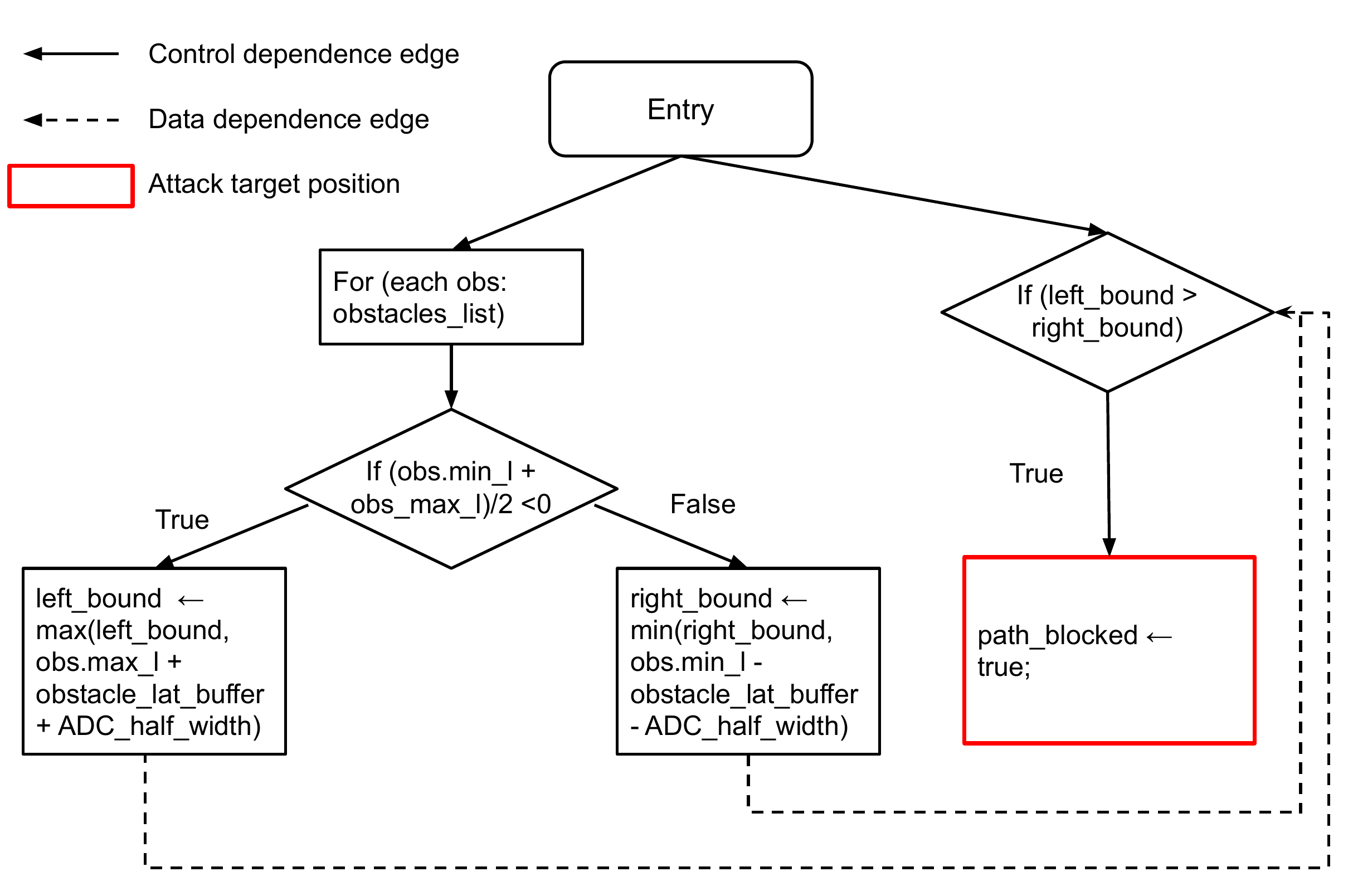}
    \caption{PDG graph of code example.  }
    \label{fig:PDG}
\end{figure}

To calculate the BP vulnerability distance, we first need to apply the \textit{attack target control} and \textit{data dependency analysis} in the offline analysis and instrumentation phase. In this phase, we first build a Program Dependence Graph (PDG) to capture all the control and data dependence. Program dependence graph is a common way to capture the dependence relationships in the program~\cite{ferrante1987program, liu2017ptrsplit}. We followed the approach described in \cite{liu2017ptrsplit} to construct the PDG. First, we construct the intra-procedural PDG by combining the results of control dependence, def-use dependence, and flow-insensitive read-after-write data dependence. We also follow their parameter tree approach to avoid whole-program-analysis for inter-procedural data dependence analysis. After getting the PDG, we traverse the dependence graph from the target position to extract all the predicates which will affect the final decision via control dependence or data dependence. We further divide the predicates into two parts: critical predicates and non-critical predicates. Critical predicates refer to the predicates which connect themselves with the attack target position with control dependence edges only. On the other side, non-critical predicates refer to the predicates that must connect with the attack target via data dependence edges. The generated PDG of the example code (\S\ref{subsec:mov_example}) is shown in Figure~\ref{fig:PDG}. In the figure, we can clearly see that predicate on line 22 is a critical predicate since it is connected with the attack target with one control dependence edge. The other predicate on line 17 is a non-critical predicate since the data dependence edge is required to connect with the attack target in PDG. Here we define the control flow distance for each predicate to measure the distance between a certain predicate and the target position. For the dependence chain between a predicate and a certain target position, the control flow distance (CFD) is defined as 
\begin{equation}
\footnotesize
    CFD(p, T) = \min_{T_i\in T} \sum_{c\in Dep(p, T_i)} (CFD(c))
\end{equation}
where $p$ is a predicate, $T$ is the set of the target positions. $Dep(p, T_i)$ is the dependence chain between predicate $p$ and the target position $T_i$ and $c$ is a certain control dependence edge in the dependence chain between $p$ and $T_i$. $CFD(c)$ will calculate distance on control flow between two nodes of this control dependence edge. The control flow distance is purely calculated in the offline phase. 

The overall BP vulnerability distance is the sum of the vulnerability distance of each single predicate. For the critical predicates, our distance metric want to guide the execution flow to the branch which is reachable from the attack target or the closer branch if two branches are reachable. So besides the static control flow distance, we also need to get runtime information to calculate the data flow distance for each single predicate. Considering the PDG in Figure \ref{fig:PDG}, the predicate on line 13 is directly connected with the attack target position with a control dependence edge. The CFD of this predicate is 1. The CFD of predicate on line 7 is 2 since the path from the predicate to target position contains 2 control dependence edges and 1 data dependence edge. 

For the critical predicates, we can directly infer which branch is reachable to the target positions or closer to the target positions. We record the execution number of each branch as $n_1$ and $n_2$. $n_1$ is the number of reachable branches and $n_2$ is the execution number of the unreachable or farther branch. The data flow distance can be calculated as follow:
\begin{equation}
\footnotesize
 DFD(p) = \begin{cases}
  1 & \;p \; not\; executed \\
  \frac{n_2}{n_1+n_2} \frac{Min_{op}}{Max_{history}} & otherwise
 \end{cases}
\end{equation}
where $Max_{history}$ is the maximal value of difference between two operands of this predicate in the evolutionary testing history and $Min_{op}$ is the minimal difference between two operands in this specific execution. We want to use this term to measure the distance between two operands in this predicate such that the fitness function can be smaller when the operands are approaching and finally change the result of this predicate towards a closer execution trace.

For the non-critical predicates, since we do not know which branch is the closer branch towards the target positions, we simply minimize the difference between two operands to get more diverse execution traces.  
\begin{equation}
\footnotesize
 DFD(p) = \begin{cases}
  1 & p \; not\; executed \\
  \frac{n_1}{n_1\!+\!n_2}\!
  \frac{Min_{op}^1}{Max_{history}^1}\! +\! \frac{n_2}{n_1\!+\!n_2} \!\frac{Min_{op}^2}{Max_{history}^2}  &  otherwise
 \end{cases}
\end{equation}
In the above equation, we use number 1 and 2 to refer to different branches. 
Finally, the overall BP vulnerability distance is defined as 
\begin{equation}
\footnotesize
\begin{split}
    BPVD(T) &= \sum_{p\in P_{crit}} CFD(p, T)DFD(p, T) \\
   &+ c\sum_{p\in P_{non-crit}} CFD(p, T)DFD(p, T)\\
   \end{split}
\end{equation}
where $P_{crit}$ and $P_{non-crit}$ is the set of critical predicates and non-critical predicates. $c$ is a constant number to give smaller weight to the non-critical predicates since we cannot get the precise information of which branch is closer to the target positions for the non-critical predicates.

We use the motivating example in \S\ref{subsec:mov_example} to demonstrate the design of BP vulnerability distance. As shown in Figure~\ref{fig:distance_demo}, imaging that the static obstacles are placed increasingly closer to the victim AD vehicle's driving lane until the DoS vulnerability is triggered. At the starting point, no obstacle is around the driving lane. Then the lateral distance between two static obstacles will become closer until the DoS vulnerability is triggered. If we directly use the output of the planner as the fitness value, we can only get a sharp drop when the DoS vulnerability is triggered, since before that the planning behavior is unchanged. If we applied the idea of directed greybox fuzzing~\cite{bohme2017directed} and use their defined distance, we can only get two sharp drops since that is the only changes of the control flow graph. However, benefit from the fine-grained control and data dependence distance metrics, our BP vulnerability distance design can successfully capture the approaching tendency of the static obstacles during the mutation, since the distance of the predicates on line 7 and 13 is getting smaller. At the beginning, since there is no obstacles around, both predicates are not executed and the DFD of two predicates are 1. The BP vulnerability distance is $1 + 2c$. When the static obstacles are placed closer, both predicates are executed. The DFD of predicate on line 7 is getting smaller since the distance between obstacles and the center line is smaller. The DFD of predicate on line 13 is getting smaller since the distance between two obstacles is smaller. Once the DoS vulnerability is triggered, the DFD of the predicate 13 becomes 0 since the correct branch is executed.

\section{Details of each Discovered Vulnerability}\label{sec:appendix_PIsummary}
\subsection{V1. Apollo Lane Following} \label{sec:appendix_v1_apollo_lane_follow}
\input{Detailed_vuln/V1}

\subsection{V2. Apollo Lane Changing} \label{sec:appendix_v2_lane_change}
\input{Detailed_vuln/V2}

\subsection{V3. Apollo Lane Borrow (In-lane Static Object)} \label{sec:appendix_v3_lane_borrow_off}
\input{Detailed_vuln/V4}

\subsection{V4. Apollo Lane Borrow (Off-lane Static Obstacles)} \label{sec:appendix_v4_lane_borrow_in}
\input{Detailed_vuln/V3}

\subsection{V5. Apollo Intersection with Traffic Signal (Standing Pedestrian)} \label{sec:appendix_v5_crosswalk_standing}
\input{Detailed_vuln/V5}

\subsection{V6. Apollo Intersection with Traffic Signal (Walking Pedestrian)} \label{sec:appendix_v6_crosswalk_walking}
\input{Detailed_vuln/V6}

\subsection{V7. Apollo Intersection with Stop Sign} \label{sec:appendix_v7_stop_sign}
\input{Detailed_vuln/V7}

\subsection{V8. Autoware Lane Following (Static Objects)} \label{sec:appendix_v8_autoware_static}
\input{Detailed_vuln/V8}

\subsection{V9. Autoware Lane Following (Dynamic Objects)} \label{sec:appendix_v9_autoware_dynamic}
\input{Detailed_vuln/V9}
\end{appendices}






%% file: Detailed_vuln/V1.tex
We have described the decision logic, the discovered vulnerability, and exploits in details in \S\ref{subsec:mov_example}.\cut{
\begin{figure}
    \centering
\hrule
\begin{minted}[fontsize=\footnotesize,xleftmargin=18pt,linenos,escapeinside=||,mathescape=true]{python}
# Pre-defined lateral safety buffer
obstacle_lat_buffer |$\leftarrow$| 0.4f # unit: meter
# AD vehicle width (default value in Apollo)
ADC_width |$\leftarrow$| 2.06f # unit: meter 
# Initialize the left and right boundaries of drivable space
left_bound |$\leftarrow$| left_lane_boundary
right_bound |$\leftarrow$| right_lane_boundary

# Iterate over static obstacles at this longitudinal position  
for (each obs in obstacles_list) 
  if ((obs.min_l + obs.max_l)/2 < 0):
    # The obstacle is on the left side of the lane center
    left_bound|$\leftarrow$|max(left_bound,obs.max_l+obstacle_lat_buffer)
  else:
    # The obstacle is on the right side of the lane center
    right_bound|$\leftarrow$|min(right_bound,obs.min_l-obstacle_lat_buffer)
  
# Check whether there exists drivable space laterally
if (right_bound - left_bound < ADC_width):
  # Conclude that the lane is blocked
  path_blocked |$\leftarrow$| true
\end{minted}
\hrule
  \caption{Simplified pseudo code for a BP (Behavior Planning) DoS vulnerability our system discovered from Baidu Apollo, an industry-grade AD system~\cite{apollo}}. 
  \label{fig:app_v1_code}
\end{figure}
}

%% file: Detailed_vuln/V2.tex
\cut{ 
\begin{figure}
    \centering
\hrule
\begin{minted}[fontsize=\footnotesize,xleftmargin=18pt,linenos,escapeinside=||,mathescape=true]{python}
# Iterate over the obstacle list
for (each obs : obstacle_list):
  # Judge based on the lateral position of a vehicle
  # start_l & end_l are vehicle's left & right boundaries
  if (obs.end_l < -2.5 or obs.start_l > 2.5):
    continue
  # The backward safe buffer
  BackwardSafeBuffer |$\leftarrow$| 4.0f
  # Check whether the vehicle is close 
  if (ego_start_s - obs.end_s < BackwardSafeBuffer):
    IsClearChangeLane |$\leftarrow$| false
\end{minted}
\hrule
  \caption{Simplified vulnerable pseudo code of V2 (lane changing) in Apollo.}
  \label{fig:app_v2_code}
\end{figure}
}

\begin{figure}[htbp]
    \centering
    \includegraphics[width=0.95\linewidth]{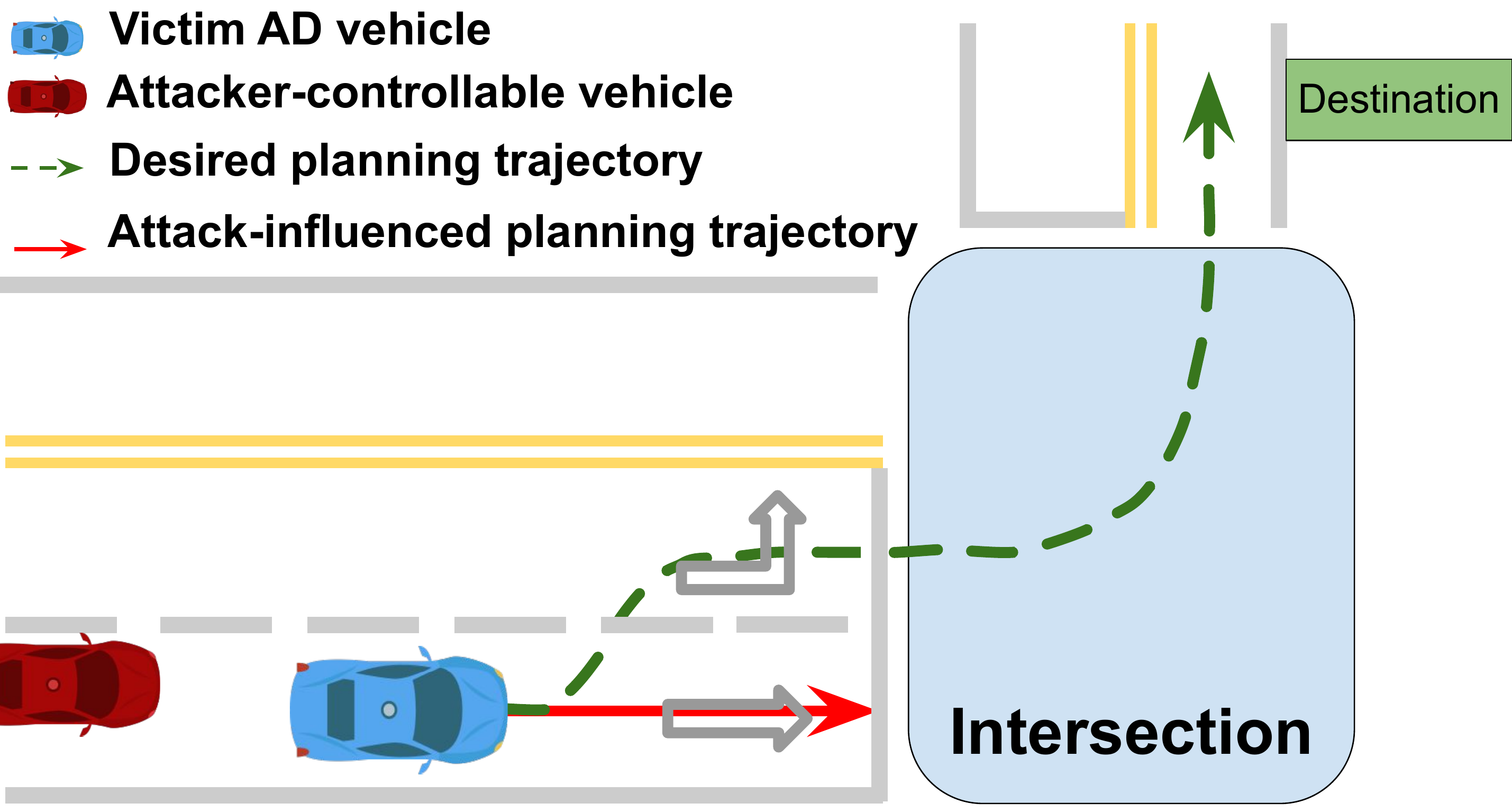}
    \caption{Attack scenario setup: Block the AD vehicle from changing lane with a following vehicle.} 
    \label{fig:app_v2_scenario}
\end{figure}

\textbf{Vulnerable code.} We presented the pseudo code of the vulnerability in Fig.~\ref{fig:app_v2_code}. The whole program shown in the figure is used to decide whether it is safe to perform lane changing given the positions of surrounding vehicles. All the position variables in the code are in the Frenet coordinate based on the target lane of the lane changing. The first check on line 5 is used to check whether a surrounding vehicle is located on the target lane. If not, this vehicle will be ignored. The second check on line 10 is used to check the longitudinal distance between the AD vehicle and the vehicle. If the distance is smaller than a backward safety distance, then Apollo will come to the conclusion that it is not a good time to change the lane since another vehicle is driving on the target lane and the longitudinal distance between them is less than a safety distance. The vulnerable code is on line 5. Since the lateral buffer is too large, a vehicle following the AD vehicle is also considered as a blocking vehicle (usually the half lane width is smaller than 2m even on the highway~\cite{lane_width}).

\textbf{Attack scenario setup.}
As shown in Fig.~\ref{fig:app_v2_scenario}, we design an attack scenario where the attacker can force the victim AD vehicle to give up lane changing during the whole driving by exploiting V2. The victim AD vehicle needs to change the lane so that it can turn left at the intersection to reach to the destination. The attacker just needs to drive a car following the victim AD vehicle so that the check on line 5 can not be satisfied. As a result, the victim AD vehicle always considers the attacker's vehicle as a blocking vehicle and it is not safe to change the lane. In the end, the victim AD vehicle has to re-route to get to the destination or stop in the current lane to wait until it thinks it is clear to change the lane. If the AD vehicle has to re-route, the overall travelling time and energy consumption will greatly increase and damage the passengers' experience. Stopping to wait for lane changing will also block the normal traffic operations and is considered traffic violation in some cases~\cite{calihandbook}. As a result, both
behaviors will increase the traffic load, which contradicts with one of the benefits of autonomous driving: efficiency and convenience, based on a report from United States department of transportation~\cite{AVbenefitl}.

\textbf{End-to-end result.}
We construct the attack scenario in LGSVL \cite{lgsvl}. The whole scenario happens in front of an intersection where the victim AD vehicle has to turn left to get to the destination.
By carefully controlling the trajectory of a attacker-controllable vehicle with the API provided by LGSVL, we are able to force the victim AD vehicle to stay on the right lane while our controlled vehicle is always driving on the right lane. In benign scenario (left figure in Fig.~\ref{fig:case_study_lane_change}, we control an vehicle following the AD vehicle and our controlled car is driving slightly on the right side of the lane (to make sure the vehicle is skipped by line 5). Apollo does not consider the vehicle as blocking the road and changes the lane successfully. For the scenario under attack (right figure in Fig.~\ref{fig:case_study_lane_change}), we controlled the vehicle to follow the car and drive on the left side of the lane to pass the lateral position check on line 5. As a result, our controlled car can block the victim AD vehicle so that it fails to change the lane until it reaches the intersection.

%% file: Detailed_vuln/V4.tex
\begin{figure}
    \centering
\hrule
\begin{minted}[fontsize=\footnotesize,numbersep=5pt,xleftmargin=10pt,linenos,escapeinside=||,mathescape=true]{python}
# Iterate over the obstacle list
for (obs : obstacle_list):
  if (obs.id == current_obs.id):
    continue
  #Check lateral intersection
  if (obs.start_l > current_obs.end_l or
      obs.end_l < current_obs.start_l):
    # Skip if no lateral intersection
    continue
  delta_s |$\leftarrow$| obs.start_s - current_obs.end_s
  if(delta_s<0 or delta_s> Threshold)
    continue
  # Make decision that current_obs is temporarily 
  # blocked by another obstacle and will move  
  # forward once its blocking obstacle is cleared
  return false
# Make decision that current_obs is non-movable
return true
\end{minted}
\hrule
  \caption{Simplified vulnerable pseudo code of V3 (lane borrow + on-lane objects) in Apollo. This is part of a function which decides whether the blocking obstacle is movable or not. }
  \label{fig:app_v4_code}
\end{figure}

\begin{figure}[htbp]
    \centering
    \includegraphics[width=0.95\linewidth]{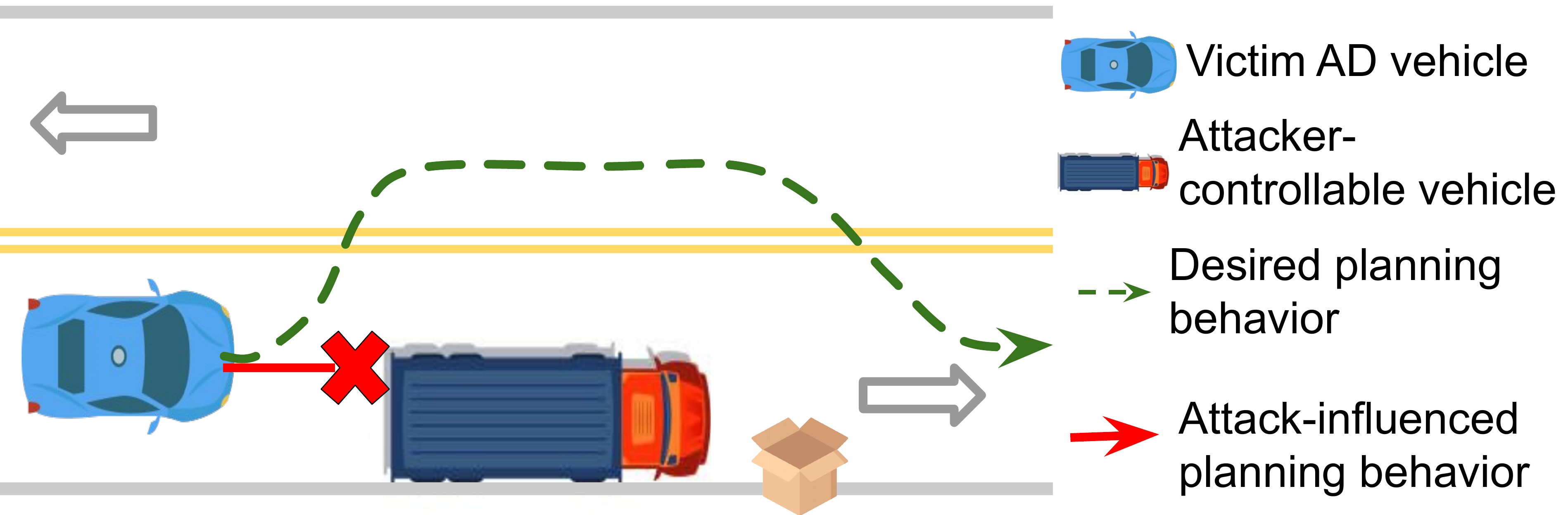}
    \caption{Attack scenario setup: A static box blocks the victim AD vehicle under a lane borrow scenario.} 
    \label{fig:app_v4_scenario}
\end{figure}

\textbf{Decision logic.}
This part is executed to make decision for lane borrow behavior, which happens in a single-lane road scenario and there is a blocking vehicle or other object blocks the current lane. As a result, the AD vehicle has to borrow the reverse lane to bypass the blocking object. The decision logic shown in Fig. \ref{fig:app_v4_code} is designed to recognize the pattern of traffic queue caused by checking whether the front vehicle is blocked by another one. If so, the decision logic thinks the front vehicle is temporarily blocked by another one and will move forward later. Since there exists a queue on the current lane and the AD vehicle should wait on the current lane. This is a very common scenario, for example, a line of vehicles wait statically due to a traffic congestion. In this case, the AD vehicle should not borrow the reverse lane since, which will potentially make the traffic congestion even worse. This part of decision logic is necessary since the lane borrow behavior is designed for non-movable obstacles which can be passed by borrowing the reverse lane, which should not be applied when there is a traffic queue.

\textbf{BP DoS vulnerability.}
Our testing system is able to discover a vulnerability in the aforementioned decision logic. The triggering condition is that if we put another static physical object in front of the blocking vehicle and with lateral intersection, the decision logic will mistakenly think there exists a traffic queue and the blocking vehicle is waiting for the newly-added object. As a result, the AD vehicle should wait statically until others objects to move. However, all of the blocking objects are static and they will not move. This will directly cause permanent stop of the AD vehicle.   

\cut{
The root cause behind this is the implementation error. We notice that in the code comment, the Apollo's developer clearly describe the function of this part of decision logic~\cite{is_perception_block_comment} is to estimate if an obstacle in certain range in front of ADV blocks too much space perception behind itself. If the decision is made separately for each obstacle, then clearly the variable \texttt{left\_most\_angle}, \texttt{right\_most\_angle}, \texttt{right\_most\_found} should updated for each obstacle. 
}

\textbf{Potential exploitation.} To exploit this vulnerability in real world, the attacker has to first create a lane borrow scenario. This requires a single-lane road and a blocking object on the lane. Since the current lane is blocked by the attacker, so the victim AD vehicle has no choice except borrowing the reverse lane. The attacker can control a vehicle parking on the lane with the emergence light on. This is very common in the real world, for example, a taxi is waiting for its passenger and a truck is unloading on the roadside. The attacker also has to control a static object in front of the blocking vehicle with lateral intersection to trigger the vulnerable logic and make the blocking vehicle movable. As a result, the victim AD vehicle will wait for the blocking vehicle to move and thus permanently stop. Note that there is no specific requirement for the second object, the only requirement is that it has to have lateral intersection with the blocking vehicle and has to be static. In the piratical attack scenario we create in Fig.~\ref{fig:app_v4_scenario}, the attacker can park the blocking vehicle slightly right so that a small part of the vehicle is off-road. Then the attacker can put a static object, such as an easy-to-carry cardboard box purely off-road to change the decision of the AD vehicle and make it stop permanently.   

%% file: Detailed_vuln/V3.tex
\begin{figure}
    \centering
\hrule
\begin{minted}[fontsize=\footnotesize,breaklines,linenos,escapeinside=||,mathescape=true, numbersep=5pt,xleftmargin=10pt]{python}
#scanning range centering at AD vehicle heading
search_range |$\leftarrow \pi$|
#length of scanning beam in meter
search_beam_length |$\leftarrow$| 20 
#resolution of the beam scanning
search_beam_intensity |$\leftarrow$| 0.08
#checking if perception is blocked by obstacle
block_angle_threshold |$\leftarrow$| 0.5
left_most_angle |$\leftarrow$| Normalize(adv_heading + 0.5 * search_range)
right_most_angle |$\leftarrow$| Normalize(adv_heading - 0.5 * search_range)
right_most_found |$\leftarrow$| false

#Iterate over the obstacle list
for (obs : obstacle_list):
  # Start beam scanning
  for (search_angle=0; search_angle<search_range; 
      search_angle+=search_beam_intensity):
    #Calc beam heading based on AD veh heading
    beam_heading |$\leftarrow$| adv_heading - 0.5 * search_range + search_angle
    #Check if obstacle has intersection with beam
    overlap |$\leftarrow$| obs.Overlap(adv_pos, beam_heading, beam_length)
    #Update right_most_angle when beam first encounter obs
    if (not(right_most_found) and overlap):
      right_most_found |$\leftarrow$| true
      right_most_angle |$\leftarrow$| beam_heading
    #Update left_most_angle if no longer overlap with obs
    if (right_most_found and not(overlap)):
      left_most_angle |$\leftarrow$| beam_heading - search_angle
      break
  if (not(right_most_found)):
    continue
  # Compare angle range with threshold to make decision
  if (abs(Normalize(left_most_angle - right_most_angle))>block_angle_threshold): 
    # Make decision whether the perception is blocked
    return true
# false means perception is blocked by an obstacle
return false
  
\end{minted}
\hrule
  \caption{Simplified vulnerable pseudo code of V4 (lane borrow + off-road objects) in Apollo. This is part of a function which estimates if an obstacle in front of AD vehicle blocks too much space of perception by beam scanning. }
  \label{fig:app_v3_code}
\end{figure}
\begin{figure}[htbp]
    \centering
    \includegraphics[width=0.95\linewidth]{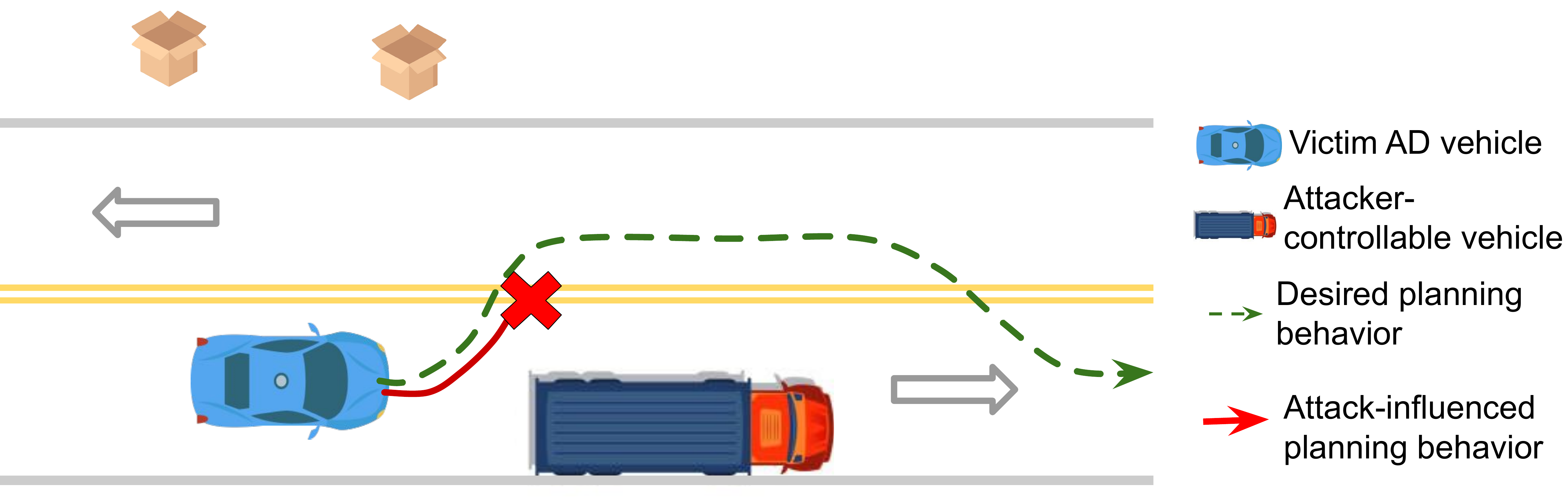}
    \caption{Attack scenario setup: Two static off-road boxes block the victim AD vehicle under a lane borrow scenario.} 
    \label{fig:app_v3_scenario}
\end{figure}

Fig. \ref{fig:app_v3_code} shows the simplified pseudo code for a BP DoS vulnerability our system discovered from the 5.0 version of Apollo.

\textbf{Decision logic.}
As shown in Fig.~\ref{fig:app_v3_code}, the overall goal of this code snippet is to calculate the blocked perception angle by each obstacles in front of the AD vehicle and determine whether the perception is blocked by an obstacle. Specifically, it maintains two variables \texttt{left\_most\_angle} and \texttt{right\_most\_angle}, which are resolved by beam scanning to calculate the angle range of blocked perception for a particular obstacle. The decision logic will uniformly sample the beam angle in front of the AD vehicle and check whether the obstacle has intersection with the beam on that direction. Once the beam first touches the obstacle, the \texttt{right\_most\_angle} will be updated (line 25). And when the beam no longer touches the obstacle, the \texttt{left\_most\_angle} will be updated (line 28). If the angle blocked by a certain obstacle is larger than the threshold \texttt{block\_angle\_threshold}, then the decision logic will come to the conclusion that the current perception is blocked by this obstacle. As a result, the AD vehicle will stop during the lane borrow since it may be dangerous to take an aggressive driving behavior (borrowing the reverse lane) when the perception is blocked.

\textbf{BP DoS vulnerability.}
Our testing system is able to discover a DoS vulnerability in the aforementioned decision logic. Fig. \ref{fig:app_v3_scenario} provides a malicious physical objects setup to trigger this vulnerability. The code snippet first performs the beam scanning for the right cardboard box and updates the two angle boundaries since the right box is in front of the AD vehicle. Note that the flag \texttt{right\_most\_found} is set to be true. While the code performs the beam scanning for the left cardboard box, line 28 will be executed immediately since \texttt{right\_most\_found} is true. As a result, the \texttt{left\_most\_angle} will be set to be \texttt{adv\_heading-0.5*search\_range}. When the decision is made on line 33, the difference between \texttt{left\_most\_angle} and \texttt{right\_most\_angle} will be large since one is on the most left side, while the other one is on the most right side. As a result, the decision logic make the decision that perception is blocked due to a static object even not in the search range. 

The root cause behind this is the implementation error. We notice that in the code comment, the Apollo's developer clearly describe the function of this part of decision logic~\cite{is_perception_block_comment} is to estimate if an obstacle in certain range in front of ADV blocks too much space perception behind itself. If the decision is made separately for each obstacle, then clearly the variable \texttt{left\_most\_angle}, \texttt{right\_most\_angle}, \texttt{right\_most\_found} should updated for each obstacle. 

\textbf{Potential exploitation.} To exploit this vulnerability in real world, the attacker can prepare two easy-to-carry static objects such as cardboard boxes, which are not too uncommon in road regions. The attacker needs to put the two objects in the off-road area and visible to the LiDAR and camera sensors of the AD vehicle. 

%% file: Detailed_vuln/V5.tex
\begin{figure}
    \centering
\hrule

\begin{minted}[fontsize=\footnotesize,breaklines,linenos,escapeinside=||,mathescape=true,numbersep=5pt,xleftmargin=8pt]{python}
# Blocking condition is set as AD vehicle's width
threshold |$\leftarrow$| adc_width
# Check if the obs is static
if (obs.is_static or obs.trajectory.empty):
  # Checks relationship between driving path & obs
  if(reference_line.IsBlock(obs.perception_box, threshold)):
    #Make decision that obs crosses driving path
    obs.is_path_cross |$\leftarrow$| true
\end{minted}
\hrule
  \caption{Simplified vulnerable pseudo code of V5 (lane borrow + standing pedestrian) in Apollo. This is part of the code used to check whether the AD vehicle should stop in front of a crosswalk due to the pedestrian.}
  \label{fig:app_v5_code}
\end{figure}

\begin{figure}[htbp]
    \centering
    \includegraphics[width=0.95\linewidth]{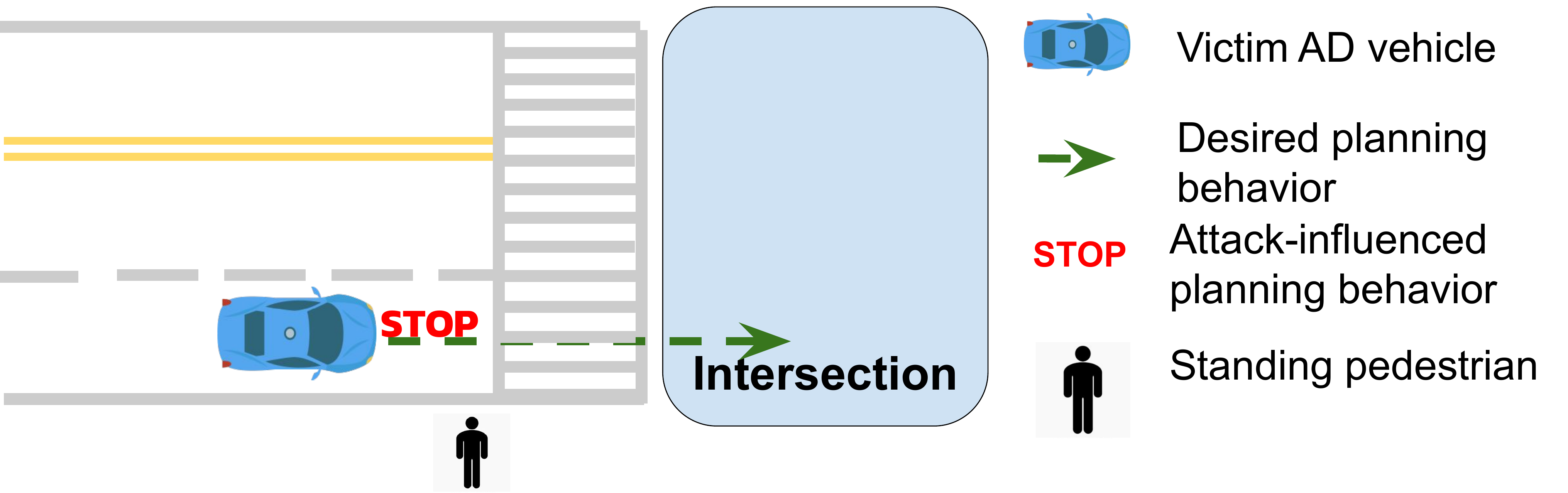}
    \caption{Attack scenario setup: An off-road standing pedestrian force the victim AD vehicle to stop in front of the crosswalk.} 
    \label{fig:app_v5_scenario}
\end{figure}

\textbf{Decision logic.}
As shown in Fig.~\ref{fig:app_v5_code}, the decision logic determines whether a certain obstacle overlaps with the driving path (or the driving reference line) on a crosswalk . Note that the AD vehicle will stop in front of the crosswalk if the \texttt{is\_path\_cross} flag is true for any pedestrian. The decision logic shown in Fig.~\ref{fig:app_v5_code} is designed specific for the static obstacles. Thus, line 4 skips all the dynamic obstacles. Line 6 calls the function \texttt{IsBlock} to determine the relationship between perception box of the obstacle and driving reference line. This function takes two arguments: the obstacle bounding box and a threshold. It will calculate the minimum distance between reference line and the bounding box and returns true if the minimum distance is smaller than the threshold.  

\textbf{BP DoS vulnerability.}
Our system discover a DoS vulnerability caused by a standing off-road pedestrian. The reason that the off-road pedestrian is that the threshold set on line 6 is overly-conservative such that a pedestrian off-road is still within the range. 

\textbf{Potential exploitation.} To exploit this vulnerability, the attacker needs to control a pedestrian which keep standing near a crosswalk. As a result, the victim AD vehicle will got stuck in front of the crosswalk. Note that crosswalk is very common in intersection, the victim AD vehicle could greatly block the normal traffic and violate the traffic rule~\cite{cvc22500} if exploited at an intersection.

%% file: Detailed_vuln/V6.tex
\begin{figure}
    \centering
\hrule

\begin{minted}[fontsize=\footnotesize,breaklines,linenos,escapeinside=||,mathescape=true,numbersep=5pt,xleftmargin=10pt]{python}
# A lateral distance configuration 
strict_l_distance |$\leftarrow$| 6 # unit: meter
# Compare lateral distance of obstacle with configured value
if (l_distance < strict_l_distance):
  # Get vector between positions of AD vehicle and obstacle
  obs_to_adc |$\leftarrow$| adc_path_point - obstacle.position
  # Compare inner product with a pre-defined threshold
  if (InnerProduct(obstacle.v, obs_to_adc) > 1e-6):
    # Make decision that pedestrian is moving towards AD veh
    stop |$\leftarrow$| true
  
\end{minted}
\hrule
  \caption{Simplified vulnerable pseudo code of V6 (intersection + walking pedestrian) in Apollo. This is part of the code used to check whether the AD vehicle should stop in front of a crosswalk due to the pedestrian. }
  \label{fig:app_v6_code}
\end{figure}

\begin{figure}[htbp]
    \centering
    \includegraphics[width=0.95\linewidth]{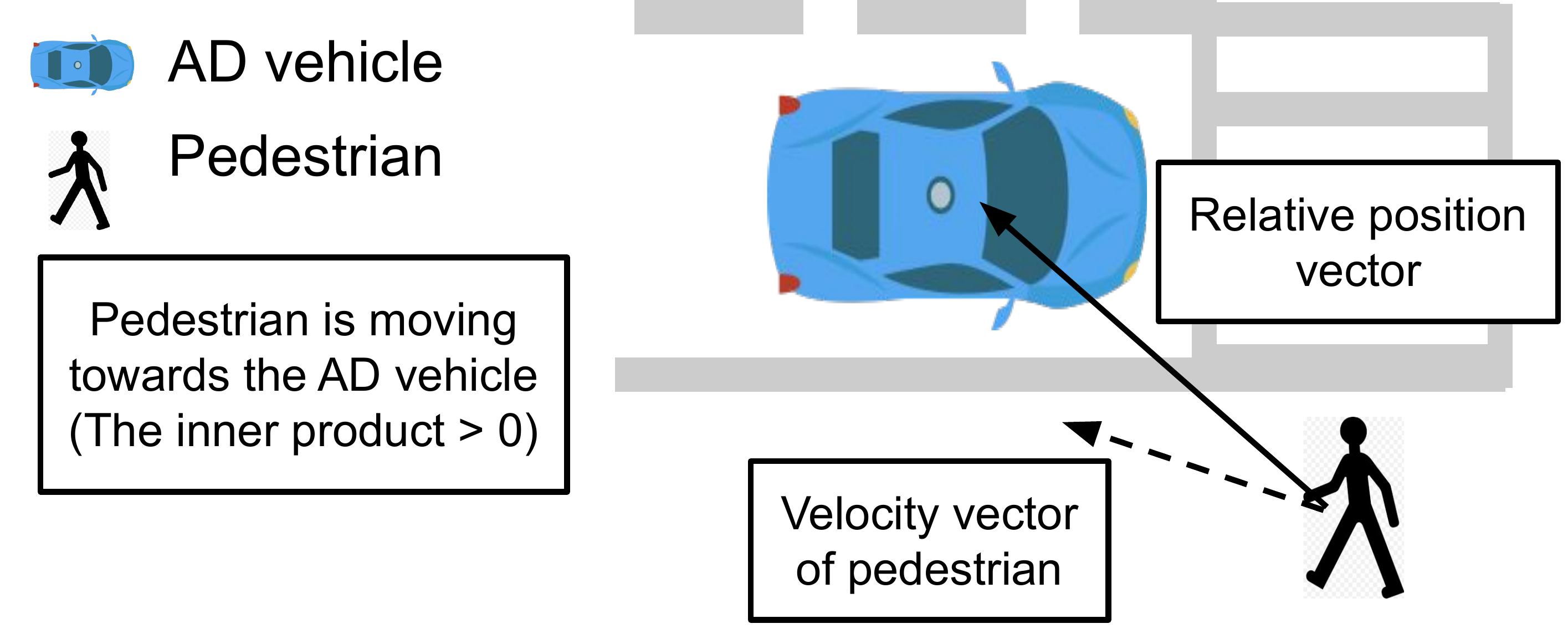}
    \caption{An explanation of how decision logic determines the pedestrian is moving towards the AD vehicle} 
    \label{fig:app_v6_explain}
\end{figure}

\begin{figure}[htbp]
    \centering
    \includegraphics[width=0.95\linewidth]{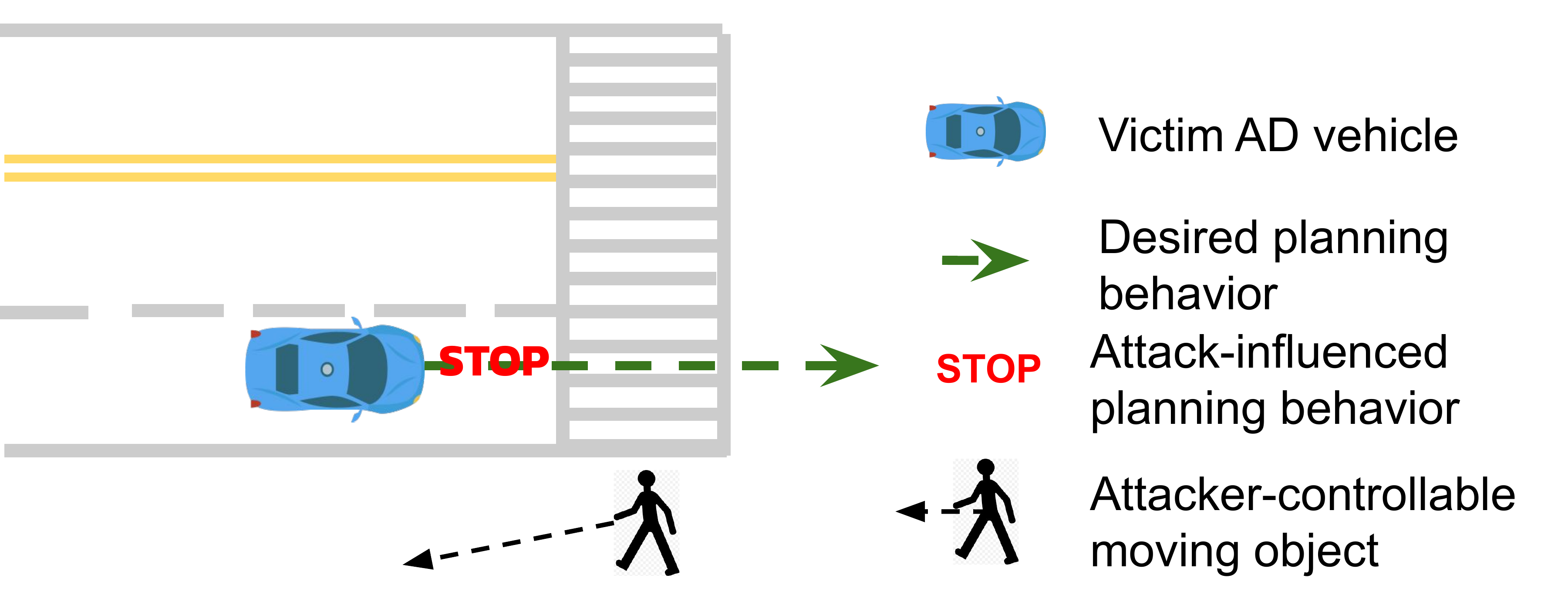}
    \caption{Attack scenario setup: An off-road moving pedestrian force the victim AD vehicle to stop in front of the crosswalk.} 
    \label{fig:app_v6_scenario}
\end{figure}

\textbf{Decision logic.}
As shown in Fig. \ref{fig:app_v6_code}, the AD vehicle will stop in front of the crosswalk if two requirements are satisfied: (1) the lateral distance between the obstacle and the driving reference line is within a predefined range, and (2) the pedestrian is moving towards the AD vehicle. Once the lateral distance requirement is satisfied, the decision logic calculates the inner product between the velocity of the pedestrian and its relative position from the AD vehicle to determine whether the pedestrian has a moving tendency towards the AD vehicle, as illustrated in Fig.~\ref{fig:app_v6_explain}. The principle behind this is that when the inner product calculated on line 8 is positive, the angle between the velocity and the relative position is an acute angle and when the pedestrian moves along its current velocity direction, the distance between the AD vehicle and the pedestrian will become closer. Thus, in this case, the AD vehicle decides to stop to yield the pedestrian.

\textbf{BP DoS vulnerability.}
Our system discovers a semantic DoS vulnerability in the above decision logic. When the AD vehicle is approaching to a crosswalk, a pedestrian who just passes the crosswalk and leaves the intersection along the road will make the victim AD vehicle to stop since the angle between the velocity and relative position is an acute angle. 

\textbf{Potential exploitation.}
To exploit this vulnerability in the real world, as shown in Fig.~\ref{fig:app_v6_scenario}, the attacker could keep a moving angle such that she is leaving the intersection while keeping a acute angle between vector vector and relative position vector. As a result, the AD vehicle will always make decision to stop in front of the crosswalk during this process. Also, the change of the angle is a dynamic process, the attacker may carefully choose a moment to change the angle to acute such that there is enough space for the victim AD vehicle to decelerate. The victim AD vehicle has to brake with the maximal deceleration, which will greatly damage the passengers' experience and potentially cause traffic accident, such as rear-end collision. Moreover, due to the overly conservative logic in Apollo, it will treat an unknown moving object as a pedestrian. A more interesting attack scenario is that the attacker may use multiple rovers, which can not be recognized by the current AD system and classified into unknown type to exploit this vulnerability. The attacker can carefully design the moving trajectory of the rovers and make sure there is always one rover can exploit this vulnerability. As a result, the victim AD vehicle will stop permanently.

%% file: Detailed_vuln/V7.tex
\begin{figure}
\centering
\hrule
\begin{minted}[fontsize=\footnotesize,breaklines, linenos,escapeinside=||,mathescape=true,numbersep=5pt,xleftmargin=10pt]{python}
# Iterate over all the obstacles
for (each obs : obstacle_list):
  # skip all the static objects
  if (obs.type!=BICYCLE and obs.type!=VEHICLE):
    continue
  # Get the closest lane within 5m. 
  obstacle_lane |$\leftarrow$| GetClosestLane(obs.position, 5.0)
  # Check if obstacle_lane is an associated
  # lane guarded by a stop sign
  if(obstacle_lane != NULL and obstacle_lane in asso_lanes):
    if (obs.distance_to_stop_line < max_stop_distance):
      #Add obstacle into watch list AD veh
      #has to wait until watch list is empty
      watch_list.add(obstacle_lane)
\end{minted}
\hrule
  \caption{Simplified vulnerable pseudo code of V7 (stop sign) in Apollo.}
  \label{fig:app_v7_code}
\end{figure}
\begin{figure}[htbp]
    \centering
    \includegraphics[width=0.95\linewidth]{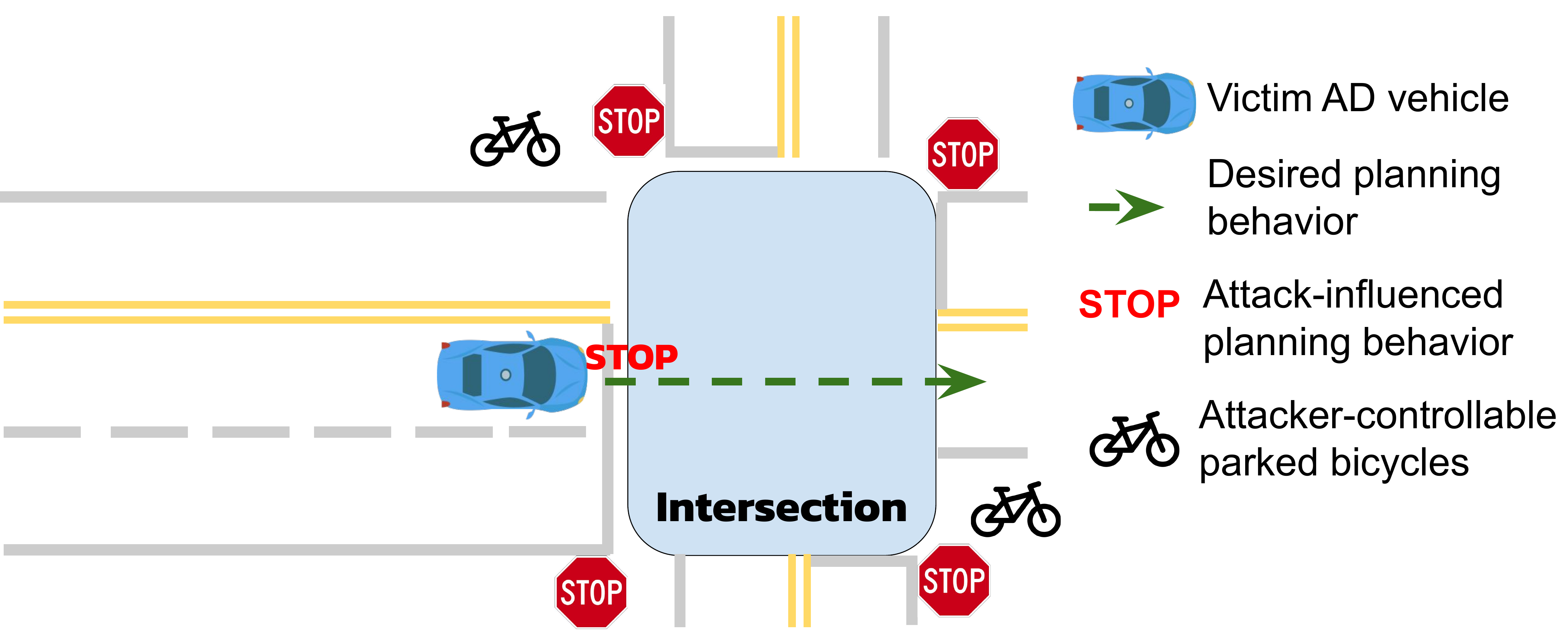}
    \caption{Attack scenario setup: Force the AD vehicle to permanently stop with 2 off-road static bicycles.} 
    \label{fig:app_v7_scenario}
\end{figure}

\textbf{Vulnerable code.} We presented the vulnerable code logic in Fig.~\ref{fig:app_v7_code}. Due to the traffic rule that a vehicle must wait until all the vehicles arrive at the intersection earlier have left, Apollo has to maintain a watch list. When the AD vehicle is approaching to an intersection with stop sign, an object will be put into the watch list if it is a dynamic object (line 4) and it is located on a lane guarded by a stop sign (line 10). Also, the object needs to stop in front of the stop line (line 11). After satisfying all the conditions above, that object will be added into the watch list. The problem is that on line 4, Apollo does not skip a static bicycle and on line 7, the distance buffer is too large such that off-road objects are also considered to be on the lane. As a result, a static bicycle without any cyclist off the road will be mistakenly put into the watch list. Apollo BP in fact has a timeout design, where the AD vehicle will still move forward when the watch list contains only one vehicle or bicycle and the timeout duration has passed. However, such design can be easily bypassed if the attacker puts two static bicycles off the road.

\textbf{Attack scenario setup.}
As shown in Fig.~\ref{fig:app_v7_scenario}, we design an attack scenario with a 4-way stop sign intersection. The attacker needs to put two static bicycles within 5m of an associated lane guarded by the stop sign. In our setup, we put two static bicycles to bypass the timeout design. As a result, the victim AD vehicle will stop in front of the stop line permanently. 

%% file: Detailed_vuln/V8.tex
\begin{figure}[tbh]
    \centering
\hrule
\begin{minted}[fontsize=\footnotesize,breaklines, linenos,escapeinside=||,mathescape=true,numbersep=5pt,xleftmargin=10pt]{python}
# Lateral threshold to decide the blocking object
critical_lateral_distance |$\leftarrow$| car_width/2 + 1.2f
# Longitudinal distance threshold to ignore a veh
minFollowingDistance |$\leftarrow$| 35.0f
for (each obs : obstacle_list)
  # Iterate all the candidate trajectories
  for (each traj : candidate_trajectories):
    # Iterate all the contour points of an object
    for (each point : obs.contourPoints):
      # Calculate the lateral/longitudinal distance 
      # between trajectory and contour point
      lateraldist|$\leftarrow$|LateralDistance(traj,point)
      longitudinaldist |$\leftarrow$| LongitudinalDistance(traj, point)
      # Check whether the trajectory is blocked
      if (lateraldist <= critical_lateral_distance  
          and longitudinaldist >= -car_length/1.5
          and longitudinaldist < minFollowingDistance):
        traj.isBlocked |$\leftarrow$| true
# Boolean to indicate if AD veh is fully blocked
FullyBlocked |$\leftarrow$| true
for (each traj: candidate_trajectories):
  if (traj.isBlocked == false):
    FullyBlocked |$\leftarrow$| false
\end{minted}
\hrule
  \caption{Simplified vulnerable pseudo code of V8 (lane follow) in Autoware.}
  \label{fig:app_v8_code}
\end{figure}

\textbf{Vulnerable Code.} We presented the pseudo code of the vulnerable planner logic in Autoware in Fig.~\ref{fig:app_v8_code}. This part of code is designed to handle static objects. Autoware generates a list of parallel candidate planning trajectories for the AD vehicle and check whether each trajectory is blocked by the static obstacle one by one.
Specifically, Line 9 iterates all the contour points of a physical objects to get precise relationship between the candidate trajectory and the physical object. $lateraldist$ and $longitudinaldist$ are calculated based on the current candidate trajectory and represent the lateral and longitudinal distances between the AD vehicle center and the contour point. On line 15, if the lateral distance is within the critical lateral distance range and it is ahead of the AD vehicle within the following distance, the object is considered as a blocking static object on line 18. 
After that, it then determines whether the AD vehicle is fully blocked by merging the decisions on the candidate trajectories (line 20--23). A fully blocking decision is made if all candidate trajectories are blocked by the static object.
We can exploit this logic by putting static objects off the road and block all the candidate trajectories since the $critical\_lateral\_distance$ is much larger than the half lane width~\cite{lane_width}. 

\textbf{Attack scenario setup} We create an attack scenario where the attacker put two cardboard boxes off the lane to block all the candidate trajectories of the victim AD vehicle. As a result, the victim AD vehicle stops in the middle of the lane permanently. The permanent stop block the normal traffic operations and may directly violate the traffic rules. The improper stop can further cause severe consequences if this vulnerability is triggered inside a tunnel or in front of a fire station~\cite{cvc22500, calihandbook}.

%% file: Detailed_vuln/V9.tex
\begin{figure}
    \centering
\hrule
\begin{minted}[fontsize=\footnotesize,breaklines, linenos,escapeinside=||,mathescape=true,numbersep=5pt,xleftmargin=10pt]{python}
#Pre-defined lateral distance threshold
critical_lateral_d=AD_half_width+1.2f #unit: meter
#Iterate over each point in predicted obj trajectory
for (pred_point : pred_traj):
  #Iterate over each point in planning path of AD veh
  for (path_point : path):
    #Compute distance between two points
    collision_distance |$\leftarrow$| distance(pred_point, path_point)
    #Compare distance with pre-defined threshold
    if (collision_distance < critical_distance):
      #Make decision: candidate trajectory is blocked
      blocked |$\leftarrow$| true
\end{minted}
\hrule
  \caption{Simplified vulnerable pseudo code of V9 (lane following + dynamic vehicles) in Autoware. This is part of the code used to check whether current lane is blocked by dynamic objects. }
  \label{fig:app_v9_code}
\end{figure}
\begin{figure}[htbp]
    \centering
    \includegraphics[width=0.95\linewidth]{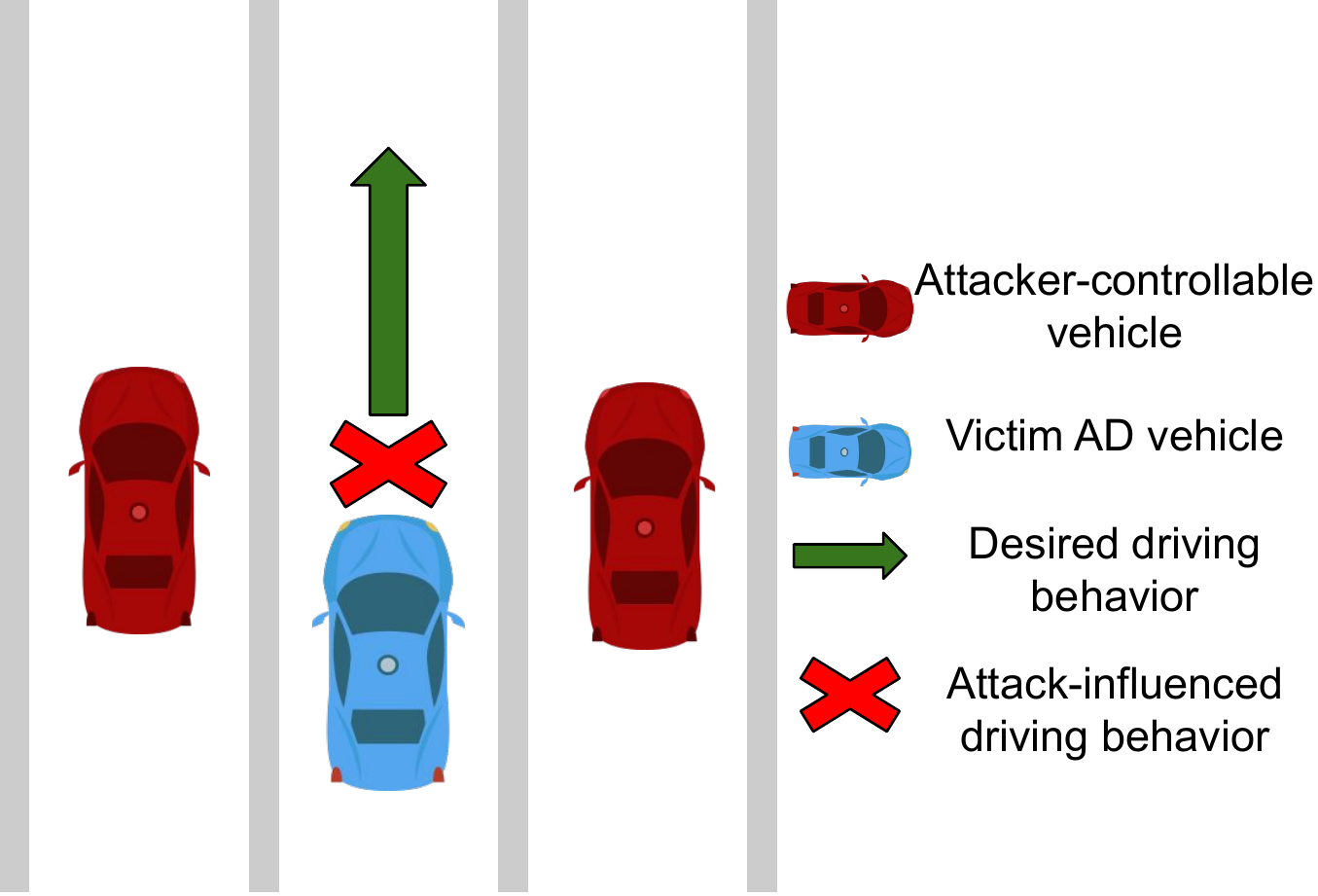}
    \caption{Attack scenario setup: Force the AD vehicle to sharp brake with 2 driving vehicles on adjacent lanes.} 
    \label{fig:app_v9_scenario}
\end{figure}

\textbf{Decision logic.}
As shown in Fig.~\ref{fig:app_v9_code}, the decision logic is designed to check whether a candidate planning path will be blocked by a dynamic physical object with the prediction trajectory \texttt{pred\_traj}. Similar to the decision logic in Fig.~\ref{fig:app_v8_code}, it makes decision whether the current lane is blocking by checking whether all the candidate planning trajectories are blocked. Specifically, on line 8, \texttt{collision\_distance} is calculated between each point on the prediction trajectory of the physical object and the point on planning path. Once this distance is smaller than the pre-defined threshold \texttt{critical\_lateral\_d}, the decision logic will consider this as a potential collision point and makes the decision that the planning path is blocked. 

\textbf{BP DoS vulnerability.}
Our system discover a DoS vulnerability caused by driving vehicles on adjacent lanes. The reason that vehicles on adjacent lanes could change the desired planning behavior into an overly-conservative decision is the large lateral distance threshold defined on line 2. Given that the common urban lane width is smaller 3.6m~\cite{lane_width}, the vehicles on adjacent lane can easily be within the range defined by the lateral threshold and thus affect the driving decision.  

\textbf{Potential exploitation.} As shown in Fig.~\ref{fig:app_v9_scenario}, a possible exploitation scenario is that the attacker controls two vehicles driving on the two adjacent lanes near the victim AD vehicle. The attacker can carefully choose the timing such that both the attacker-controllable vehicles are observed by the victim AD vehicle at the same time and within the blocking decision range. For example, two attacker-controllable vehicles can accelerate and appear in front of the victim AD vehicle at the same time. As a result, the victim AD vehicle will come to the conclusion that all the candidate trajectories are blocked by the two vehicles and it should decelerate to avoid collision. This will lead to a sharp brake, which will greatly damage the passengers' experience and cause potential rear-end collision.

\cut{
\subsection{Performance of BPFuzz}
We evaluate the performance numbers of BPFuzz. We measure the time to generate/mutate and test 10000 inputs on lane following scenarios. The time used by input generation and genetic algorithm is 131s while the time used for planning execution is 1249s.
}